\documentclass{aa}
\pdfoutput=1
\usepackage[varg]{txfonts}
\usepackage{graphicx}
\usepackage{soul}
\usepackage{ifthen}
\usepackage{natbib}
\bibpunct{(}{)}{;}{a}{}{,}
\usepackage{hyperref}
\hypersetup{colorlinks=true,citecolor=blue,linkcolor=blue,urlcolor=blue}
\usepackage{url}
\usepackage{amsmath}
\usepackage{amssymb}
\usepackage{bm}
\usepackage{amsfonts}
\usepackage{enumitem}
\usepackage{tabularx}
\usepackage{multirow}
\def\be{\begin{equation}}
\def\ee{\end{equation}}
\def\bea{\begin{eqnarray}}
\def\eea{\end{eqnarray}}

\defcitealias{Planck.Cosmo.2016}{Planck Collaboration XIII 2016}
\defcitealias{Planck.DE.2016}{Planck Collaboration XIV 2016}
\defcitealias{LSST.2009}{LSST Science Collaboration 2009}
\defcitealias{DES.WL.2017}{DES Collaboration 2017}
\defcitealias{BOSS.2014}{Anderson et al. 2014}
\defcitealias{BKG.etal.2014}{Betoule et al. 2014}
\defcitealias{KIDS.WL.2017}{Hildebrandt et al. 2017}
\defcitealias{CFHTLenS.WL.2013}{Heymans et al. 2013}

\begin{document}

\title{Breaking degeneracies in modified gravity with higher (than 2nd) order weak-lensing statistics}
\titlerunning{Breaking degeneracies in MG with higher-order WL statistics}

\author{Austin Peel\inst{1}
   \and Valeria Pettorino\inst{1}
   \and Carlo Giocoli\inst{2,3,4}
   \and Jean-Luc Starck\inst{1}
   \and Marco Baldi\inst{2,3,4}}
\authorrunning{A. Peel et al.}

\institute{AIM, CEA, CNRS, Universit{\'e} Paris-Saclay, Universit{\'e} Paris Diderot, 
           Sorbonne Paris Cit{\'e}, F-91191 Gif-sur-Yvette, France\\
           \email{austin.peel@cea.fr}
      \and Dipartimento di Fisica e Astronomia, Alma Mater 
           Studiorum Universit\`{a} di Bologna, via Gobetti 
           93/1, 40129 Bologna, Italy
      \and Astrophysics and Space Science Observatory Bologna, 
           via Gobetti 93/2, 40129, Bologna, Italy
      \and INFN - Sezione di Bologna, viale Berti Pichat 6/2, 
           40127, Bologna, Italy}

\date{Received 23 May 2018 / Accepted 31 July 2018}

\abstract{General relativity (GR) has been well tested up to solar system scales, but it is much less certain that standard gravity remains an accurate description on the largest, that is, cosmological, scales. Many extensions to GR have been studied that are not yet ruled out by the data, including by that of the recent direct gravitational wave detections. Degeneracies among the standard model ($\Lambda$CDM) and modified gravity (MG) models, as well as among different MG parameters, must be addressed in order to best exploit information from current and future surveys and to unveil the nature of dark energy. We propose various higher-order statistics in the weak-lensing signal as a new set of observables able to break degeneracies between massive neutrinos and MG parameters. We have tested our methodology on so-called $f(R)$ models, which constitute a class of viable models that can explain the accelerated universal expansion by a modification of the fundamental gravitational interaction. We have explored a range of these models that still fit current observations at the background and linear level, and we show using numerical simulations that certain models which include massive neutrinos are able to mimic $\Lambda$CDM in terms of the 3D power spectrum of matter density fluctuations. We find that depending on the redshift and angular scale of observation, non-Gaussian information accessed by higher-order weak-lensing statistics can be used to break the degeneracy between $f(R)$ models and $\Lambda$CDM. In particular, peak counts computed in aperture mass maps outperform third- and fourth-order moments.}

\keywords{dark energy -- gravitation -- gravitational lensing: weak}
\maketitle

\section{Introduction}\label{sec:intro}
Current observations from the cosmic microwave background (CMB) anisotropies, baryon acoustic oscillations, and supernova luminosity distances all support a phase of accelerated cosmic expansion during the present epoch \citepalias[e.g.][]{Planck.Cosmo.2016,Planck.DE.2016,BOSS.2014,BKG.etal.2014}. The cause of such evolution is still unknown and should amount to approximately 70\% of the total energy density of the Universe. Whether it is simply a fundamental constant of nature, a physical fluid, or instead a modification of general relativity (GR) itself at cosmological scales, the accelerated expansion remains one of the biggest puzzles in modern cosmology. 

The simplest cosmological model consistent with the data assumes a constant contribution ($\Lambda$) and a cold dark matter component (CDM). The cosmic expansion of a $\Lambda$CDM model, however, can be mimicked by vast range of models, which are also in agreement with current observations. These scenarios typically assume either a fluid component, in other words, dark energy (DE), or a modification of general relativity at large scales. Even when considering linear perturbations and the impact on CMB, dark energy and modified gravity (MG) models are still viable \citepalias{Planck.DE.2016} and may be advocated to solve tensions currently present between Planck data and late time probes, such as weak lensing \citepalias{KIDS.WL.2017,DES.WL.2017} and, in smaller measure, redshift space distortions \citepalias{Planck.DE.2016}.

The recent direct detection of gravitational waves (GW) with an electromagnetic counterpart has contributed to the exclusion of a range of MG models in the space of general Horndeski scalar-tensor theories \citep{LT.2016,LL.2017,CV.2017,BBF.etal.2017,EZ.2017,SJ.2017}. Using measurements in \cite{Sakstein.2015}, \citet{DV.2018} put further constraints on models beyond Horndeski, which make use of the Vainshtein mechanism to protect high density regions where GR is well tested. A large range of models is still viable, however, including those discussed in \citetalias{Planck.DE.2016} and in particular scalar-tensor theories with a universal coupling, that is the $f(R)$ models discussed in this paper. We refer to \citet{CK.2017} (and to references above) for a general formulation of viable models which assume a universal coupling. We note also that theories such as coupled DE \citep{Amendola.2000,PB.2008,Pettorino.2013} are still viable after GW constraints. As they do not introduce derivative couplings in the action, they therefore do not modify the speed of GWs in tensor equations; see for example \citet[][]{BEH.etal.2017}, where terms leading to anomalous speeds have been identified.

Distinguishing between $\Lambda$CDM and MG scenarios that mimic $\Lambda$CDM at the linear level, but which may be different at non-linear scales, is the next challenge for future galaxy surveys. In this paper, we investigate whether various observables based on statistics of the weak-lensing signal can be used to discriminate among different theoretical models that are strongly degenerate in the matter power spectrum on linear and possibly also on non-linear scales. 
In particular, we focus on the strong degeneracy between $f(R)$ modified gravity and $\Lambda $CDM that occurs when the latter models include the effects of massive neutrinos \citep[see][]{MSY.2013,He.2013,BVNV.etal.2014,WWK.2017}. Given such a degeneracy, we are motivated to explore non-Gaussian weak-lensing probes that access higher-order information than two-point correlation functions or their associated power spectra. Our goal is to determine which statistics are most promising and how the discrimination efficiency between models depends on redshift and angular scale.

Within $\Lambda$CDM, there is evidence that weak lensing alone can break the degeneracy between $\sigma_8$ and $\Omega_\mathrm{m}$ by a combination of higher-order (than second) convergence moments and shear tomography \citep{VCM.etal.2018}. In terms of MG, \citet{LLZ.etal.2016} have shown that non-Gaussian statistics, in particular weak-lensing peak counts, contain significant cosmological information and can constrain $f(R)$ model parameters. In addition, \citet{HS.2016,SNL.etal.2017} have studied the effects of $f(R)$ gravity on statistical properties of the weak-lensing field, including the convergence power spectrum and bispectrum, peak counts, and Minkowski functionals. Unlike these analyses, the matter power spectrum degeneracy in our case is facilitated (and the constraining power of WL observation is challenged) by the presence of massive neutrinos, a component known to play an important role during structure formation in the real universe.

The deflection of light coming from distant galaxies represents an important tool to indirectly map the projected matter density distribution along the line of sight. When light bundles travel from distant galaxies to an observer, they are continuously deflected by the inhomogeneities of the density field of the intervening non-linear structures. As a consequence, images of background galaxies appear slightly stretched and distorted, the phenomenon known as weak gravitational lensing. Considering that by definition the weak-lensing effect is small, it is necessary to average over a large number of background sources in order to quantify it. Future wide-field experiments like LSST \citepalias{LSST.2009} and the ESA space mission Euclid\footnote{\href{https://www.euclid-ec.org}{https://www.euclid-ec.org}} \citep{Euclid.2011} will both increase the number and redshift range of the background source population used to measure the weak-lensing signal.

Given vastly larger datasets and substantially improved image quality, next-generation surveys will thus permit the most stringent tests yet on a variety of different theoretical scenarios \citep{Euclid.2016}. In these tests, weak lensing will be a primary tool in our effort to understand the dark universe, and in particular the nature of the late-time acceleration phase. In this work, we present the first high-order moment analysis of weak-lensing fields of non-standard MG cosmologies with massive neutrinos. Our aim is to quantify the capability of future galaxy surveys to distinguish non-standard cosmological features using weak lensing.

The paper is organised as follows. In Sect.~\ref{sec:theory} we describe the theoretical cosmologies used in this paper as a proof of concept of the methodology we propose. In Sect.~\ref{subsec:weaklensing} we briefly recall the basic weak lensing definitions that will be needed in the analysis. In Sect.~\ref{sec:sims} we discuss the $N$-body simulations used for the analysis described in Sect.~\ref{sec:analysis}. Our results are presented and discussed in Sect.~\ref{sec:results}, and we draw our conclusions in Sect.~\ref{sec:conclusion}.

\section{\texorpdfstring{$\lowercase{f}(R)$}{f(R)} cosmology: The Hu-Sawicki model}\label{sec:theory}
Among the classes of theories that modify gravitational attraction at large scales in order to have cosmic acceleration, a large class of (still viable) models is given by $f(R)$ theories. In these cosmologies, the dependence of the action on the Ricci scalar $R$ is modified with respect to GR, and depends on a generic function $f$ of $R$, such that:
\be\label{eq:action_def}
S=\int d^4x \sqrt{-g} \left[\frac{R +f(R)}{2 \kappa_G^2} + {\cal L}_{\rm m}\right],
\ee
where ${\cal L}_{\rm m}$ is the matter Lagrangian and $\kappa_G^2 \equiv 8 \pi G$. (We use a  subscript $G$ to avoid confusion with the weak lensing convergence $\kappa$.) It is possible to show that these theories are a subclass of scalar-tensor theories, meaning they include a universal fifth force mediated by a scalar field, that adds to the metric tensor already present in GR. Since this force is universal, it affects baryons as well as dark matter: as a consequence, $f(R)$ theories typically need some screening mechanism that restores GR in high density regions, such as in the solar system, where observations show no significant deviation from GR. In addition, the choice of $f(R)$ should provide a cosmology that is close enough to $\Lambda$CDM in the high-redshift regime (or else it would affect the CMB in a way that would have already been observed) and provide cosmic acceleration at late times.

One of the few known functional forms of $f(R)$ able to satisfy solar system constraints is the \cite{HS.2007} model in which
\be\label{eqn:HuSawicki}
f(R) \equiv -m^2 \frac{c_1(R/m^2)^n}{c_2(R/m^2)^n+1} \qquad \mathrm{for} \quad n > 0,
\ee
and 
\be\label{eqn:HuSawicki2} m^2\equiv \frac{\kappa_G^2\, \bar{\rho}_0}{3}=(8315\,{\rm Mpc})^{-2}\left(\frac{\Omega_\mathrm{m} h^2}{0.13}\right),
\ee
where $\bar{\rho}_0$ is the average density at present time, and $c_1$ and $c_2$ are dimensionless parameters. 
The sign of $f(R)$ is chosen such that its second derivative $f_{\rm RR}$ is positive for $R \gg m^2$, leading to stable solutions in the high curvature regime \citep{HS.2007}. 
It has been shown that for 
\be
\frac{c_1}{c_2} \approx 6 \frac{\Omega_\Lambda}{\Omega_\mathrm{m}},
\ee
this model mimics the background evolution of a $\Lambda$CDM model with a cosmological constant relative density of $\Omega_\Lambda$ and a relative matter density of $\Omega_\mathrm{m}$. In the \cite{HS.2007} model however, there is no cosmological constant and cosmic acceleration is given by the modification of gravity. 
Furthermore, the parameter $c_2$ is usually expressed in terms of $f_{R0} \equiv d f / dR (z = 0)$.

In these cosmologies, therefore, two additional parameters are present: $n$ and $f_{R0}$. This set of functional forms are such that $f(R)/m^2$ have a transition from zero (for $R \rightarrow 0$) to a constant, for $R \gg m^2$. The sharpness of this transition increases with n, while $f_{R0}$ determines when the transition occurs. A detailed analysis of the impact of such cosmologies at different scales has been presented already in the proposal paper of \cite{HS.2007} and as well in more recent papers \citep{Koyama.2016,Lombriser.2014}. In \cite{HS.2007} (see their Fig.~9), for solar system to galaxy scales, assuming a galactic density $\rho_g = 10^{-24}$ g cm$^{-3}$, the two parameters of the theory have to satisfy
\be 
|f_{R0}| < 74\, (1.23 \times 10^6)^{n-1} \left[\frac{R_0}{m^2}\frac{\Omega_\mathrm{m} h^2}{0.13}\right]^{-(n+1)}.
\ee

\cite{HS.2007} models have also been shown to satisfy background observations \citep{MMA.2009} and have been tested against some cosmological constraints \citep[e.g.][]{LSB.etal.2012,BGM.etal.2014,HRR.etal.2016}. (See \citet{Koyama.2016} and \citet{Lombriser.2014} for an overview and detailed list of constraints.) CMB and large-scale structure provide relatively weak constraints, giving $|f_{R0}|< 10^{-2}$ to $10^{-4}$. Solar system constraints typically require $|f_{R0}|< 10^{-4}$ to $10^{-6}$, depending on environmental assumptions of the Milky Way, while galaxy clusters give the slightly stronger bound of $|f_{R0}|< 10^{-5}$. Strong lenses \citep{Smith.2009} or dwarf galaxies and Cepheids may reach $|f_{R0}|< 10^{-6}$ to $10^{-7}$  \citep[][]{JVS.2013,VSD.etal.2014}, with bounds that depend, however, on several assumptions and approximations, for example the shape of density profiles. \cite{HS.2007} cosmologies automatically satisfy constraints recently derived by \citep{Sakstein.2015} for any choice of the parameters.

\begin{figure}
    \includegraphics[width=\columnwidth]{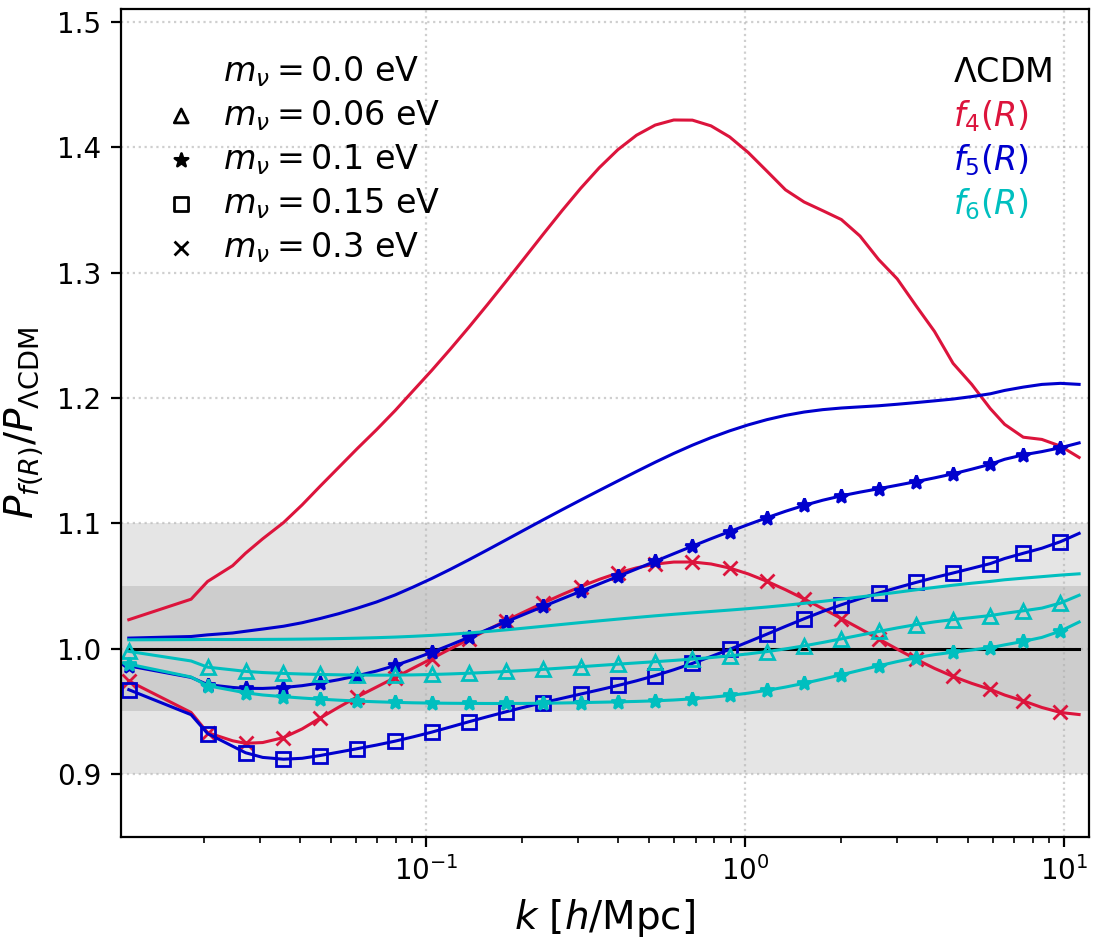}
    \caption{Ratio of the matter power spectra $P(k)$ of different $f(R)$ models, including neutrinos of different mass, with respect to $\Lambda$CDM. The different $f(R)$ models, labelled as $f_4(R)$, $f_5(R)$, and $f_6(R)$, are described in the text. $f_6(R)$ with any neutrino mass and $f_5(R)$ with $m_{\nu}=0.15~\mathrm{eV}$ all have a matter power spectrum within 10\% of the $\Lambda$CDM one. In the following we illustrate how different statistics can be used (or not) to discriminate these models, showing complete results in particular for $f_5(R)$ for illustration.}
    \label{fig:power_ratio}
\end{figure}

In the following we fix $n=1$ and consider different values of $f_{R0}$, labelled as follows.
\begin{itemize}[leftmargin=*]
  \item[]  $f_4(R)$ (or equivalently fR4) labels $f(R)$ with $f_{R0}=-10^{-4}$,
  \item[] $f_5(R)$ (or equivalently fR5) labels $f(R)$ with $f_{R0}=-10^{-5}$,
  \item[] $f_6(R)$ (or equivalently fR6) labels $f(R)$ with $f_{R0}=-10^{-6}$.
\end{itemize}
This choice of range allows us to check typical values used in the literature for $f(R)$ models. A model like $f_4(R)$ may be at the more extreme limit of solar system observations, possibly requiring further screening at those scales, whereas the other models may instead be more in tension with strong lenses, dwarf galaxies and Cepheids, depending on assumptions in deriving the constraints. Assuming that a screening mechanism is in place, the values chosen here may still be viable at cosmological scales, where much weaker bounds are in place. 

The ratios of the power spectra for these models with respect to $\Lambda$CDM are shown in Fig.~\ref{fig:power_ratio}. As we see from the figure, the inclusion of neutrinos suppresses the matter power spectrum on scales smaller than their free streaming length. This suppression is degenerate with the value of $f_{R0}$, which also introduces a transition scale: the higher the value, the more the linear perturbations grow below the Compton wavelength $\lambda_{f_R}$ and the more the power spectrum is enhanced. The connection between $\lambda_{f_R}$ and $f_{R0}$ can be seen directly by defining $\lambda_{f_R} \equiv m^{-1}_{fR}$, where
\be 
m^2_{f_R} \equiv \frac{\partial^2 V_\mathrm{eff}}{\partial f^2_R} = \frac{1}{3}\left(\frac{1+f_R}{f_{RR}} - R\right),
\ee
and again $f_R$ and $f_{RR}$ are, respectively, the first and second derivatives of $f$ with respect to $R$. In this expression, the effective potential $V_{\rm eff}$ is defined in such a way that the equation of motion for $f_{\rm R}$ can be written as $\Box f_{\rm R} \equiv \frac{\partial V_{\rm eff}}{\partial f_{\rm R}}$; specifically, this happens for $\frac{\partial V_{\rm eff}}{\partial f_{\rm R}} = \frac{1}{3} (R-f_{\rm R} R) +2f = \kappa_{\rm G}^2 \rho$, where $\rho$ is the total energy density. Values of $\lambda_{f_R}$ large enough to be in the non-linear regime may therefore lead to effects which are observable in structure formation.

Models such as $f_4(R)$ without massless neutrinos would lead to matter power spectra that are strongly amplified with respect to $\Lambda$CDM (solid red curve in Fig.~\ref{fig:power_ratio}). On the other hand, the inclusion of massive neutrinos as large as 0.3~eV (crossed red line) would compensate such a choice of $f(R)$, bringing it back to within 10\% from $\Lambda$CDM even down to the smallest scales \citep[see][for an early investigation of such degeneracy in the non-linear regime]{BVNV.etal.2014}. 

The degeneracy between $f_{R0}$ and neutrino masses leads similar choices to have matter power spectra within about 10\% from $\Lambda$CDM, making it difficult to discriminate between $f(R)$ and a cosmological constant, or between different $f(R)$ models, purely on the basis of observations of second order statistics. In the following, we will investigate whether higher-order weak-lensing observables can help to break such degeneracies. Although we have tested the results for different choices of $(f_{R0}, m_{\nu})$, we will show results for the $f_5(R)$ use case, which is an intermediate one among the ones shown in the figure and available in simulations.

\section{Weak lensing}\label{subsec:weaklensing}
Weak gravitational lensing describes small distortions in the observed shapes of distant galaxies induced by the gravitational fields of massive structures in the universe along the line of sight. Significant information about the distribution and masses of cosmic structures is imprinted on galaxy images as the statistical alignment of their observed ellipticities. This signal has been used successfully in the data analysis of numerous galaxy surveys to constrain cosmological parameters \citepalias[e.g.][]{CFHTLenS.WL.2013,KIDS.WL.2017,DES.WL.2017} and is potentially a key probe to test gravity and models beyond $\Lambda$CDM in future weak lensing surveys such as Euclid \citep{Euclid.2016}. For this purpose, the combination of weak lensing with other probes, such as galaxy clustering \citepalias{Planck.DE.2016}, helps to break degeneracies in measuring the gravitational potentials. On the other hand, it is worth investigating whether one can break degeneracies in parameter space relying on weak lensing only, independently of other late-time probes. This is an important test that can help maximise the information we extract from the large amount of data that is going to be available for this probe. Furthermore, it can help identify whether tensions present in the data \citepalias{Planck.DE.2016} between weak lensing and the cosmic microwave background will point to a signature of new physics or rather to systematics.

The basic weak lensing quantities are the shear $\boldsymbol{\gamma}(\boldsymbol{\theta})$ and convergence $\kappa(\boldsymbol{\theta})$ fields. Shear is a spin-2 field that quantifies the anisotropic distortion of source galaxy images due to the tidal gravitational fields of foreground structures. The shear measured at some position $\boldsymbol{\theta}$ on the sky depends on the amplitude of density fluctuations along the line of sight to the source galaxy, as well as on the relative distances between the observer, deflectors, and source. Convergence, a scalar field, is derivable from the shear up to a constant \citep{KS.1993} and reflects isotropic changes in the observed shapes of galaxies. 

The connection with cosmic density fluctuations is straightforward with $\kappa$, as it can be interpreted as the projected total matter density along the line of sight:
\be\label{eqn:kappa_integral}
\kappa(\boldsymbol{\theta},\chi)=\frac{3H_0^2\Omega_\mathrm{m}}{2c^2}\int_0^{\chi}\,\mathrm{d}\chi'\,\frac{f_K(\chi')f_K(\chi-\chi')}{f_K(\chi)}\frac{\delta(f_K(\chi')\boldsymbol{\theta},\chi')}{a(\chi')}.
\ee
In this expression, $H_0$ is the present-day Hubble parameter, $c$ is the speed of light, $a$ is the universal scale factor, and $\delta$ is the 3D density contrast defined as $\delta=(\rho-\bar{\rho})/\bar{\rho}$, where $\bar{\rho}$ is the spatially averaged matter density. The radial coordinate $\chi$ is comoving, and the geometrical function $f_K$ determines the comoving angular distance, whose form depends on the (constant) curvature of 3D space. We refer to \citet{Kilbinger.2014}, for example, for a recent review of weak lensing cosmology that includes an introduction to the formalism. We deal directly with $\kappa$ maps in this work as derived from our simulated cosmologies (cf. Sect.~\ref{subsec:mapmaking}).

\section{Numerical setup}
\label{sec:sims}
\subsection{The DUSTGRAIN-{\em pathfinder} simulations}
We make use of the {\small DUSTGRAIN}-{\em pathfinder} simulations (see \citet{GBM.2018} for a detailed description), a suite of cosmological collisionless simulations specifically designed to sample the joint parameter space of $f(R)$ gravity and massive neutrino cosmologies in order to investigate their main observational degeneracies and to devise strategies to break them. These simulations followed the evolution of $768^{3}$ dark matter particles of mass $m^p_{\rm CDM}= 8.1\times 10^{10}$ M$_{\odot}/h$ (for the case of $m_{\nu}=0$) and of as many neutrino particles (for the case of $m_{\nu}>0$) within a periodic cosmological box of $750$ Mpc$/h$ per side, under the effect of an $f(R)$ gravitational interaction defined by Eq.~(\ref{eq:action_def}).

The simulations were performed with the {\small MG-Gadget} code \citep[][]{PBS.2013}, which is a modified version of the {\small GADGET} code \citep[][]{Springel.2005} that implements the extra force and the chameleon screening mechanism \citep[see][]{KW.2004} characterising $f(R)$ gravity theories. As was already done in \citet{BVNV.etal.2014}, such an MG solver was combined with the particle-based implementation of massive neutrinos developed for {\small GADGET} by \citet{VHS.2010}. Therefore, massive neutrinos were included in the simulations as a separate family of particles with its specific transfer function and velocity distribution, so that both CDM and neutrino particles contributed to the density source term that enters the calculation of the $f_{R}$ extra degree of freedom.

Initial conditions were produced by generating two separate but fully correlated random realisations of the linear density power spectrum for CDM and massive neutrino particles as computed by the linear Boltzmann code {\small CAMB} \citep[][]{CAMB} at the starting redshift of the simulation $z_{i}=99$. Following the approach of, for example, \citet{ZBVN.etal.2017} and \citet{VNBD.etal.2017}, we then computed the scale-dependent growth rate $D^{+}(z_{i},k)$ for the neutrino component in order to correctly account for neutrino gravitational velocities. Apart from these, neutrino particles also received an additional thermal velocity extracted from the neutrino momentum distribution for each value of neutrino mass under consideration.

In the present work, we use a subset of the full {\small DUSTGRAIN}-{\em pathfinder} runs consisting of nine simulations whose parameters are summarised in Table \ref{tab:sims}. All simulations share the same standard cosmological parameters, which are set in accordance with the Planck 2015 constraints \citepalias{Planck.Cosmo.2016}, namely $\Omega_\mathrm{m}=\Omega_\mathrm{CDM}+\Omega_\mathrm{b}+\Omega_\nu = 0.31345$, $\Omega_\mathrm{b}=0.0481$, $\Omega_\Lambda= 0.68655$, $H_{0}= 67.31$ km s$^{-1}$ Mpc$^{-1}$, ${\cal{A}}_\mathrm{s}= 2.199\times 10^{-9}$, and $n_\mathrm{s}=0.9658$.

\begin{table*}
\caption{The subset of the {\small DUSTGRAIN}-{\em pathfinder} simulations considered in this work with their specific parameters.}
\centering
{\renewcommand{\arraystretch}{1.35}
\begin{tabular}{lccccccc}
\hline\hline
Simulation Name & Gravity type & $f_{R0}$ & $m_{\nu}$ [eV] & $\Omega _{\rm CDM}$ & $\Omega _{\nu }$ & $m^{p}_{\rm CDM}$ [M$_{\odot}$/h] & $m^{p}_{\nu}$ [M$_{\odot}$/h]\\
\hline
$\Lambda $CDM & GR & -- & 0 & 0.31345 & 0 & $8.1\times 10^{10}$  & 0 \\
fR4 & $f(R)$  & $-1\times 10^{-4}$ & 0 & 0.31345 & 0 & $8.1\times 10^{10}$  & 0\\
fR5 & $f(R)$  & $-1\times 10^{-5}$ & 0 & 0.31345 &0  & $8.1\times 10^{10}$  & 0\\
fR6 & $f(R)$  & $-1\times 10^{-6}$ & 0 & 0.31345 & 0 & $8.1\times 10^{10}$  & 0\\
fR4-0.3eV & $f(R)$  & $-1\times 10^{-4}$ & 0.3 & 0.30630 & 0.00715 & $7.92\times 10^{10}$ & $1.85\times 10^{9}$\\
fR5-0.15eV & $f(R)$  & $-1\times 10^{-5}$ & 0.15 & 0.30987 & 0.00358 & $8.01\times 10^{10}$ & $9.25\times 10^{8}$\\
fR5-0.1eV & $f(R)$  & $-1\times 10^{-5}$ & 0.1 & 0.31107 & 0.00238 & $8.04\times 10^{10}$ & $6.16\times 10^{8}$\\
fR6-0.1eV & $f(R)$  & $-1\times 10^{-6}$ & 0.1 & 0.31107 & 0.00238 & $8.04\times 10^{10}$ & $6.16\times 10^{8}$\\
fR6-0.06eV & $f(R)$  & $-1\times 10^{-6}$ & 0.06 & 0.31202 & 0.00143 & $8.07\times 10^{10}$ & $3.7\times 10^{8}$\\
\hline
\end{tabular}}
\label{tab:sims}
\end{table*}

\subsection{Map making}\label{subsec:mapmaking}
For each cosmological simulation, we constructed different lens planes from the various stored snapshots using the \textsc{MapSim} pipeline \citep{GMB.etal.2015}. The particles were distributed onto different planes according to their comoving distances with respect to the observer. For each simulation, we used the particles stored in 21 different snapshots to construct continuous past-light-cones from $z=0$ to $z=4$, with a square sky coverage of 25 deg${}^2$. From the stored snapshots, we were able to construct $27$ lens planes of the projected matter density distribution.

In \textsc{MapSim}, the observer was placed at the vertex of a pyramid whose square base was set at the comoving distance of $z=4$. We constructed $256$ different light-cone realisations within each simulation by randomising the various comoving boxes. This included changing signs, inverting, as well as redefining the centre of the coordinate system. By construction, these procedures preserve the clustering properties of the particle density distribution at a given simulation snapshot \citep{RMB.etal.2007}. 

On each pixel of the lens plane, with coordinate indexes $(i,j)$, we can define the particle surface mass density
\begin{equation}
\Sigma_l(i,j) = \frac{\sum_k m_k}{A_{l}}\,,
\end{equation}
where $m_k$ is the mass of the $k^\mathrm{th}$ particle within the pixel, and $A_{l}$ represents the comoving pixel area of the $l$-lens plane. Since gravitational lensing is sensitive to the projected matter density distribution along the line-of-sight, we projected onto each lens plane all particles between two defined comoving distances from the observer, and in the simulations with massive neutrinos we consistently also accounted for this component, as well as for the proper Hubble function and the comoving distance calculation.

As was done by \citet{PHM.2016,PHM.2017,GDMM.etal.2017,GMB.etal.2018,CQG.etal.2018}, we constructed the convergence maps by weighting the lens planes by the lensing kernel and assuming the Born approximation \citep{SHK.etal.2012,GJM.etal.2016,CQG.etal.2018}, which represents an excellent estimation for weak cosmic lensing down to very small scales ($\ell \geq 10^4$). From $\Sigma_l$ we can write down the convergence map $\kappa$ at a given source redshift 
$z_s$ as
\begin{equation}
\kappa = \sum_l \frac{\Sigma_l}{\Sigma_{\rm{crit},l,s}},
\end{equation}
where $l$ varies over the different lens planes with redshift $z_l$ smaller than $z_s$. The critical surface density $\Sigma_{\rm{crit},l,s}$ at the lens plane $z_l$ for sources at redshift $z_s$ is given by
\begin{equation}
\Sigma_{\rm{crit},l,s} = \frac{c^2}{4 \pi G} \frac{D_l}{D_s D_{ls}}\,,
\end{equation}
where $c$ indicates the speed of light, $G$ Newton's constant, and $D_l$, $D_s$, and $D_{ls}$ the angular diameter distances between observer-lens, observer-source  and source-lens, respectively. We followed this approach also in constructing the convergence maps in the MG models, since the definition of the lensing potential in $f(R)$ gravity models remains unchanged with respect to the GR case of the standard $\mathrm{\Lambda CDM}$ model. The resulting maps for our analysis each contain $2048^2$ pixels, giving a pixel scale of approximately 8.8 arcsec.

\section{Analysis}\label{sec:analysis}
Statistics of the aperture mass $M_\mathrm{ap}(\vartheta)$ have been used in many weak-lensing analyses as a probe of the matter distribution in the Universe and to constrain cosmological parameters \cite[e.g.][]{vWMR.etal.2001,JBF.etal.2003,HMS.etal.2003,KS.2005,CSA.etal.2006,HSS.etal.2007,SHJ.etal.2010,KFH.etal.2013}. In particular, the variance, or second central moment $\langle M^2_\mathrm{ap}(\vartheta) \rangle$, is commonly used, which at a certain scale $\vartheta$ measures the lensing power spectrum within a narrow window in $l$-space. Furthermore, the non-linear evolution of density fluctuations in the low-redshift Universe gets imprinted as non-Gaussian features in the weak-lensing signal, which can be accessed through higher-order moments of the aperture mass. For example, \citet{PLS.2012} have shown that the capture of weak-lensing non-Gaussianities is able to break the degeneracy between $\sigma_8$ and $\Omega_\mathrm{m}$ within $\Lambda$CDM. 

We explore in this work the ability of various statistics of $M_\mathrm{ap}$ to break the degeneracy between $\Lambda$CDM and $f(R)$ models that include non-vanishing neutrino mass. We use maps free of noise throughout most of our analysis; that is, the lensing field is the true one as derived from the simulations. A discussion of galaxy shape noise and its impact on our results via a simple prescription has been included in the Appendix.

\subsection{Aperture mass calculation}\label{subsec:apmass}
Given a convergence map $\kappa(\boldsymbol{\theta})$ for a particular model realisation, we compute the aperture mass map \citep{Schneider.1996,SvWJ.etal.1998} as 
\be\label{eq:ap_mass}
M_\mathrm{ap}(\boldsymbol{\theta};\vartheta) = \int\,\mathrm{d}^2\theta' U_\vartheta(|\boldsymbol{\theta}'-\boldsymbol{\theta}|)\,\kappa(\boldsymbol{\theta}'),
\ee
where $\boldsymbol{\theta}$ is a two-dimensional position vector within the map, $U_\vartheta(|\boldsymbol{\theta}|)$ is a circularly symmetric filter function, and $\vartheta$ is the aperture radius. The aperture mass is by design insensitive to the mass-sheet degeneracy, which describes the fact that a shear signal is unchanged by the presence of a uniform mass sheet along the line of sight between the observer and the source. To achieve this, the function $U$ should be compensated in 2D, or in other words satisfy
\be\label{eq:U_function}
\int\,\mathrm{d}\theta\, \theta\, U_\vartheta(\theta) = 0,
\ee
where we have written the angular separation as $\theta=|\boldsymbol{\theta}'-\boldsymbol{\theta}|$.
It is also typically assumed that $U_\vartheta$ has either finite support or goes to zero smoothly at sufficiently small $\theta$ in order that the integral in Eq.~(\ref{eq:ap_mass}) be computable on a finite data field.

Various definitions of the filter function $U$ have been adopted in previous studies. For example, a useful parameterised family of polynomial functions was first introduced by \citet{SvWJ.etal.1998}. Nearly concurrently, \citet{vanWaerbeke.1998} provided an alternative definition based on a Gaussian function and resembling wavelets in order to have better Fourier space properties. This latter $U$ was subsequently used (in slightly modified form) by, for example, \citet{CNP.etal.2002,ZPZ.etal.2003,JBJ.2004}. A third type of filter was derived in \citet{SEH.etal.2007} which approximates a Navarro-Frenk-White (NFW) density profile \citep{NFW.1996}. As it mimics the shear profile of a spherically symmetric dark matter halo, this form has been used for the detection of individual mass concentrations in real data. 

\begin{figure*}[t]
	\includegraphics[width=\textwidth]{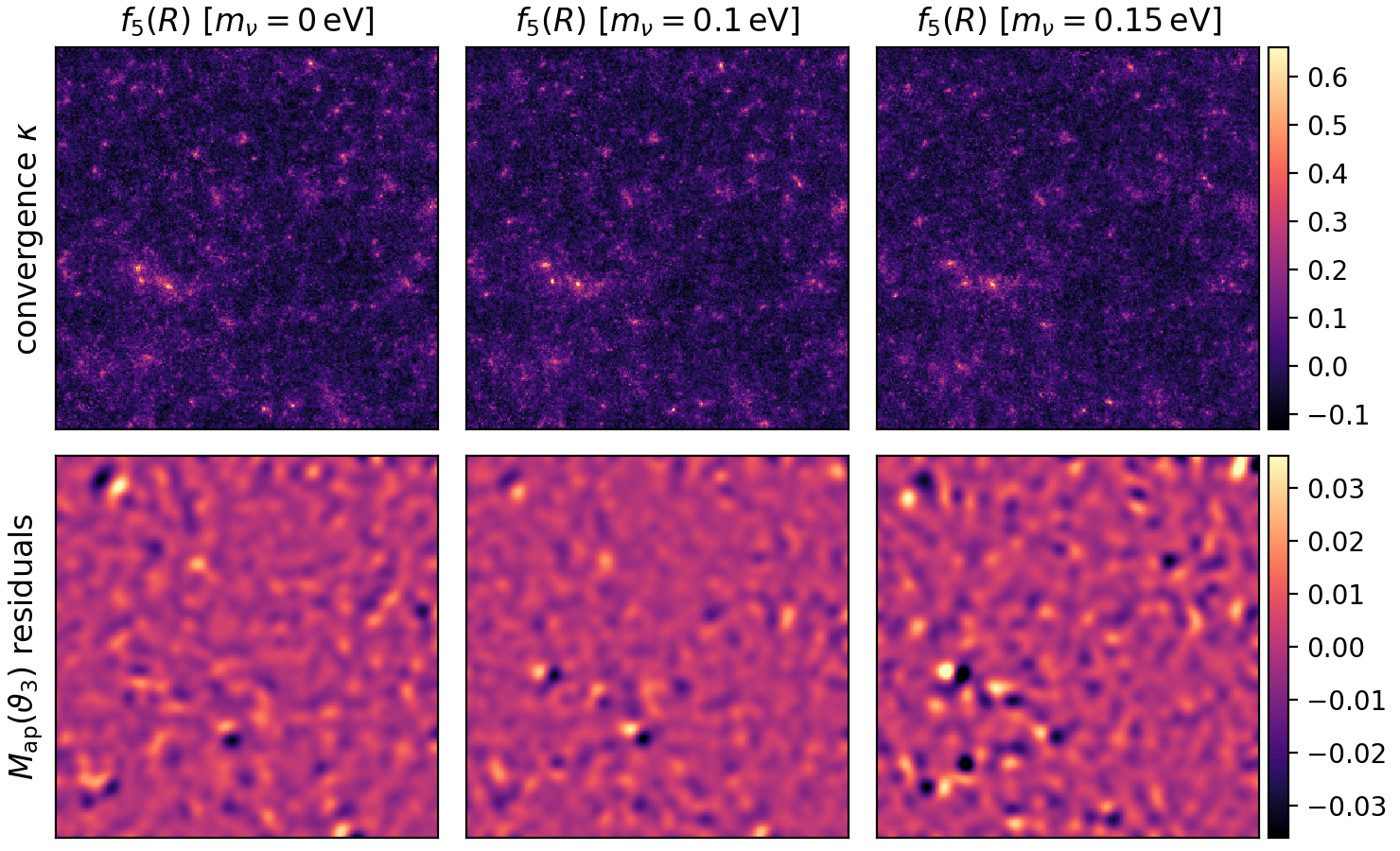}
    \caption{Example convergence fields (\textit{top row}) extracted from the three $f_5(R)$ simulations with neutrino mass sums of 0~eV, 0.1~eV, and 0.15~eV. Each map shows the same line of sight for sources at redshift $z_s=2.0$. Essentially the same structures are seen across all three maps due to identical initial conditions and the common field of view. Subtle differences, however, due to the different model evolutions can also be detected in the peak positions and their morphologies. Aperture mass residuals (\textit{bottom row}) with respect to the $\Lambda$CDM map (not shown) at filtering scale $\vartheta_3=1.17'$ are shown for each corresponding map above. The compensated starlet filter has the effect of highlighting features with an approximate size of $\vartheta_3$. Our aim is to distinguish between these models using differences in the statistics of such aperture mass maps computed at different scales.}
    \label{fig:kappa_maps}
\end{figure*}

Equation (\ref{eq:ap_mass}) is equivalently expressible in terms of the tangential component of shear $\gamma_t$ about the position $\boldsymbol{\theta}$, where the convolution kernel is then a filter $Q_\vartheta$ derivable analytically from $U_\vartheta$ \citep{KSF.etal.1994, Schneider.1996}. This form is often used in the literature for convenience \citep[e.g.][]{DH.2010,DES.SV.2016,MSH.etal.2018}, since the actual weak-lensing observable is not the convergence, but instead the (reduced) shear via galaxy ellipticity measurements. Borders and missing data due to observational masks are also treated more accurately in shear space, although at the expense of significantly reduced algorithm speeds. Because the output of ray-tracing in our $N$-body simulations is the convergence field directly, we compute $M_\mathrm{ap}$ on pixellated $\kappa$ maps throughout this paper.

For high resolution fields like our $2048 \times 2048$ maps, the convolution is typically carried out in Fourier space, since direct-space convolution can be too time-consuming. We adopt a slightly different approach here for the practical computation of aperture mass maps than direct application of equation (\ref{eq:ap_mass}). We instead compute $M_\mathrm{ap}$ by means of the starlet transform \citep{SFM.2007}, which is a wavelet transform that simultaneously produces a set of maps filtered at apertures of increasing dyadic (powers of two) scales. In other words, with a single wavelet transform of $\kappa(\boldsymbol{\theta})$ with $O(n \log n)$ time complexity, we obtain $\{M_\mathrm{ap}(\boldsymbol{\theta}; \vartheta_j)\}_{j=1,2,...,j_{\max}}$, where $\vartheta_j=2^j$ pixels. With a pixel scale of 8.8 arcsec, this corresponds to filter scales of $0.293, 0.586, 1.17,...$ arcmin for values $j=1, 2, 3,...$ in our simulated maps. The maximum $j$ is determined by the map size as $\log_2 N$ for an $N \times N$ map.

It was demonstrated in \citet{LPS.2012} that the aperture mass is formally identical to a wavelet transform at a specific scale. Different transforms are associated with different effective filter functions. The particular transform we adopt is the isotropic undecimated wavelet transform, also called the starlet transform, whose wavelet function at a given scale is defined as the difference between $B_3$-spline functions at neighboring resolutions. We refer to \citet{SFM.2007} for more details of the formalism and to \citet{LPS.2012} for the explicit form of $U_\vartheta$ corresponding to this transform.

In contrast to the above mentioned filter functions, the starlet transform filter presents several significant simultaneous advantages: (i) it is non-oscillatory both in angular space and in Fourier space, meaning that it can better isolate features of the signal represented in either domain; (ii) it has compact support in direct space, which is useful to control the systematics due to a mask or the image borders;
(iii) it is compensated, or has a zero mean, so it is not sensitive to the mass sheet degeneracy problem (cf. equation \ref{eq:U_function}); and (iv) it presents an exact reconstruction, meaning that we can reconstruct the decomposed image from its starlet coefficients. We do not make use of this last property in this paper, but it could be very convenient in the case where we need to fill gaps due to missing data.
The aperture mass maps resulting from the starlet transform then represent only information at a particular scale, with little leakage from other frequencies. Finally, we note that although the starlet transform returns maps filtered only at dyadic scales, which we find sufficient for our purposes here, this is not a general restriction of the wavelet formalism. Alternative wavelet transforms are possible that allow filtering at any intermediate scale.

As an illustration of the weak lensing fields in our simulations, as well as of our aperture mass computation, we show typical maps corresponding to three different models in Fig.~\ref{fig:kappa_maps}. In the top row are $\kappa(\boldsymbol{\theta})$ maps extracted from the three $f_5(R)$ simulations with, from left to right, neutrino mass sums of $0\,\mathrm{eV}$, $0.1\,\mathrm{eV}$, and $0.15\,\mathrm{eV}$. All simulations share the same initial phases in the random realisation of the matter power spectrum, and these maps were generated by ray-tracing from the same observer position. The same physical structures can therefore be seen across the three convergence fields. Subtle differences in the peak positions are also apparent by eye, which is due to the different gravitational interaction in MG compared to GR and the presence or lack of neutrinos affecting structure growth.

Shown in the lower panels of Fig.~\ref{fig:kappa_maps} are aperture mass residuals $M^{f_5(R)}_\mathrm{ap}-M^\mathrm{\Lambda CDM}_\mathrm{ap}$ corresponding to the maps above and filtered at $\vartheta_3=1.17'$. The $\Lambda$CDM $\kappa$ and aperture mass maps (not shown) are hardly distinguishable from any of the $f_5(R)$ maps by eye, but differences in the positions and amplitudes of structures can be clearly seen among the models by their residuals, especially for the most prominent peaks. However, it is not a priori obvious that the statistics of such maps will be significantly different enough to distinguish the models from each other. Whether this is possible, for which statistics and for which filtering scales, are the questions we seek to answer in the following sections.

We compute all aperture mass maps in this paper using the \texttt{mr\_transform} binary of the publicly available Interactive Sparse Astronomical Data Analysis Packages (i\textsc{SAP}\footnote{http://www.cosmostat.org/software/isap}). This is a C{}\verb!++! code with many optional multi-resolution wavelet transforms, which, given the large size of our mass maps, affords a significant speed increase over direct-space 2D convolution. Various statistics of the maps, which are discussed in the next section, are computed using an independent Python code that we have validated on maps from the \textsc{CoDECS} simulations used in \citet{GMB.etal.2015}. In particular, we are able to reproduce Figs.~8 and 9 of that paper using a further independent aperture mass calculation that implements the \citet{SvWJ.etal.1998} filter.

\subsection{Statistics}\label{subsec:stats}
We consider the following statistics of aperture mass maps as a function of filter scale and source galaxy redshift. To avoid edge effects, we exclude from all calculations pixels whose distance to the map border is smaller than the filter diameter.

\subsubsection{Variance}
The aperture mass variance $\langle M^2_\mathrm{ap} \rangle(\vartheta)$ quantifies fluctuations in lensing strength along different directions in the sky, meaning that it also measures the amplitude of fluctuations in the matter density contrast. As a second-order statistic, it can be expressed as an integral over the convergence power spectrum $P_\kappa(\ell)$
\be\label{eq:apmass_var_integral}
\langle M^2_\mathrm{ap} \rangle(\vartheta) = \frac{1}{2\pi}\int\,\mathrm{d}\ell\,\ell\,P_\kappa(\ell)\,W^2(\ell\vartheta),
\ee
where the window function $W$ is the Fourier transform of $U_\vartheta$ \citep{SvWJ.etal.1998}. Aperture mass variance is therefore mostly sensitive to the Gaussian information in the distribution of matter. This integral form is convenient when comparing measurements with theoretical predictions, since $P_\kappa$ is readily calculated (at least in $\Lambda$CDM) for a given set of cosmological parameters. Because we have simulated maps for all the models of interest in this work, we instead compute $\langle M^2_\mathrm{ap} \rangle$ directly on filtered convergence maps:
\be\label{eq:apmass_var_sum}
\langle M^2_\mathrm{ap} \rangle(\vartheta) = \frac{1}{N}\sum_k \left[M_\mathrm{ap}(\boldsymbol{\theta}_k;\vartheta) - {\overline{M_\mathrm{ap}}(\vartheta)}\right]^2,
\ee
where $\boldsymbol{\theta}_k$ refers to the $k$th pixel position.

\subsubsection{Skewness}
As a third-order statistic, skewness is complementary to variance in that it probes non-Gaussian information contained in the lensing observables. 
The skewness $S$ is zero if the data are symmetrically distributed around the mean. If a tail extends to the right (resp. left), $S$ is positive (resp. negative). In particular, skewness has been shown to be a sensitive probe of $\Omega_\mathrm{m}$ and can break the degeneracy with $\sigma_8$ if combined with two-point statistics \citep{BvWM.1997,JS.1997}. Analogously to the variance, skewness can be written as a bandpass filter over the convergence bispectrum \citep[e.g.][]{KS.2005}, the Fourier transform of the three-point correlation function. Following \citet{GMB.etal.2015}, we compute skewness as the (non-standardised) third-order moment of our filtered maps:
\be\label{eq:apmass_skew}
\langle M^3_\mathrm{ap} \rangle(\vartheta) = \frac{1}{N}\sum_k \left[M_\mathrm{ap}(\boldsymbol{\theta}_k;\vartheta) - {\overline{M_\mathrm{ap}}(\vartheta)}\right]^3.
\ee

\subsubsection{Kurtosis}
The kurtosis of a distribution is a measure of its symmetric broadening or narrowing relative to a Gaussian. Positive kurtosis implies a higher peak and larger wings than the Gaussian distribution with the same mean and variance. Negative kurtosis means a wider peak and shorter wings. Again following \citet{GMB.etal.2015}, we compute kurtosis as the (non-standardised) fourth-order moment of the aperture mass: 
\be 
\langle M^4_\mathrm{ap} \rangle(\vartheta) = \frac{1}{N}\sum_k \left[M_\mathrm{ap}(\boldsymbol{\theta}_k;\vartheta) - {\overline{M_\mathrm{ap}}(\vartheta)}\right]^4.
\ee

\subsubsection{Peak counts}
The number count distribution of lensing peaks provides another probe of non-Gaussianity. Peaks represent local regions of high convergence, and their abundance for a given cosmological model carries significant information about its matter content and clustering amplitude. The number of cosmological studies based on peak statistics in convergence and aperture mass maps, both from simulated and real lensing data, has surged in the past decade \cite[e.g.][]{KS.1999,KS.2000,DH.2010,KHM.2010,FSL.2010,YKW.etal.2011,MFM.2011,MSH.etal.2012,HOS.etal.2012,SKC.etal.2014,LK.2015,MBK.etal.2015,LLZ.etal.2016,DES.SV.2016,PLL.etal.2017,SLH.etal.2018,MSH.etal.2018,FKS.etal.2018}. The highest signal-to-noise peaks are mostly generated by single massive structures along the line of sight. On the other hand, intermediate and low signal-to-noise peaks can arise from projection effects due to many smaller structures or filaments along the line of sight, as well as simply to noise when dealing with real data \citep{YKW.etal.2011,LH.2016}.

In this work, we explore the possibility of peak counts to distinguish between GR and MG models (with and without neutrinos), which are degenerate at the level of second-order statistics. We compute peaks as a function of filter scale $\vartheta$ in maps of $M_\mathrm{ap}(\vartheta) / \sigma(\vartheta)$, where $\sigma(\vartheta)$ is the rms of that scale. A peak is defined as a pixel with larger amplitude than its eight neighbors and exceeding a given $k\sigma$ $(k=1, 2, ...)$ threshold. Throughout the rest of the paper, we display results for thresholds of $3\sigma$ and $5\sigma$.

\subsection{Model discrimination efficiency}\label{subsec:fdr}
For each cosmological model we consider, we produce 256 independent convergence maps of 25 deg${}^2$ for sources at redshifts $z_s=(0.5, 1.0, 1.5, 2.0)$ by the procedure outlined in Sect.~\ref{sec:sims}. We have sufficient area therefore to meaningfully study the statistical distributions of observables as a function of scale and redshift. In particular, we test the ability of various aperture mass statistics to distinguish between the models by employing the discrimination efficiency technique introduced and used in \citet{PSA.etal.2009, PLS.2012}. 

The concept in our context can be understood as follows. Given the distributions of any two observables, the amount of overlap between the distributions is an indicator of how likely one is to mistake one model for the other, based on measurements of that observable. Quantification of the overlap is facilitated by the notion of a false discovery rate (FDR) \citep{BH.1995}. The FDR framework allows one to set a threshold value for a distribution such that new test samples (or observations) can be classified as either belonging to the distribution or not, depending on which side of the threshold they lie. The threshold is computed according to a desired false discovery rate $\alpha$. The formalism then ensures that the fraction of false positives, or observations which indeed belong to the distribution but are incorrectly classified as detections, is less than or equal to $\alpha$. As is common, we set $\alpha=0.05$ throughout this work to ensure that the number of false detections on average does not exceed 5\%.

\begin{figure}
	\includegraphics[width=\columnwidth]{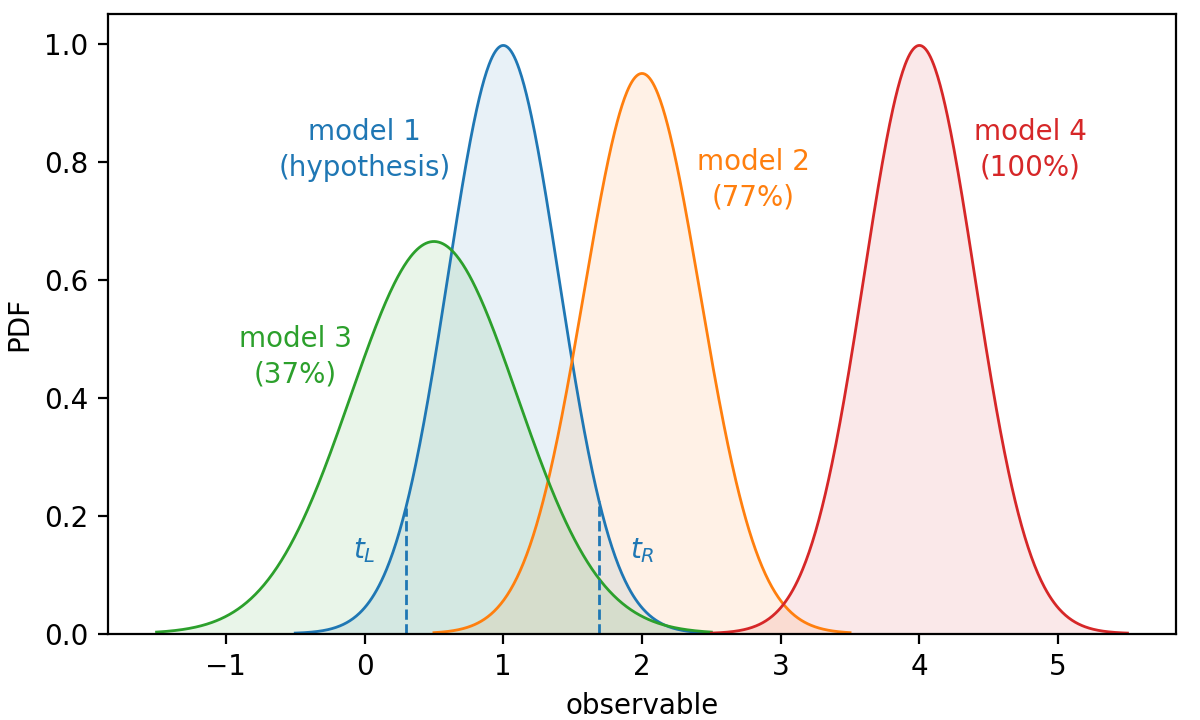}
    \caption{Toy example of discrimination efficiencies computed using the FDR formalism. Each (Gaussian) distribution represents the histogram measured for some observable given the model. The amount of overlapping area between any two distributions indicates how distinguishable the two models are according to this observable: the more the curves overlap, the more difficult it is to distinguish the corresponding models, and the higher the discrimination efficiency parameter is. In the example shown, considering model 1 as the hypothesis against which the others are tested, models 2, 3, and 4 have computed discrimination efficiencies of 77\%, 37\%, and 100\%.}
    \label{fig:fdr}
\end{figure}

As an illustration, consider the toy example in Fig.~\ref{fig:fdr}. The probability distributions of some observable, all taken to be Gaussian for simplicity, are shown for four fictitious models. Let model 1 serve as the `hypothesis' against which the other models are tested. To determine the discrimination efficiency of, for example, model 2 from model 1, we first compute the right tail threshold $t_R$ (dashed line) using FDR based on the samples of model 1. This is done by first rank-ordering the $p$-values $\{p_i\}$, $i=1,...,n$, of the $n$ model 1 samples, computed with respect to a right tail event. Next one finds the maximum $p_m$ ($1 \leq m \leq n$) for which $p_m < m \cdot\alpha / n$, where we have assumed statistical independence of the $p$-values. The threshold $t_R$ is the observation value corresponding to $p_m$, and the discrimination efficiency is the fraction of model 2 samples greater than $t_R$. Analogous reasoning holds for comparing model 3 against model 1 by computation of threshold $t_L$.

In general, then, suppose we have observations of two models, $M_1=\{x_i\}_{i=1,...,n_1}$ and $M_2=\{x_j\}_{j=1,...,n_2}$. If the centre of $M_2$ is greater than that of $M_1$, the discrimination efficiency of $M_2$ from $M_1$ is
\be\label{eq:disc_eff}
\mathcal{E}_{2,1} = \frac{\mathcal{N}[x_j > t_R\,(M_1)]}{n_2}\,,
\ee
where $\mathcal{N}[C]$ counts the number of elements satisfying the condition $C$, and the threshold $t_R$ is computed from the $M_1$ samples. If the centre of $M_2$ lies to the left of that of $M_1$, the condition $C$ becomes $x_j < t_L(M_1)$. For a more detailed description of this procedure, we refer to, for example, Appendix A of \citet{PSA.etal.2009}.

For the example in Fig.~\ref{fig:fdr}, the discrimination efficiencies for models 2, 3, and 4 from model 1 are, respectively, $\mathcal{E}_{2,1}=77\%$, $\mathcal{E}_{3,1}=37\%$, and $\mathcal{E}_{4,1}=100\%$. This corresponds with the intuition that models with fully overlapping distributions should not be considered distinguishable at all, whereas completely disjoint distributions should be 100\% distinguishable. 

There is an inherent asymmetry in discrimination efficiency between any two models that do not share the same width, or in general are not perfectly symmetric functional forms with merely different means. Considering model 2 as the hypothesis, for example, the model 1 discrimination efficiency is close to that of the reverse case at 75\%, which follows from their near symmetry. With respect to model 3, however, the model 1 efficiency drops to 9\%. This is consistent with the fact that the support of the model 1 distribution is a subset of that of model 3, and not the other way around. To take this into account in our analysis, we quote the mean of the two possible values for a given pair of models in all following results, notably in Figs.~\ref{fig:polar_fR5} and \ref{fig:polar_fR5_0.15eV} and Table~\ref{tab:max_disc_eff}. The asymmetry is largest for the third- and fourth-order moments, where the distributions deviate more from a Gaussian than do the variance and peak counts.

We have depicted continuous distributions in the example of Fig.~\ref{fig:fdr}. In practice, however, with our simulations, we have 256 samples of each distribution for a given cosmological model, statistic, filter scale, and source redshift. Given the likelihood of undersampling, especially at the tails of the distributions that are important for setting the FDR threshold, we apply a Gaussian smoothing in each case using kernel density estimation (KDE) prior to calculating discrimination efficiencies. We find that KDE smoothing improves results in terms of the agreement between the observed overlap of the distributions and the computed value.

\section{Results}\label{sec:results}

\begin{figure*}
	\includegraphics[width=\columnwidth]{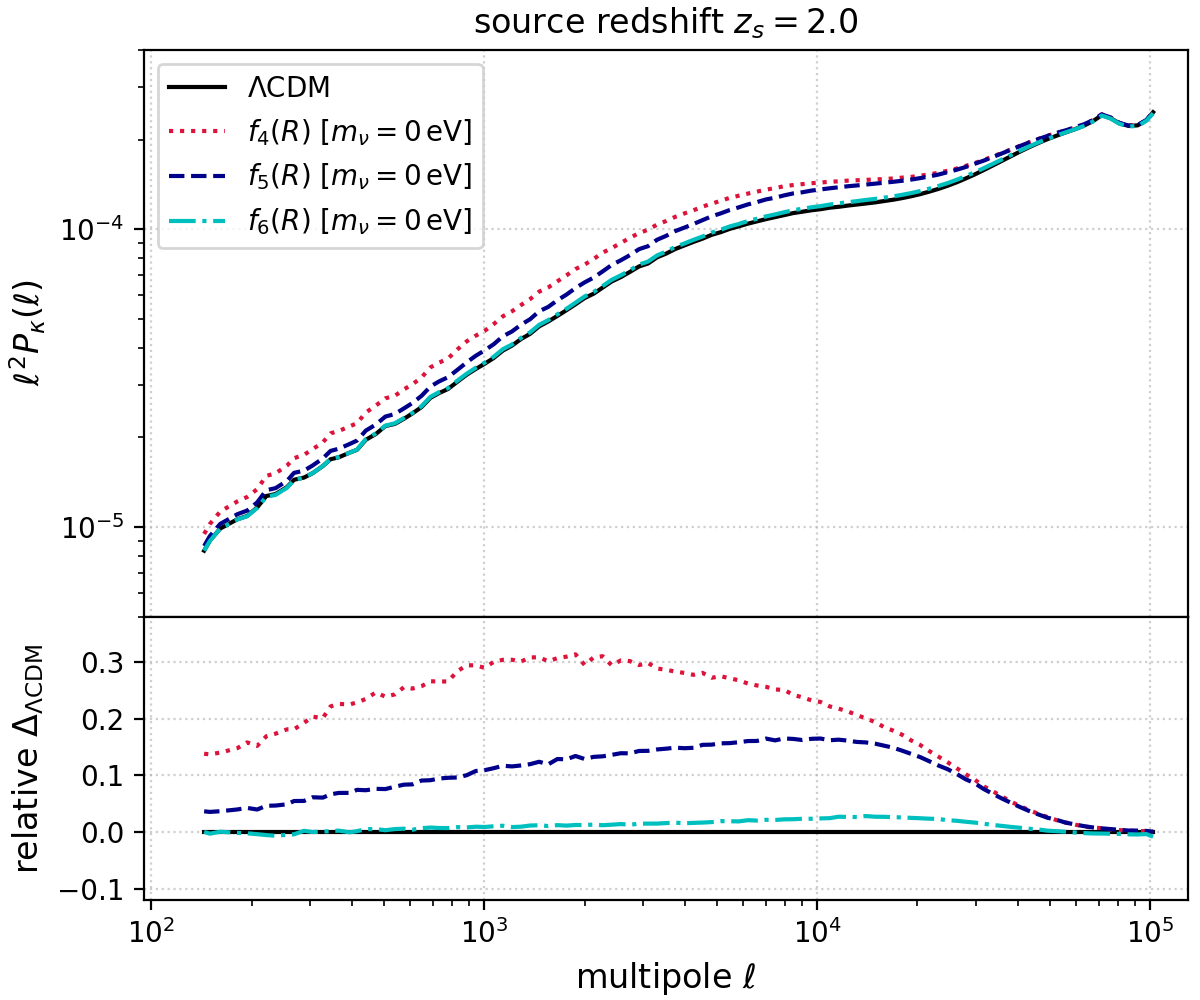}
    \includegraphics[width=\columnwidth]{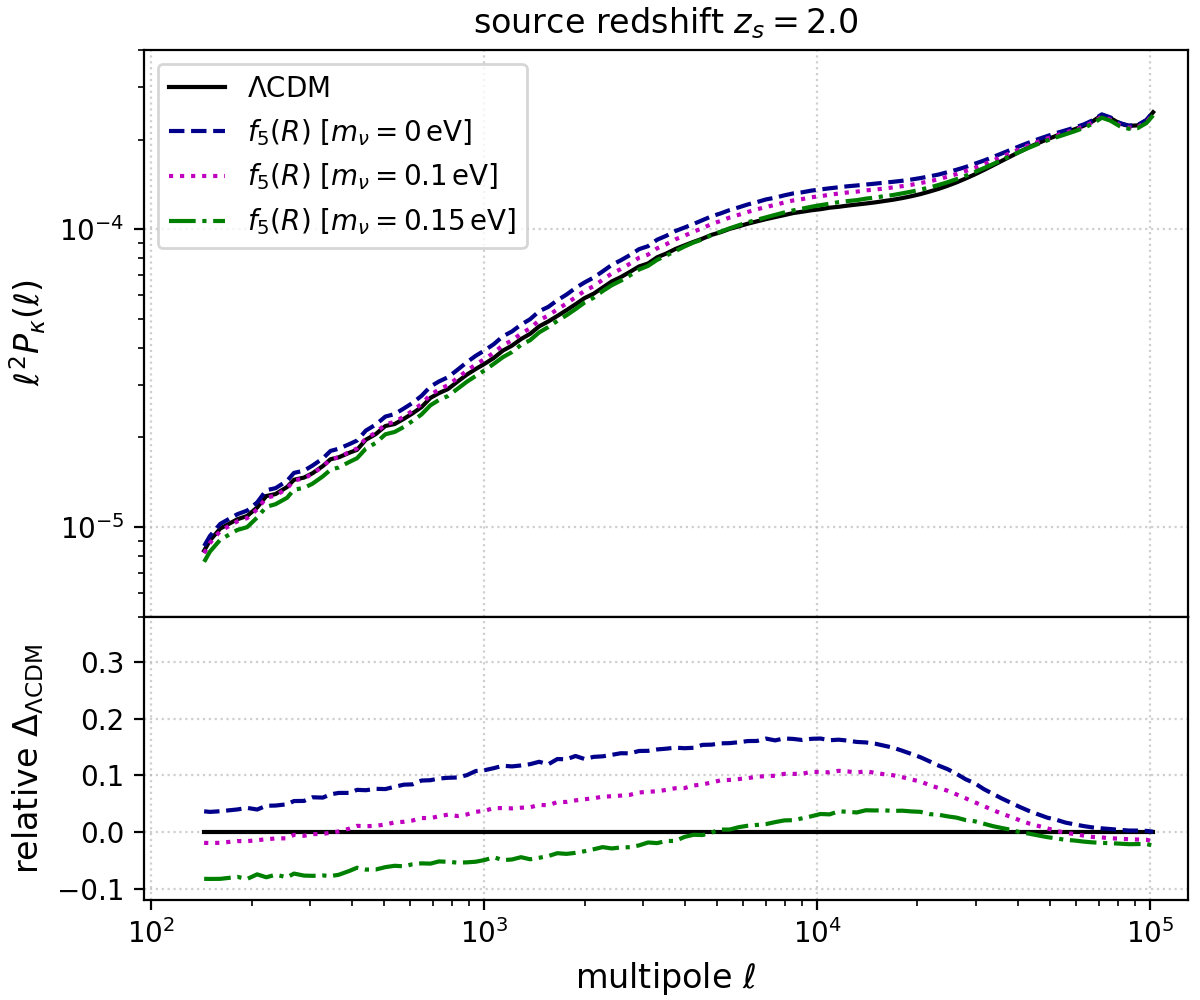}
    \caption{Convergence power spectra for $\Lambda$CDM and various MG models. The \textit{left panel} shows $f(R)$ models with $f_{R0}=(-10^{-4}, -10^{-5}, -10^{-6})$ and without neutrinos. Compared to $\Lambda$CDM, each MG model exhibits more lensing power, where the difference increases with the magnitude of $f_{R0}$. The discrepancy is most pronounced on smaller scales for $f_5(R)$ and $f_6(R)$ but shifts towards larger scales for $f_4(R)$. Shown in the \textit{right panel} are $f_5(R)$ models with varying neutrino mass sums. Neutrinos have the effect of suppressing structure growth across all scales, which is seen in the damping of the $f_5(R)$ curve with increasing $m_\nu$. In particular, the combination of parameters $(f_{R0}, m_\nu) = (10^{-5}, 0.15~\mathrm{eV})$ reproduces $\Lambda$CDM in terms of second-order statistics to within 8\% over three decades in scale. The convergent behaviour of the curves near $\ell=10^5$ reflects the resolution limit of our maps and particle noise that affects all of the models.}
    \label{fig:Pkappa}
\end{figure*}

\subsection{Second-order \texorpdfstring{$M_\mathrm{ap}$}{} statistics}\label{subsec:second-order}
The lensing power spectrum, or equivalently the angular two-point correlation function, accesses the Gaussian information in the projected matter density. As seen by their 3D matter power spectra (Fig.~\ref{fig:power_ratio}), there exist $f(R)$ models, both with and without neutrinos, that can mimic GR to better than 10\% at second order. Our goal is to look beyond second-order statistics to see if it is possible to distinguish the MG models from $\Lambda$CDM. As a check of consistency, we first compare the lensing power spectra to verify agreement between the two second-order probes.

In Fig.~\ref{fig:Pkappa} we show convergence power spectra $P_\kappa(\ell)$ computed for various $f(R)$ models along with $\Lambda$CDM for $z_s=2.0$. The left plot shows the three $f(R)$ models we consider defined by their different $f_{R0}$ values ($-10^{-4}$, $-10^{-5}$, and $-10^{-6}$), all with a neutrino mass sum $m_\nu$ equal to zero. As theory predicts, and as we have seen by their matter power spectra, $f(R)$ models more closely mimic $\Lambda$CDM the smaller the value of $f_{R0}$. In particular, $f_5(R)$ agrees with $\Lambda$CDM within 17\% and $f_6(R)$ at better than 3\% over three decades in scale. 

It is interesting to note the different scales at which the deviation from $\Lambda$CDM is maximal. For both $f_5(R)$ and $f_6(R)$, the maximum occurs around multipole $\ell\approx 10^4$, whereas for $f_4(R)$ it is around $10^3$. Power thus moves from small scales to large scales with increasing $f_{R0}$. The effect of the modified gravitational interaction is non-trivial and not simply a uniform scaling of the power spectrum across different modes. We note that all curves have been computed for sources at $z_s=2.0$, but very similar results occur for the other redshift planes, the main difference being an overall shift in amplitude.

\begin{figure*}
	\includegraphics[width=\columnwidth]{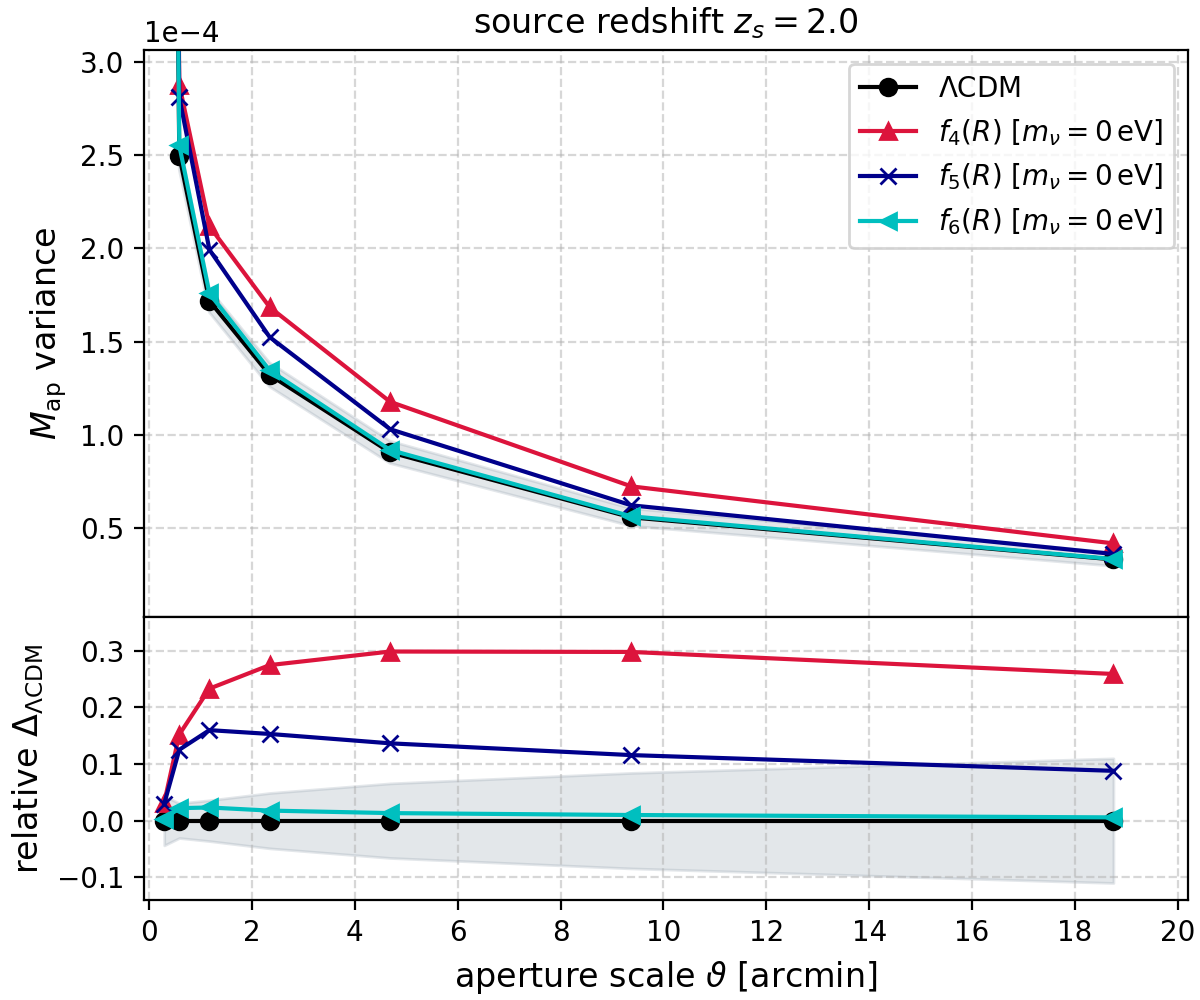}
    \includegraphics[width=\columnwidth]{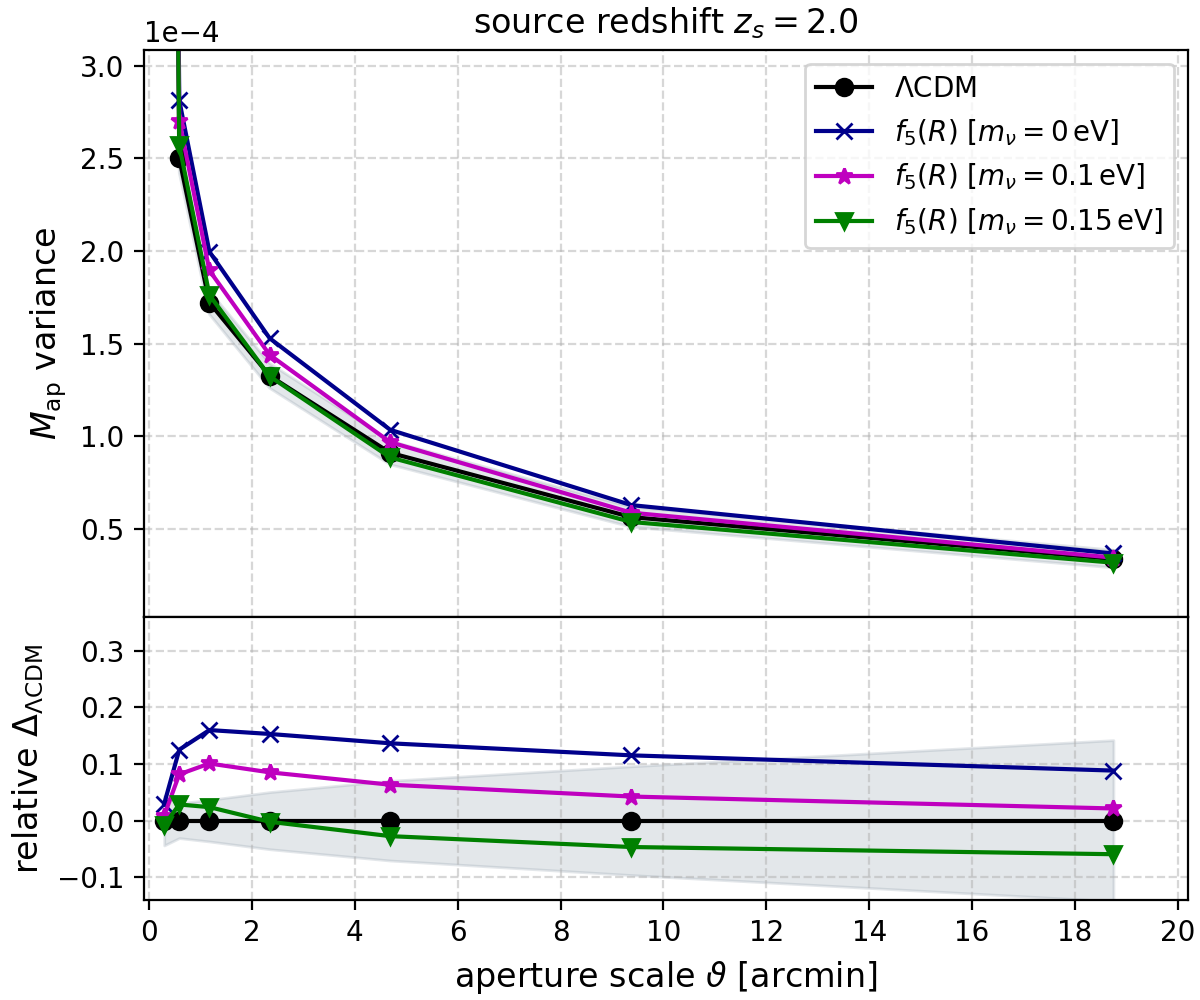}
    \caption{Aperture mass variance as a function of filtering scale for $\Lambda$CDM and MG models. Data points represent the mean over 256 realisations of each simulated cosmology and are plotted at scales $(\vartheta_1, ..., \vartheta_7) = (0.293', 0.586', 1.17', 2.34', 4.69', 9.34', 18.8')$. Shaded areas indicate the one standard deviation statistical uncertainty around $\Lambda$CDM. Shown in the \textit{left panel} are the three $f(R)$ models without neutrinos. Differences in MG variance relative to $\Lambda$CDM match those seen for the power spectra in Fig.~\ref{fig:Pkappa}. In particular, the $f_4(R)$ variance at large scales exceeds that of $\Lambda$CDM by around 30\%, while the maximum deviations of $f_5(R)$ and $f_6(R)$ occur at small scales at a level of about 16\% and 2\%, respectively. The \textit{right panel} shows the three $f_5(R)$ models with different sums of neutrino masses $m_\nu$. Trends in the curves again mirror their power spectra, and the effect of neutrinos damping structure growth is seen in the modulation of $f_5(R)$ variance with $m_\nu=0~\mathrm{eV}$ towards the curve for $m_\nu=0.15~\mathrm{eV}$.}
    \label{fig:apmass_var}
\end{figure*}

\begin{figure*}
	\includegraphics[width=\columnwidth]{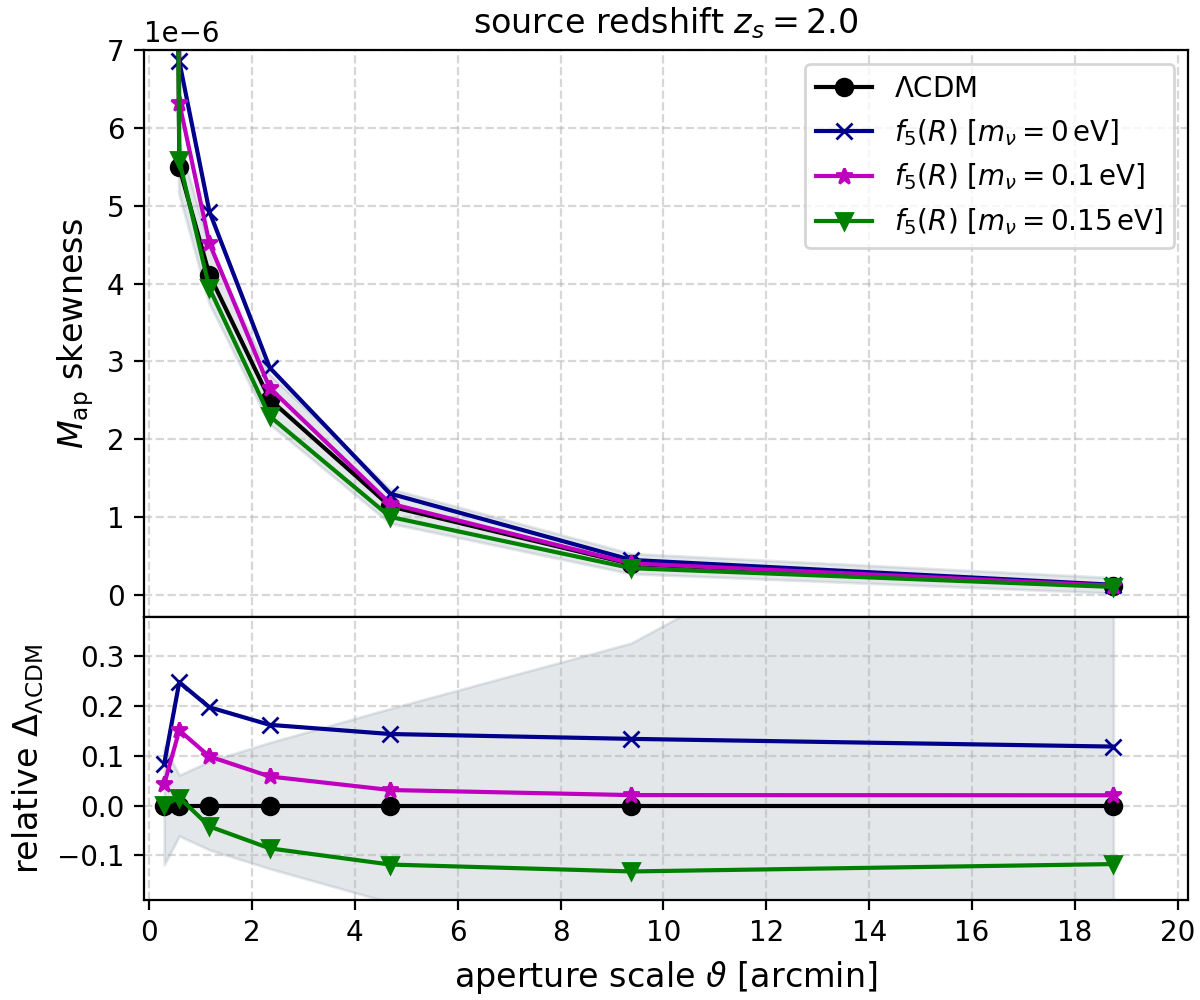}
    \includegraphics[width=\columnwidth]{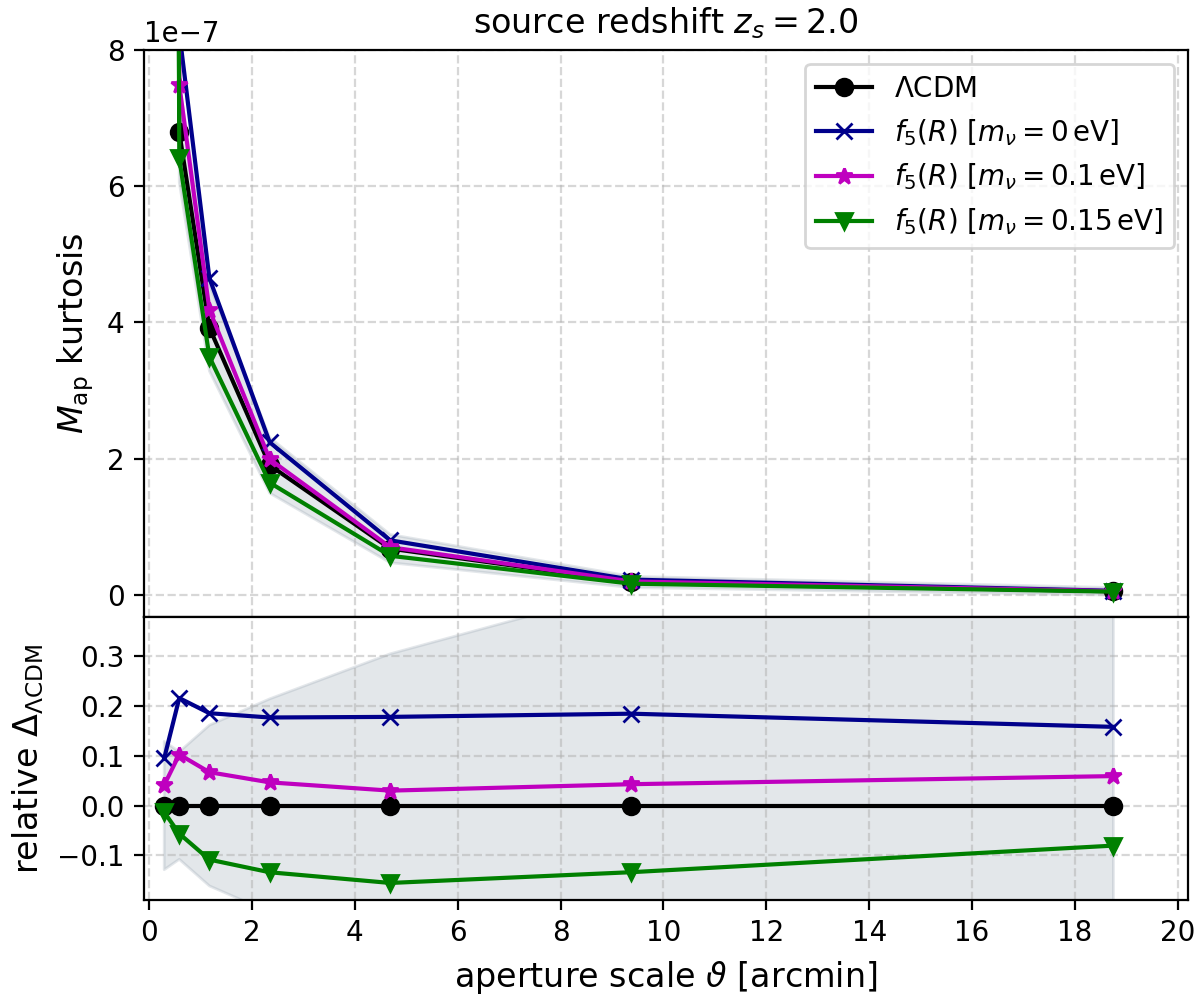}
    \caption{Higher-order aperture mass statistics for $f(R)$ models with $f_{R0}=-10^{-5}$ and a varying sum of neutrino masses $m_\nu$. Data points represent the mean over 256 realisations of each simulated cosmology and are plotted at scales $(\vartheta_1, ..., \vartheta_7) = (0.293', 0.586', 1.17', 2.34', 4.69', 9.34', 18.8')$. Shown in the \textit{left panel} is the aperture mass skewness computed as the third order moment $\langle M^3_\mathrm{ap} \rangle$. Higher neutrino mass leads to a nearly uniform reduction in skewness across all scales relative to $\Lambda$CDM. The maximum discrepancy between each model and $\Lambda$CDM is also larger than for the variance. The \textit{right panel} shows the aperture mass kurtosis computed as the fourth-order moment $\langle M^4_\mathrm{ap} \rangle$. Differences from $\Lambda$CDM are comparable to the skewness, the main distinction being that the kurtosis of the $m_\nu>0$ models approaches the $m_\nu=0$ value at larger scales.}
    \label{fig:apmass_skew_kurt}
\end{figure*}

In the right panel we focus on variations among $f_5(R)$ models with different $m_\nu$. Looking at the relative differences with $\Lambda$CDM, it is clear that the addition of neutrinos has the effect of damping power at all scales. This is consistent with the expectation that massive neutrinos should suppress structure growth. With $m_\nu=0.1~\mathrm{eV}$ ($0.15~\mathrm{eV}$), the deviation does not exceed 11\% (9\%) within the $\ell$ range considered. We see that in terms of lensing, as well as directly through the matter distribution, $f_5(R)$ models with neutrinos can produce a signal that is significantly degenerate with $\Lambda$CDM.

\begin{figure*}
	\includegraphics[width=\columnwidth]{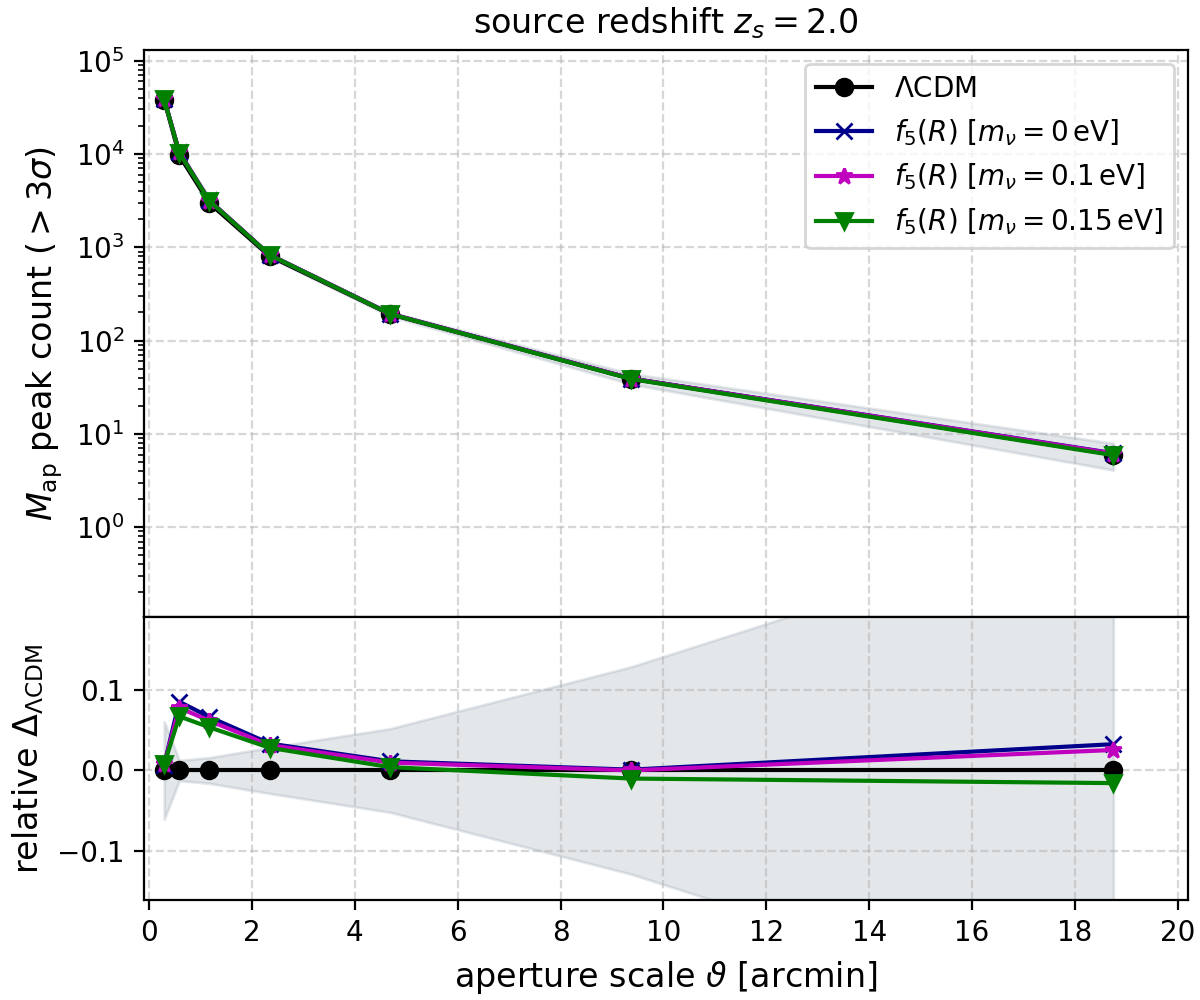}
    \includegraphics[width=\columnwidth]{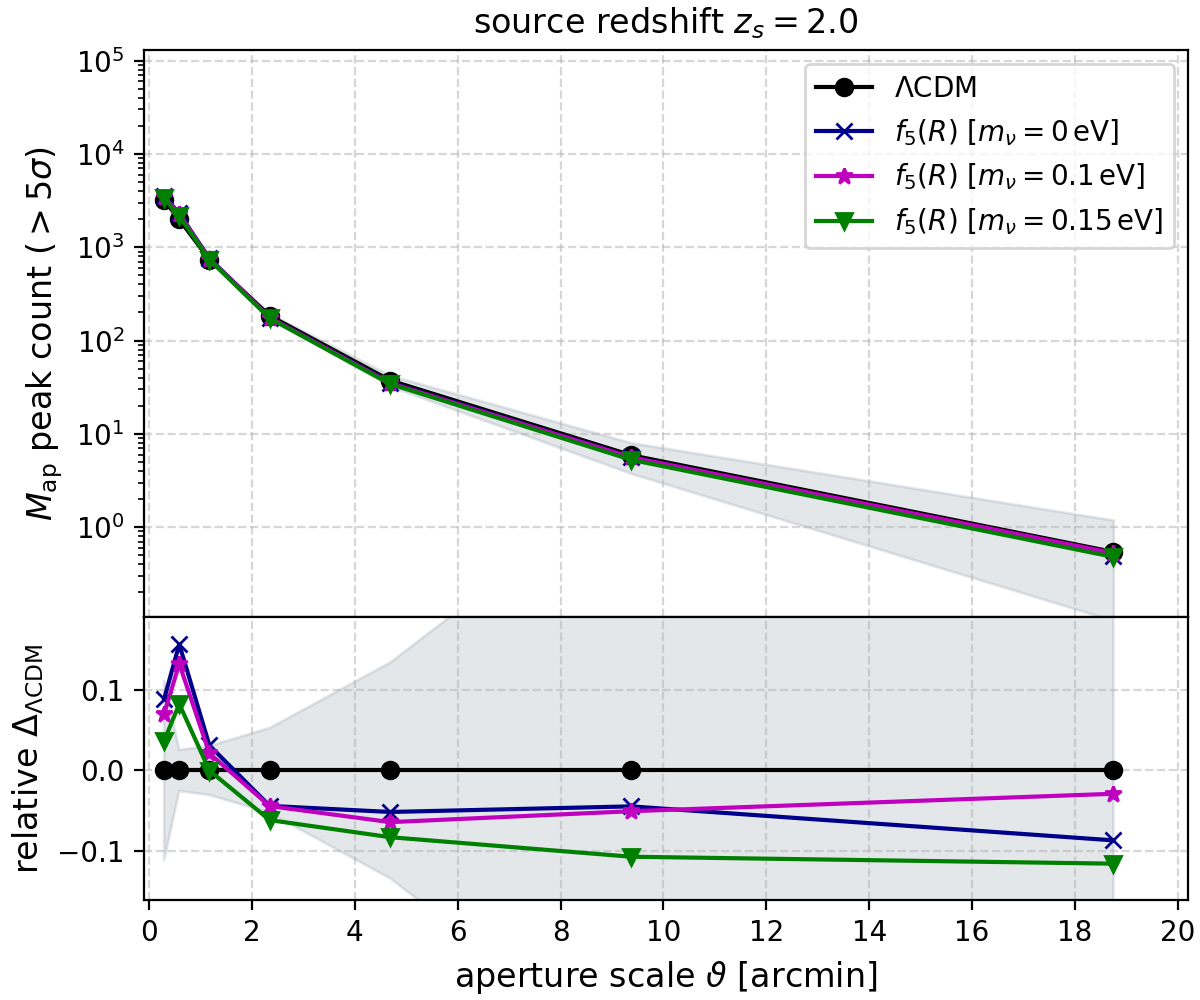}
    \caption{Aperture mass peak counts for $f(R)$ with $f_{R0}=-10^{-5}$ and a varying sum of neutrino masses $m_\nu$. Data points represent the mean over 256 realisations of each simulated cosmology and are plotted at scales $(\vartheta_1, ..., \vartheta_7) = (0.293', 0.586', 1.17', 2.34', 4.69', 9.34', 18.8')$. The abundance of peaks above a $3\sigma$ threshold is shown in the \textit{left panel}, and above a $5\sigma$ threshold in the \textit{right panel}. An order of magnitude more peaks are detected at each scale with the lower threshold, although deviations from $\Lambda$CDM, as well as among the different $m_\nu$ cases of $f_5(R)$, are more pronounced with the higher threshold. This indicates that differences between models at this redshift are mostly contained in the highest amplitude peaks. Compared to the $M_\mathrm{ap}$ moments, peak counts interestingly do not very well distinguish between models with different neutrino mass, but they do distinguish modified from standard gravity, at least at certain scales.}
    \label{fig:apmass_wpc}
\end{figure*}

Figure~\ref{fig:apmass_var} shows the mean aperture mass variance computed as a function of aperture scale for the same models as in Fig.~\ref{fig:Pkappa}. $\langle M^2_\mathrm{ap} \rangle (\vartheta)$ is plotted at filtering scales $(\vartheta_1, ..., \vartheta_7) = (0.293', 0.586', 1.17', 2.34', 4.69', 9.34', 18.8')$ corresponding to a wavelet transform with $j_{\max}=7$. The shaded bands spanning the $\Lambda$CDM curves designate the statistical uncertainty computed as $\pm 1\sigma$, where $\sigma$ is the standard deviation measured from the 256 $\Lambda$CDM maps at each aperture scale. We include this region in all plots of Figs.~\ref{fig:apmass_var}--\ref{fig:apmass_wpc} to illustrate which models are likely to be distinguishable from $\Lambda$CDM according to the statistical scatter of the observable at a given scale.

As discussed in Sect.~\ref{sec:analysis}, aperture mass variance probes the lensing power spectrum within a narrow window around the $\ell$ mode associated to scale $\vartheta$. In agreement with their power spectra, the MG models with $m_\nu=0~\mathrm{eV}$ (left), for example, have larger variance at each scale compared to $\Lambda$CDM, with the difference increasing according to the magnitude of $f_{R0}$. Moreover, the variance relative to $\Lambda$CDM for $f_4(R)$ is largest at intermediate filtering scales, whereas it is largest at small filtering scales for $f_5(R)$ and $f_6(R)$.

Plotted in the right panel of Fig.~\ref{fig:apmass_var} is $\langle M^2_\mathrm{ap} \rangle (\vartheta)$ for $f_5(R)$ models with varying neutrino mass. The trend in the curves is the same as for the power spectra seen in Fig.~\ref{fig:Pkappa}, namely the presence of neutrinos in the model can bring the aperture mass measurements into agreement with $\Lambda$CDM at better than 7\% over the full range of scales considered. We confirm therefore that by including neutrinos, MG models with importantly different gravitational interactions from GR can mimic $\Lambda$CDM at the level of two-point statistics in weak-lensing observations.

We note that the maximum variance shown has been restricted to less than the value actually attained for $\vartheta_1$ in both plots of Fig.~\ref{fig:apmass_var}. This was done merely for visualisation purposes to better exhibit the spread of the curves at larger aperture scales. The variances for all models at the smallest scale are nearly identical in any case, as the difference plots of the lower panels show.

\subsection{Higher-order \texorpdfstring{$M_\mathrm{ap}$}{} statistics}\label{subsec:higher-order}
Given that second-order $M_\mathrm{ap}$ statistics may be insufficient to distinguish certain MG models from $\Lambda$CDM, we present now results from computing higher-order statistics in these models. These include skewness, kurtosis, and peak counts, as discussed in Sect.~\ref{subsec:stats}.

The aperture mass skewness $\langle M^3_\mathrm{ap} \rangle$ provides a simple third-order statistic of the lensing field that probes the bispectrum (cf. Sect.~\ref{subsec:stats}). Results are shown in the left plot of Fig.~\ref{fig:apmass_skew_kurt} for the same $f_5(R)$ models considered in Sect.~\ref{subsec:second-order}. The effect of neutrinos is seen as the nearly uniform reduction in skewness at each aperture scale which varies with $m_\nu$ in an analogous way to the variance. That is, the curves with $m_\nu>0$ are shifted downward systematically from the curve with $m_\nu=0$. This means that the distribution of $\kappa$ becomes more Gaussian as $m_\nu$ increases, which is consistent with a higher neutrino mass more effectively suppressing the generation of the highest matter peaks. 

At the smallest scales, the skewness across all three $f_5(R)$ models is amplified relative to $\Lambda$CDM, and more so than for the variance at the same scale. For example, the $m_\nu=0~\mathrm{eV}$ model skewness is approximately 25\% larger than $\Lambda$CDM at $\vartheta_2$ compared to 12\% for the variance. In addition, for the model with $m_\nu=0.15~\mathrm{eV}$, the skewness at certain intermediate scales is up to 13\% different from $\Lambda$CDM, whereas the corresponding variance at the same scale is less than 5\%. Each $f_5(R)$ model approaches an approximately constant skewness at scales around $\vartheta\approx 10'$ and larger. Given the larger (average) differences among the models, this suggests that skewness may be a better observable to break degeneracies than the variance.

We show the aperture mass kurtosis $\langle M^4_\mathrm{ap} \rangle$ as a function of filter scale in the right plot of Fig.~\ref{fig:apmass_skew_kurt}. As with the variance and skewness, the $m_\nu=0$ curve is highest and decreases at each scale as $m_\nu$ increases. Differences with respect to $\Lambda$CDM are of the same amplitude as for the skewness within each $f_5(R)$ model. We see at fourth order, however, that neutrinos cause a more complex behaviour in the full lensing distribution than was apparent at lower orders. For example, in the relative deviation from $\Lambda$CDM (lower panel), there exists now a clear local minimum at $\vartheta\approx 5'$ for the models with $m_\nu>0$, in addition to the maximum at smaller scales. 

It is interesting to comment on the overall behaviour of the statistics of $M_\mathrm{ap}$ moments as one varies neutrino mass. The $f_5(R)$ model with $m_\nu=0~\mathrm{eV}$ overpredicts $\Lambda$CDM at each filter scale and for each statistic. Including massive neutrinos in the model, as we have discussed previously, counteracts the tendency of the modified gravitational interaction to enhance structure growth---at least on scales smaller than the neutrino free-streaming length, which in our case, for each $m_\nu$, lies above the full range of aperture scales considered. The degeneracy with $\Lambda$CDM is generally strengthened by larger $m_\nu$, but not without limit, and not uniformly, as the higher-order $M_\mathrm{ap}$ statistics reveal. We expect that $m_\nu$ values larger than we have studied would again decrease the degeneracy with $\Lambda$CDM, perhaps measurably at the two-point level. For reference, the most recent constraints from CMB data put an upper bound on the sum of neutrino masses at around $0.17~\mathrm{eV}$ \citep{CHP.etal.2017}, however this assumes a $\Lambda$CDM model that does not necessarily apply beyond this framework.

\begin{figure}
	\includegraphics[width=0.49\columnwidth]{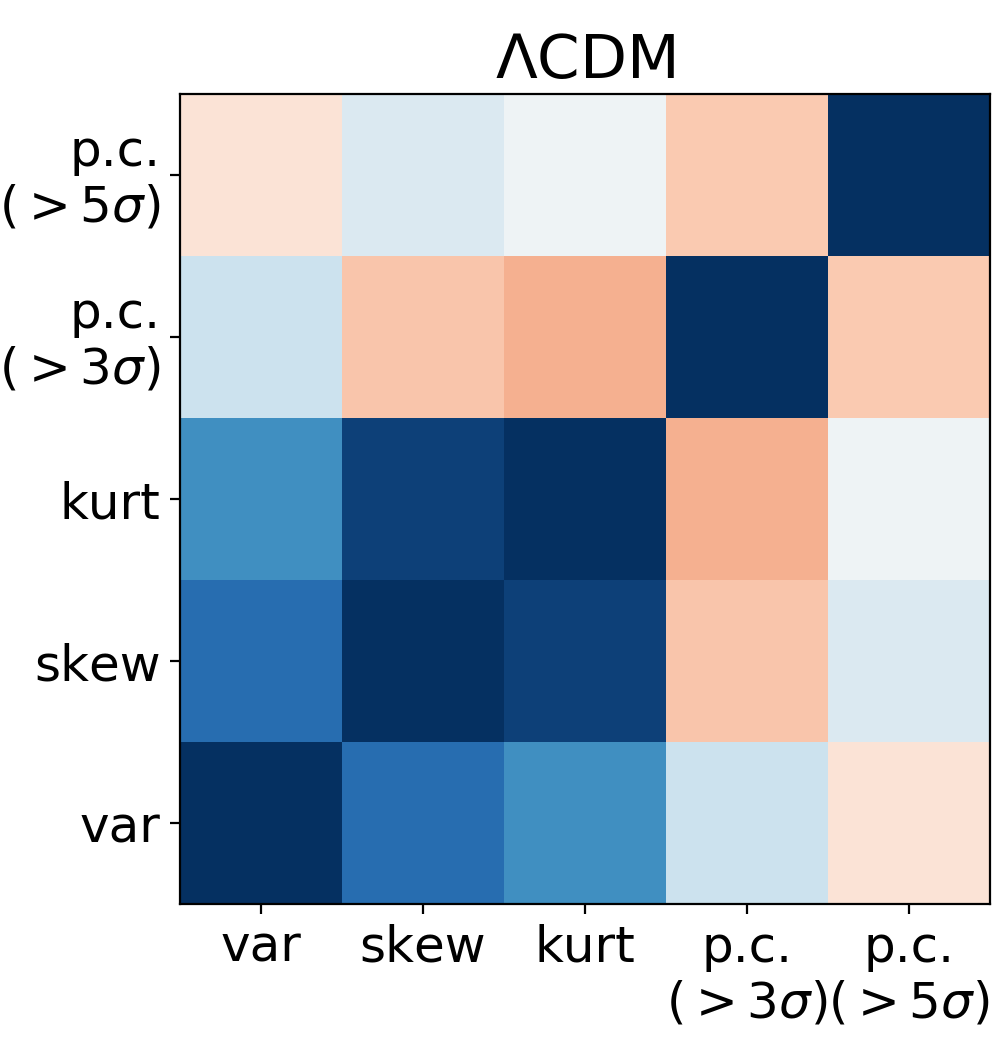}
    \includegraphics[width=0.49\columnwidth]{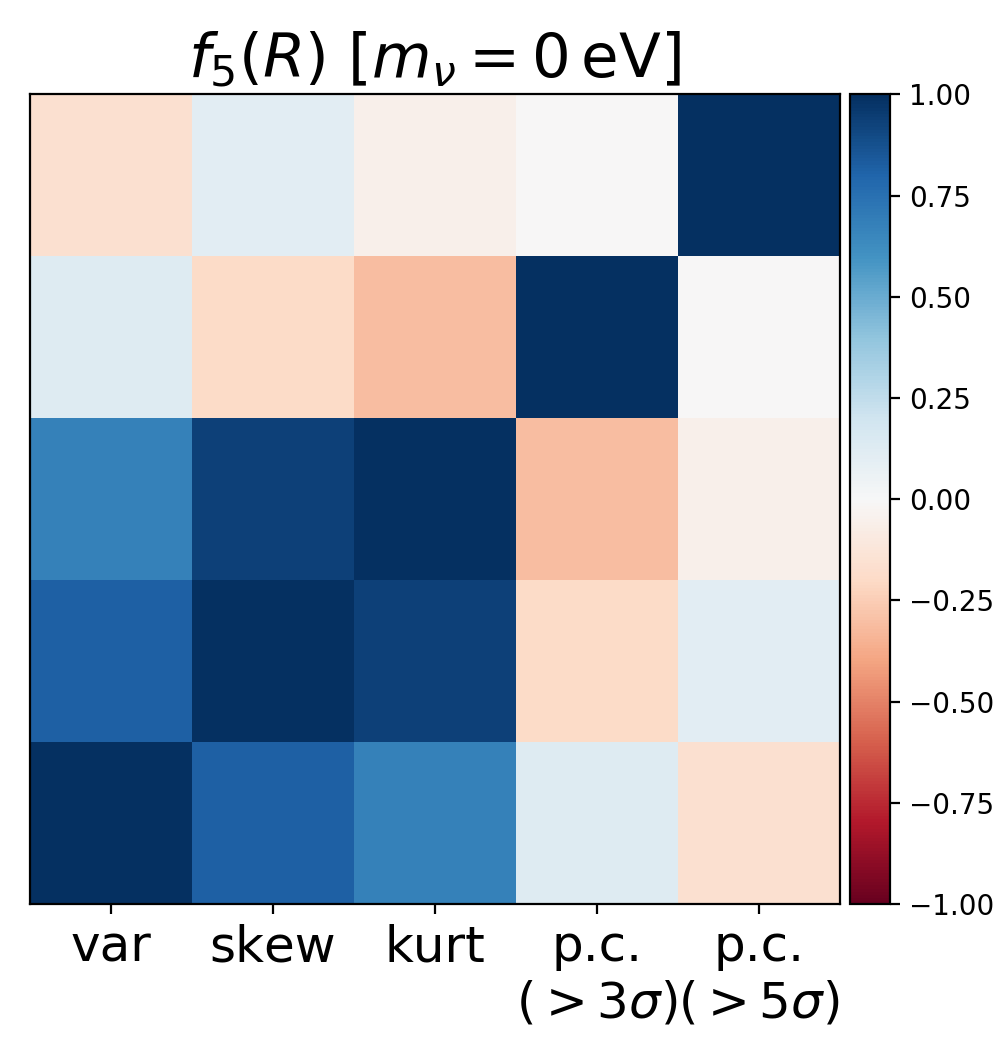}
    \caption{Correlation matrices between aperture mass moments and peak count probes for two models using maps at $z_s=2.0$ and filtering scale $\vartheta_2=0.586'$. Aperture mass moments are strongly positively correlated with each other, while peak counts show slight to moderate negative correlation between the two threshold values. The amplitude of cross correlation between moments and peaks depends on redshift and scale.}
    \label{fig:corr}
\end{figure}

The final non-Gaussian statistics we consider is the abundance of peaks counted above a given $k\,\sigma$ threshold ($k=1,2,...$). Results for $k=3$ and $k=5$ are shown in Fig.~\ref{fig:apmass_wpc}. A higher peak value cutoff means that in general fewer peaks will be detected. This is borne out in the two plots of Fig.~\ref{fig:apmass_wpc}, where the peak count at a fixed scale is about ten times smaller for $k=5$ than for $k=3$. In addition, the peak count for the smallest scale ($\vartheta_1=0.293'$) relatively close to that of $\Lambda$CDM in both cases but significantly larger than $\Lambda$CDM (up to about 15\%) by the second scale ($\vartheta_2=0.568'$). This is similar to what we have seen for all of the $M_\mathrm{ap}$ moments, in particular for $f_5(R)$ with $m_\nu=0$ and $0.1~\mathrm{eV}$.

\begin{figure*}
	\includegraphics[width=\columnwidth]{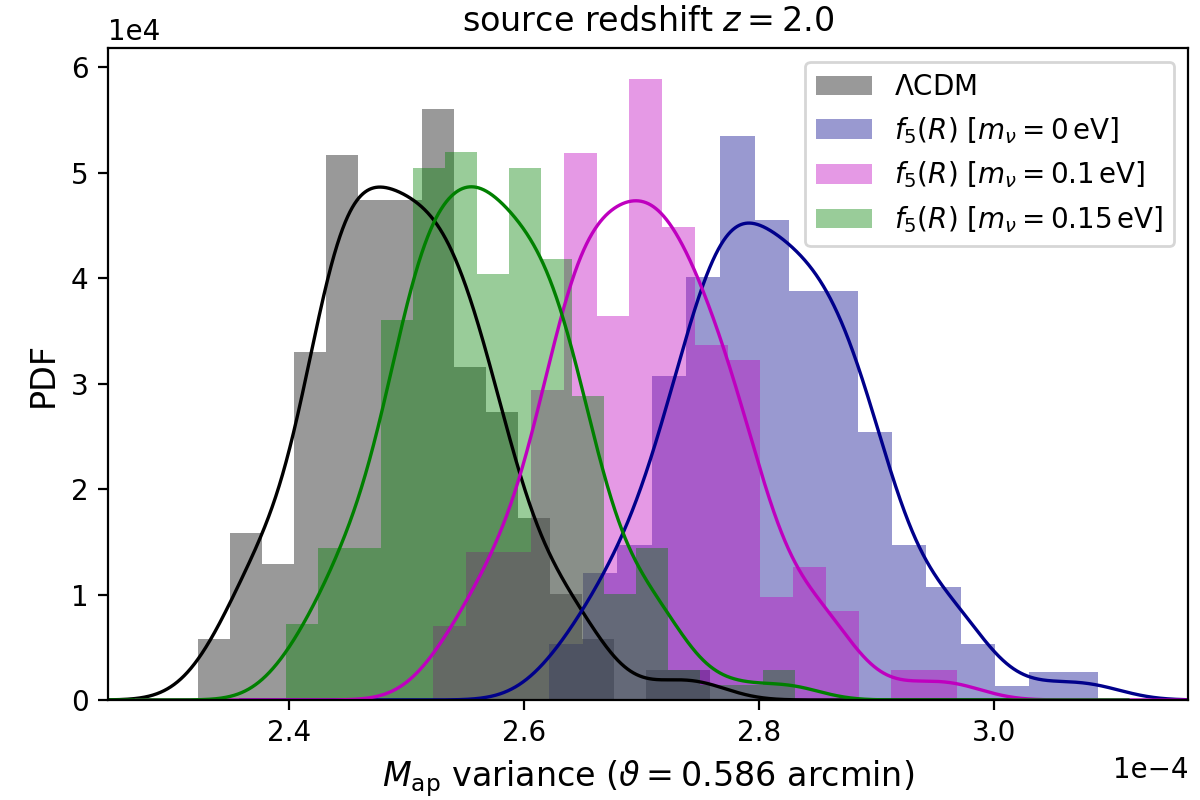}
    \includegraphics[width=\columnwidth]{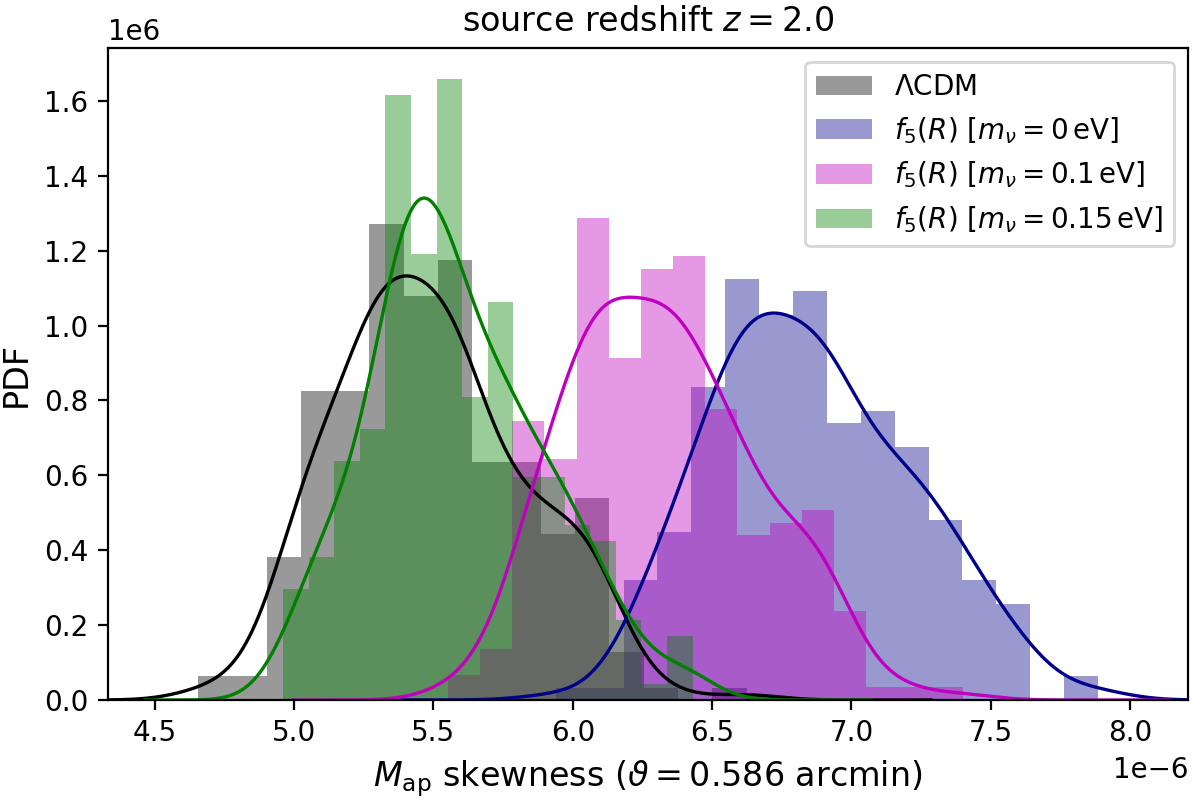}\\
    \includegraphics[width=\columnwidth]{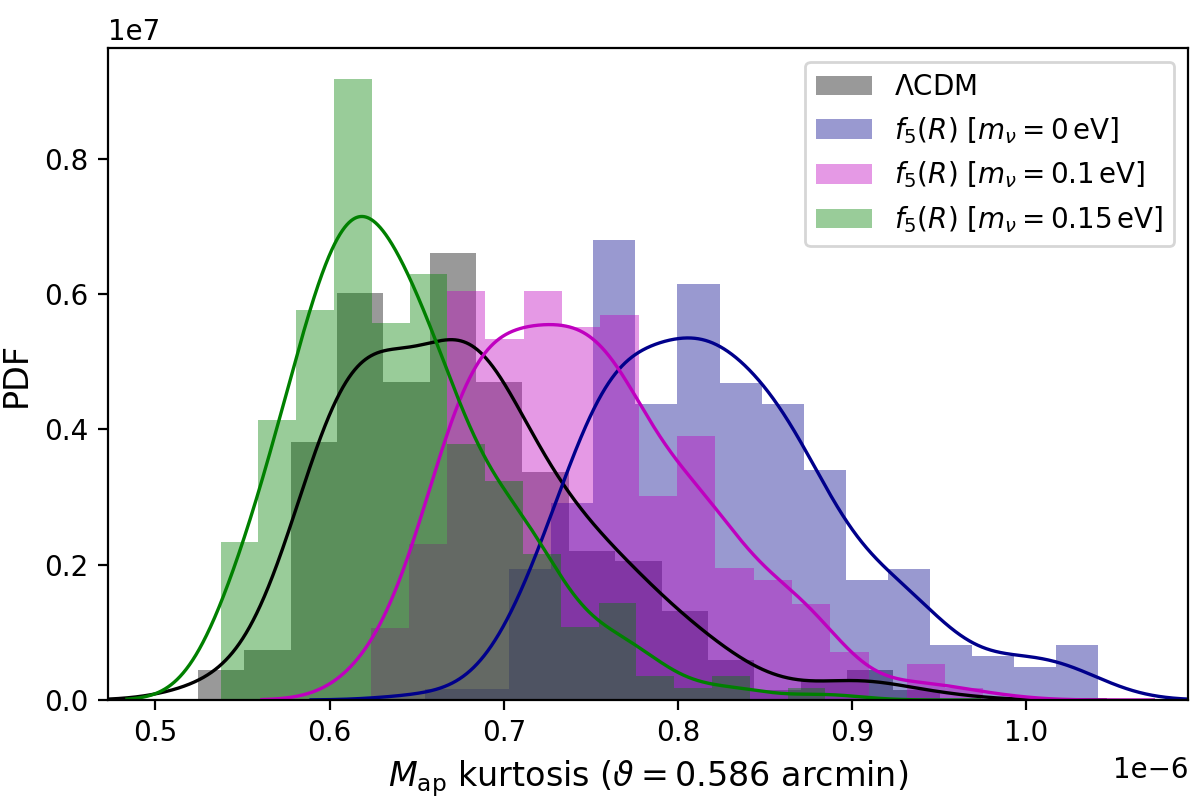}
    \includegraphics[width=\columnwidth]{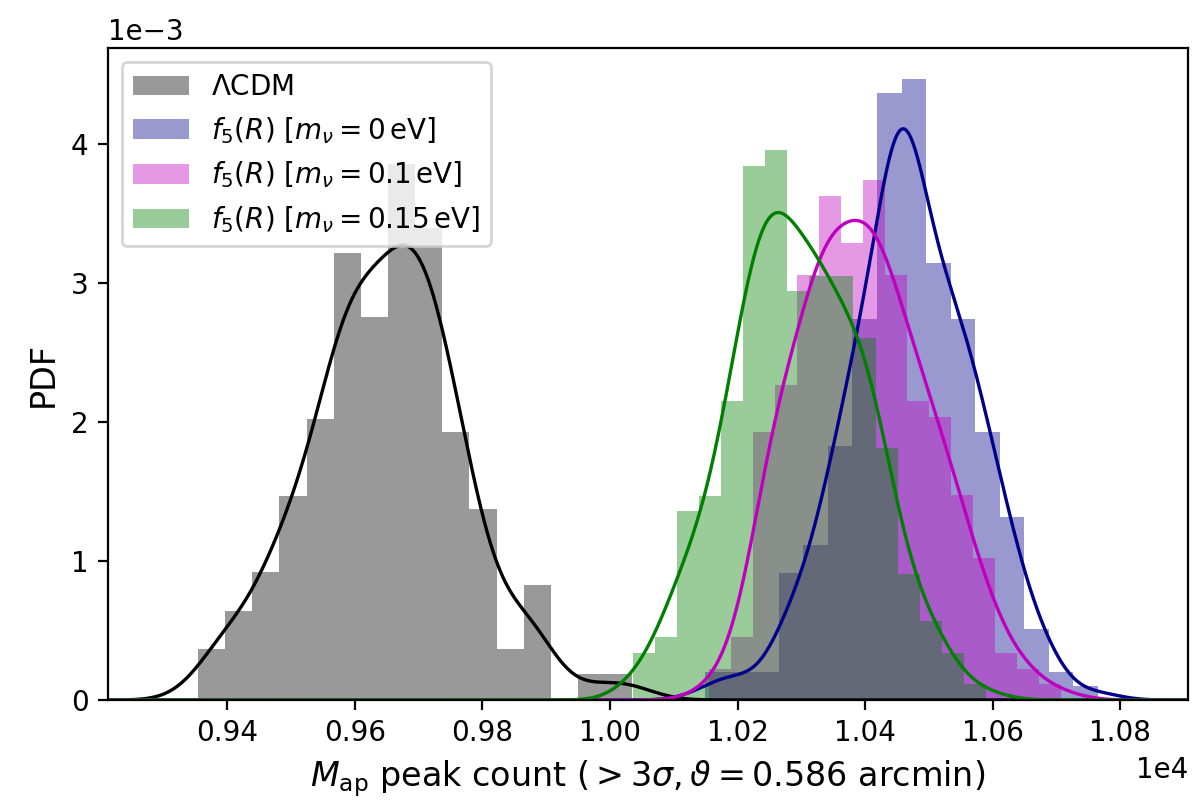}
    \caption{Histograms of aperture mass statistics for $\Lambda$CDM and $f_5(R)$ models with varying neutrino mass $m_\nu$. Each histogram, with area normalised to one, comprises 256 samples of the statistic computed at a filtering scale of $\vartheta=0.586'$ and for sources at redshift $z_s=2.0$. Solid lines represent the result of smoothing the distribution by KDE (cf. Sect.~\ref{subsec:fdr}). The central positions of the histograms correspond to the mean values of each statistic as seen in Figs.~\ref{fig:apmass_var}--\ref{fig:apmass_wpc}. Considering the most degenerate case with $\Lambda$CDM, $f_5(R)$ with $m_\nu=0.15~\mathrm{eV}$, second- and higher-order moments of $M_\mathrm{ap}$ do not appear able to distinguish the models. Peak counts, on the other hand, shown here for a $3\sigma$ threshold, cleanly separate the two distributions. Moreover, it is interesting that peak counts separate all $f_5(R)$ cases from $\Lambda$CDM by approximately the same amount, independent of $m_\nu$.}
    \label{fig:histograms}
\end{figure*}

The impact of neutrino mass on the $f_5(R)$ models is less obvious with peak counts compared to the other statistics across most scales. This is because the variation between models is much smaller than the range of peak count values between different aperture scales (up to four orders in magnitude). Peak counts are therefore much more sensitive to the scale of observation than the variance, skewness, and kurtosis. We notice as well that for a given aperture size, the spread among $f_5(R)$ models with different neutrino masses is less pronounced than for the $M_\mathrm{ap}$ moments. This suggests that peak counts could offer a robust test of modified gravity whether or not neutrinos are considered in the analysis and efficiently break the degeneracy that affects all other WL statistics.

We have found identical trends to those observed in Figs.~\ref{fig:apmass_var}--\ref{fig:apmass_wpc} for the $f_4(R)$ and $f_6(R)$ models with $m_\nu>0$. This is also the case for the other source redshift planes across all models. We further explore the evolution with redshift in Sect.~\ref{subsec:polar_plots}.

In quantifying the level at which one model is in fact distinguishable from another, visual inspection of plots is, of course, not sufficient. We therefore address this question in detail in the following sections by studying the histograms of the different observables and by computing their discrimination efficiencies.

\begin{figure*}
	\includegraphics[width=\columnwidth]{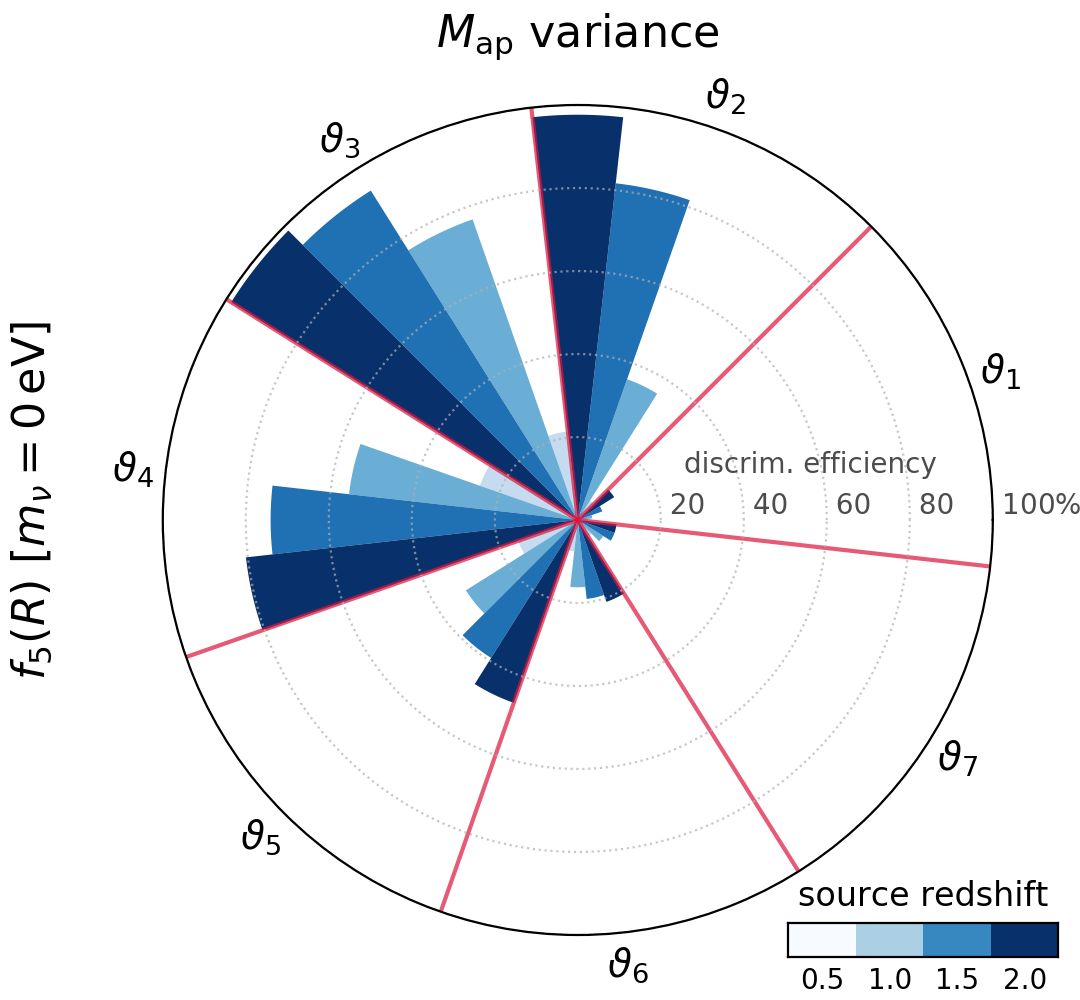}
    \includegraphics[width=\columnwidth]{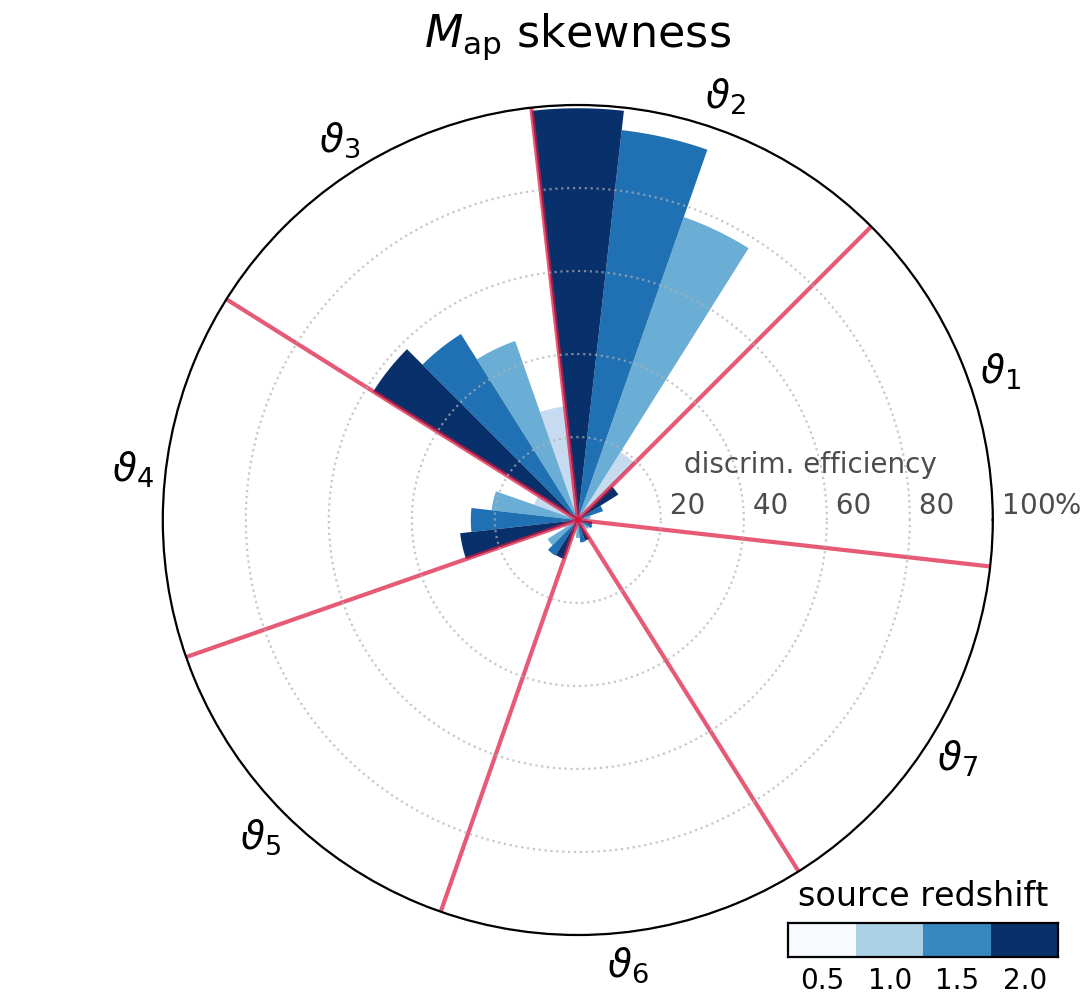}\\
    \includegraphics[width=\columnwidth]{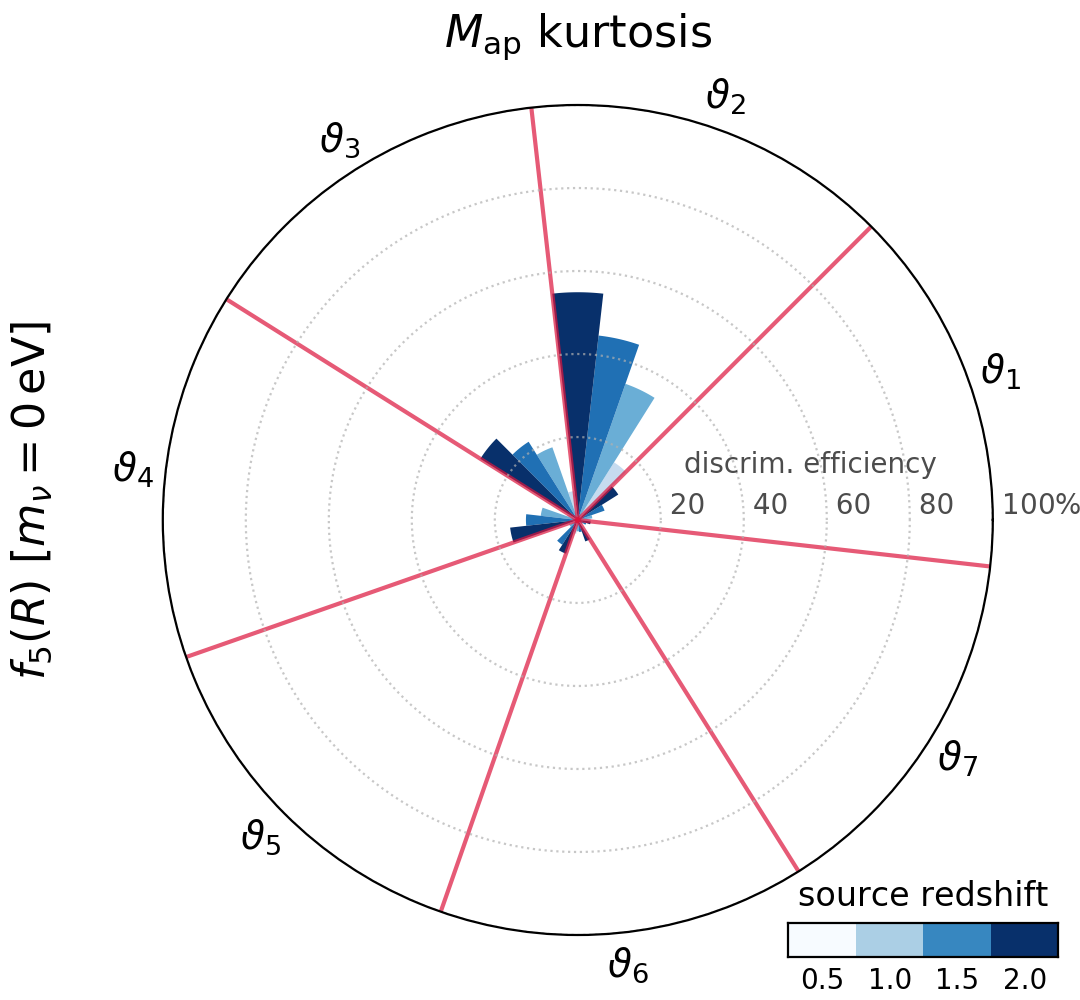}
    \includegraphics[width=\columnwidth]{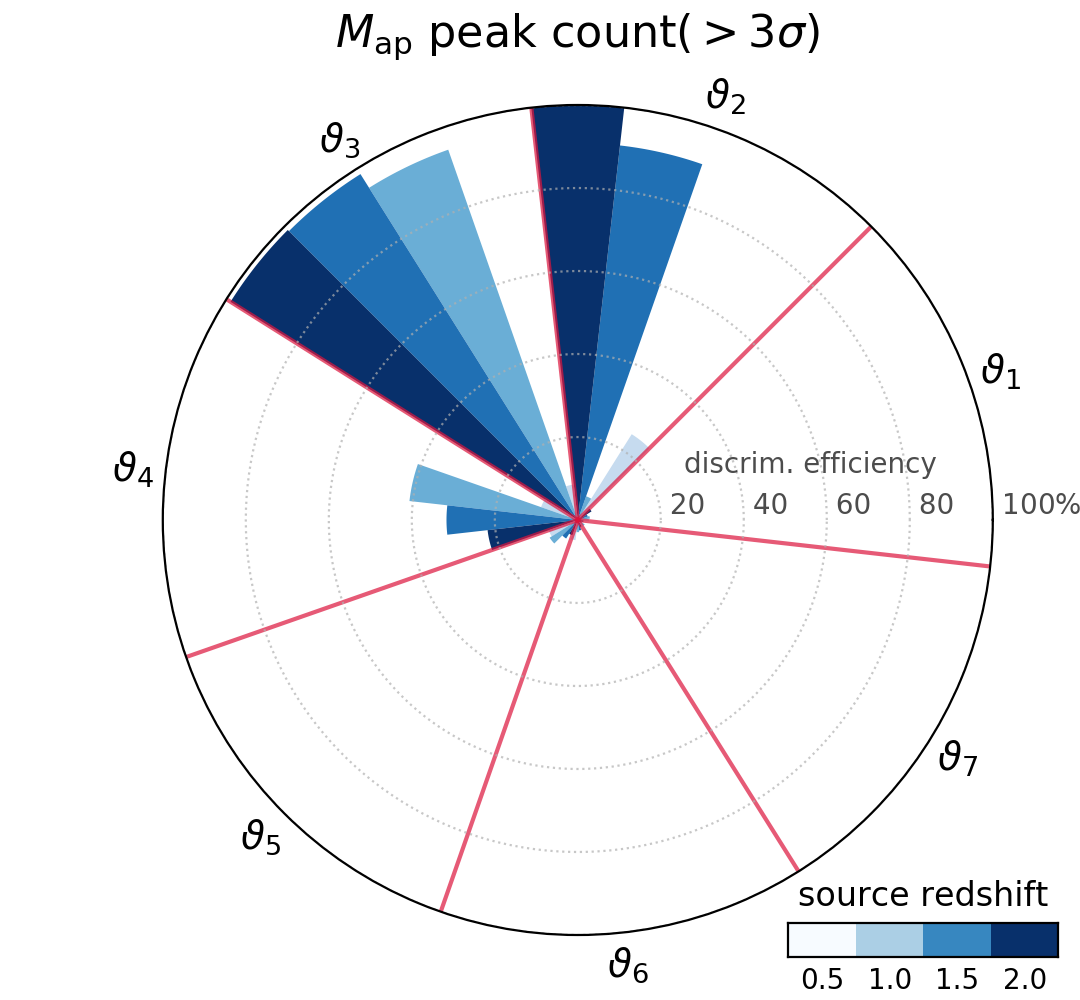}
    \caption{Discrimination efficiency with respect to $\Lambda$CDM of the MG model $f_5(R)$ with $m_\nu=0~\mathrm{eV}$ as a function of statistic, aperture scale, and source redshift. In each plot, the wedges marked by red lines indicate $M_\mathrm{ap}$ filtering by the aperture $\vartheta$ appearing at the outer edge. Numbered apertures correspond to angular sizes of $(\vartheta_1, ..., \vartheta_7) = (0.293', 0.586', 1.17', 2.34', 4.69', 9.34', 18.8')$. The radial length of a bar, shaded according to source redshift, represents the discrimination efficiency (in percent) of the statistic at the filtering scale associated to its wedge. The variance of this model suffices to distinguish it from $\Lambda$CDM at greater than 80\% at scales $\vartheta_2$ and $\vartheta_3$ for sources at $z_s\geq 1.5$. Peak counts (above a $3\sigma$ detection threshold) achieve similar results, while kurtosis proves to be a poor discriminator, not exceeding 55\% at any filter scale or redshift. Skewness shows comparably good discrimination power to the variance and peak counts only at scale $\vartheta_2$.}
    \label{fig:polar_fR5}
\end{figure*}

\begin{figure*}
	\includegraphics[width=\columnwidth]{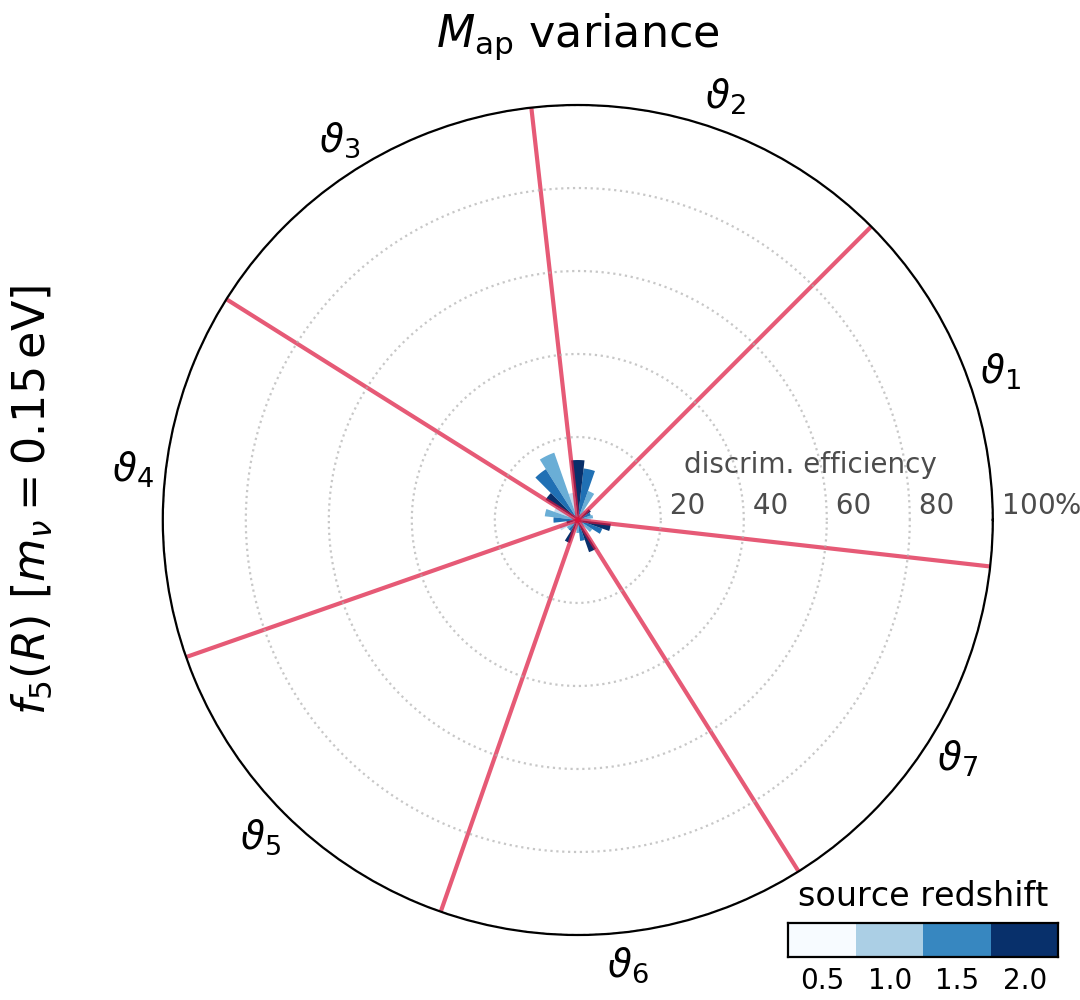}
    \includegraphics[width=\columnwidth]{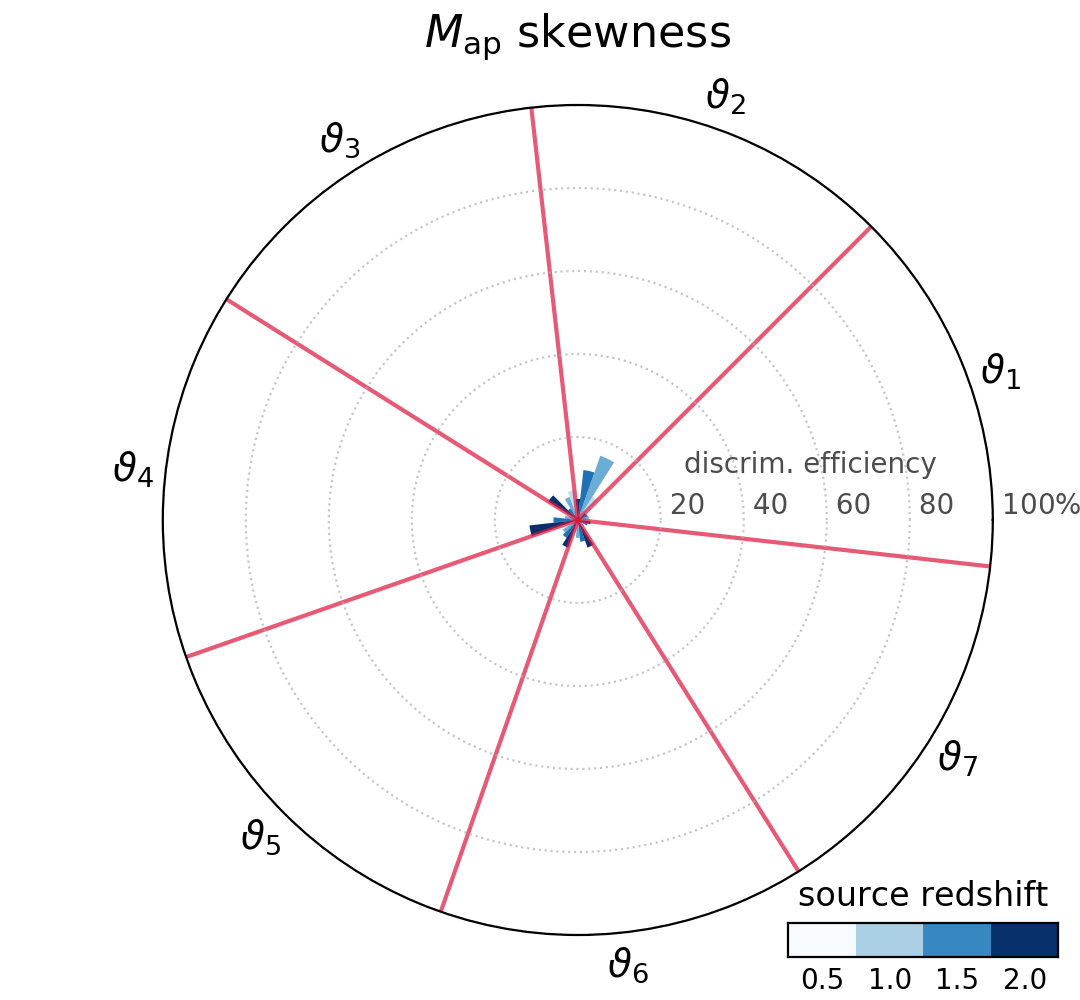}\\
    \includegraphics[width=\columnwidth]{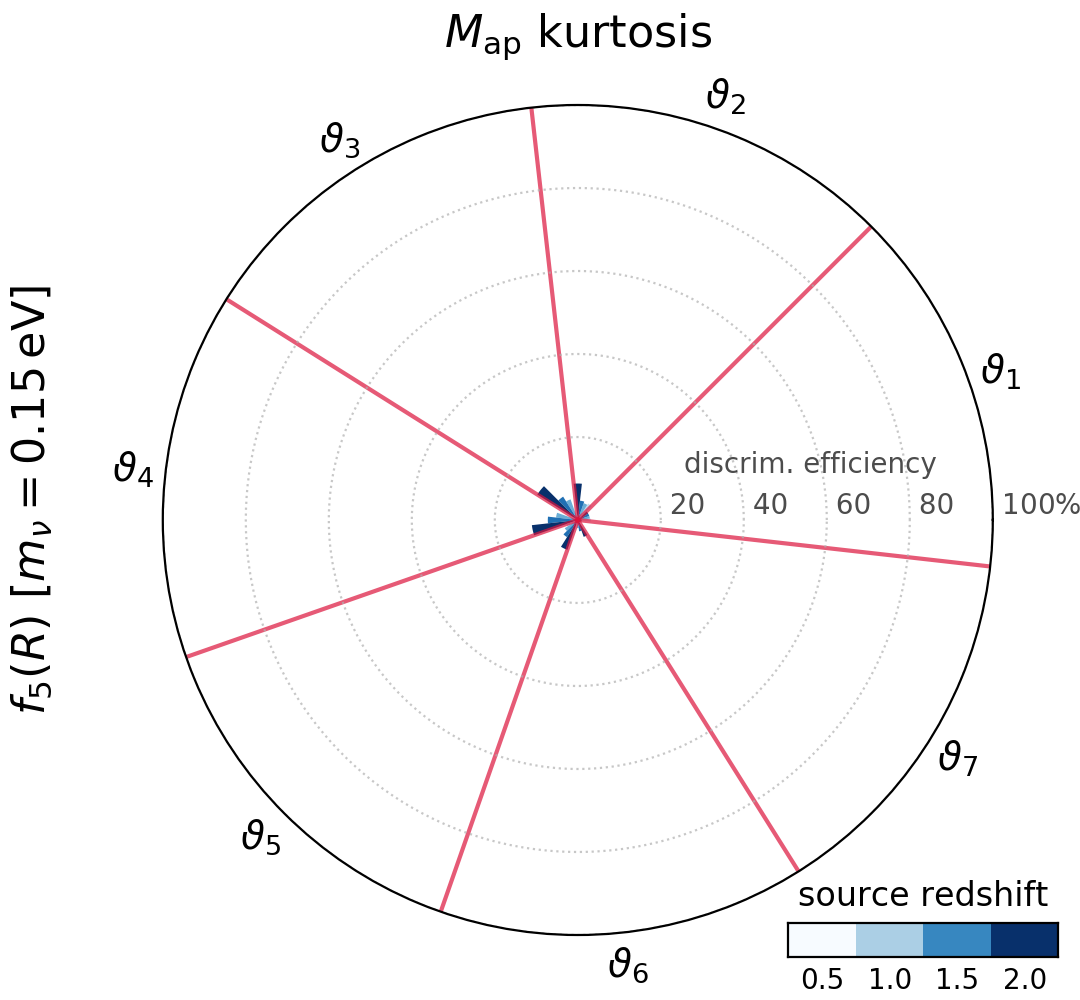}
    \includegraphics[width=\columnwidth]{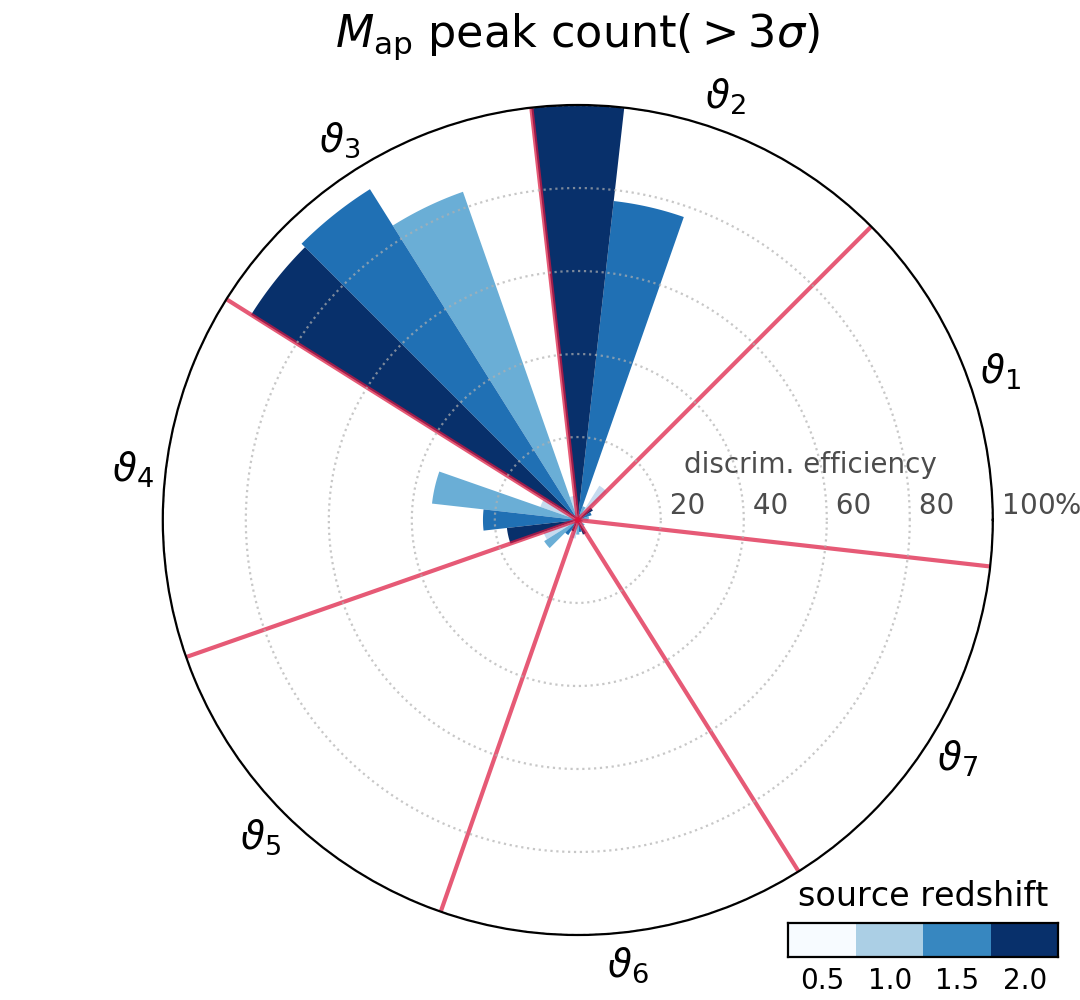}
    \caption{Discrimination efficiency with respect to $\Lambda$CDM of the MG model $f_5(R)$ with $m_\nu=0.15~\mathrm{eV}$ as a function of statistic, aperture scale, and source redshift. Plots are analogous to those in Fig.~\ref{fig:polar_fR5}. Neutrinos significantly enhance the degeneracy of this model with $\Lambda$CDM at the two-point level compared to the $m_\nu=0~\mathrm{eV}$ case. This can be seen in the variance plot (\textit{upper left}), where the discrimination efficiency is less than 18\% at all filter scales and redshifts. The higher-order moments show even less discrimination power. On the contrary, peak counts (\textit{lower right}), again above a $3\sigma$ detection threshold, achieve 100\% discrimination efficiency for $z_s=2.0$ at $\vartheta_2$, and above 92\% for $z_s\geq 1.5$ at $\vartheta_3$.}
    \label{fig:polar_fR5_0.15eV}
\end{figure*}

\subsection{Distributions of observables}
Considering seven filtering scales, four source redshift planes, and five statistical observables in our simulations (variance, skewness, kurtosis, and peak counts above two threshold levels), we have available $7 \times 4 \times 5 = 140$ observables for each simulated cosmology. These observables are not all independent. For example, correlation matrices between the different probes are shown in Fig.~\ref{fig:corr} for the two cases of $\Lambda$CDM and $f_5(R)$ without neutrinos; the cases with $m_\nu>0$ (not shown) look very similar. The statistics have been computed for maps at $z_s=2.0$ and aperture size $\vartheta_2=0.586'$. We see from the figure that aperture mass moments are highly positively correlated with each other at this scale for both models, while peak counts show a slight (moderate) negative correlation for $f_5(R)$ ($\Lambda$CDM) between the two thresholds. Correlations between moments and peak counts are strongest between kurtosis and the $3\sigma$ peak threshold, although this varies with scale and redshift.

We are interested in comparing the histograms of each of these observables between different models in order to determine which, if any, is best at breaking degeneracies. We present here only four representative plots intended to illustrate our approach and results. Shown in Fig.~\ref{fig:histograms} are the area-normalised histograms of $M_\mathrm{ap}$ variance, skewness, kurtosis, and peaks counts for $\Lambda$CDM and MG models. To compare with previous results, we focus on $f_5(R)$ models with $m_\nu=(0, 0.1, 0.15)~\mathrm{eV}$ and choose a filtering scale of $\vartheta_2=0.586'$. We will further investigate in the rest of the paper how results depend on the filtering scale and redshift, for each statistic. Solid lines indicate smoothing by KDE (cf. Sect.~\ref{subsec:fdr}). Referring to Figs.~\ref{fig:apmass_var}--\ref{fig:apmass_wpc}, the mean value of each statistic corresponds to the central positions of the histograms. For example, the variance at this scale of the $m_\nu=0~\mathrm{eV}$ $f_5(R)$ model is most distinct from $\Lambda$CDM, while the $m_\nu=0.15~\mathrm{eV}$ is least.

The histograms allow us to see qualitatively how efficient each statistic is at distinguishing between models. Considering $f_5(R)$ with $m_\nu=0.15~\mathrm{eV}$ (green), the model most degenerate with $\Lambda$CDM (black) in terms of the matter and convergence power spectra, higher order moments of $M_\mathrm{ap}$ do not appear able to break the degeneracy. The skewness and kurtosis histograms overlap with $\Lambda$CDM more than does the variance. On the other hand, peak counts, shown here for a $3\sigma$ threshold, displace the $f_5(R)$ distribution from that of $\Lambda$CDM so that they are nearly disjoint. This is the case as well for $f_5(R)$ with the other neutrino masses, supporting the result seen in Fig.~\ref{fig:apmass_wpc}.

In this example, we have chosen a combination of source redshift and filter scale that provides relatively good separation between the histograms of these models. Sources have been taken at $z_s=2.0$ here, compared to $z_s=1.0$ in Figs.~\ref{fig:apmass_var}--\ref{fig:apmass_wpc}, as the contrast between MG models and GR tends to increase with redshift for each observable. It should be noted as well that for many of the observables considered, that is for other redshifts and filtering scales, we find that the $f(R)$ model histogram is essentially fully coincident with that of $\Lambda$CDM, indicating a negligible discrimination efficiency. Only a relatively small subset of observables appear able to break degeneracies. We therefore further investigate the dependence of discrimination efficiency on redshift and filtering scale in the next subsection.

\subsection{Discrimination efficiency: variation with statistic, scale, and redshift}\label{subsec:polar_plots}
In this section we present the discrimination efficiency with respect to $\Lambda$CDM of two $f_5(R)$ models, $m_\nu=(0, 0.15)~\mathrm{eV}$, as a function of $M_\mathrm{ap}$ statistic, filtering scale, and source galaxy redshift. Figure~\ref{fig:polar_fR5} shows results for $f_5(R)$ without neutrinos. Each polar plot represents one of the statistics, where for peak counts (lower right) the threshold has been set to $3\sigma$.

Each plot is divided into seven wedges indicated by the bold red lines, where each wedge represents a filtering by the aperture $\vartheta$ appearing at the outer edge. Apertures are numbered as before according to wavelet scale and which correspond to angular sizes $(\vartheta_1, ..., \vartheta_7) = (0.293', 0.586', 1.17', 2.34', 4.69', 9.34', 18.8')$. Each wedge contains four bars, colour coded according to source redshift, where the height of each bar represents the discrimination efficiency with respect to $\Lambda$CDM (as a percent) at that redshift and filtering scale. A bar extending from the centre of the figure and touching the outer edge expresses a 100\% discrimination efficiency, and the scaling with radius is linear.

As expected from our previous results, we see that the variance can be a relatively good discriminator between $\Lambda$CDM and $f_5(R)$ without neutrinos. In particular, the discrimination efficiency is at least 80\% at scales $\vartheta_2$ and $\vartheta_3$ for sources at $z_s\geq 1.5$. The skewness performs well only for the single aperture of $\vartheta_2$ and $z_s\geq 1.5$, while the kurtosis is a poor discriminator at all filtering scales and redshifts. Considering that the kurtosis of this model can deviate from $\Lambda$CDM by up to 30\%, or $\sim$10\% more than the variance (cf. Figs.~\ref{fig:apmass_var} and \ref{fig:apmass_skew_kurt}), it is somewhat surprising that this fourth-order statistic does not offer more discrimination power compared to second order.

The final plot in Fig.~\ref{fig:polar_fR5} reveals that peak counts can discriminate at approximately the same level as the variance or better, at least for filtering scales $\vartheta_2$ and $\vartheta_3$. Indeed, peak count discrimination efficiency even exceeds that of the variance for $z_s=1.5$, for example, at these two scales. We have checked that by raising the peak detection threshold to $5\sigma$, discrimination efficiency rises to 100\% at $\vartheta_2$ for sources at $z_s\geq 1.0$. A consequence, however, is that the efficiency is reduced at $\vartheta_3$ for the higher redshift source planes. From the cases we have checked, it appears that the relationship between discrimination efficiency and peak count threshold, filtering scale, and redshift is highly non-trivial. 

We recall that our peak statistics do not take into account the full peak distribution but instead constitute a simple survival function, in other words the number of peaks remaining after a certain cut. As a result, the signal for a given threshold is dominated by the peaks with amplitudes near that threshold, of which we have only considered two. The peak information we explore is therefore sub-optimal, and results would likely improve by using the full distribution. Nonetheless, it is clear even from our simple approach that peak counts can outperform standard moments in breaking degeneracies between MG and GR using weak lensing.

Figure~\ref{fig:polar_fR5_0.15eV} is analogous to Fig.~\ref{fig:polar_fR5} for the $f_5(R)$ model with $m_\nu=0.15~\mathrm{eV}$. Comparing to the case without neutrinos, it is  apparent that no aperture mass moments up to fourth order are capable of distinguishing this model from $\Lambda$CDM. On the contrary, peak counts again show a promising result. For $\vartheta_2=0.586'$ filtering, sources at $z_s=2.0$ give a 100\% discrimination efficiency. For $\vartheta_3=1.17'$ filtering, the efficiency is 93\% for $z_s\geq 1.5$ and 84\% for $z_s=1.0$. We have checked here as well that raising the detection level to $5\sigma$ produces the same effect as it did for the previous case without neutrinos. Given that this model was designed to be degenerate with $\Lambda$CDM, it is a useful result, and indeed the main result of the paper, that there exist observables capable of discriminating between the models with high efficiency.

\begin{table}
\caption{Maximum discrimination efficiency $\mathcal{E}_{\max}$ with respect to $\Lambda$CDM attained for each MG model according to $M_\mathrm{ap}$ statistic, filtering scale $\vartheta$, and source redshift $z_s$. Results for the variance are shown first, along with the best available statistic underneath. For all models except $f_4(R)$ without neutrinos, the unique maximum discrimination efficiency is achieved with peak counts (p.c.), either above the $3\sigma$ or $5\sigma$ threshold. For $f_4(R)$ with $m_\nu=0~\mathrm{eV}$, all of the statistics reach 100\% discrimination efficiency for at least one combination of $\vartheta$ and $z_s$.}
\centering
{\renewcommand{\arraystretch}{1.35}
\begin{tabularx}{\columnwidth}{lcccc}
  \hline\hline
  Model & $\mathcal{E}_{\max}$ & statistic & $\vartheta$ & $z_s$\\
  \hline
  \multirow{2}{*}{$f_4(R)$ $[m_\nu=0~\mathrm{eV}]$} & 100\% & var. & mult. & mult. \\
                  & 100\% & all others & mult. & mult. \\
  \hline
  \multirow{2}{*}{$f_4(R)$ $[m_\nu=0.3~\mathrm{eV}]$} & 19\% & var. & $0.586'$ & 2.0 \\
                  & 38\% & p.c. ($>3\sigma$) & $0.586'$ & 2.0 \\
  \hline
  \multirow{2}{*}{$f_5(R)$ $[m_\nu=0~\mathrm{eV}]$} & 99\% & var. & $1.17'$ & 2.0 \\
                  & 100\% & p.c. (mult.) & $0.586'$ & mult. \\
  \hline
  \multirow{2}{*}{$f_5(R)$ $[m_\nu=0.1~\mathrm{eV}]$} & 76\% & var. & $1.17'$ & 2.0 \\
                  & 100\% & p.c. (mult.) & $0.586'$ & mult. \\
  \hline
  \multirow{2}{*}{$f_5(R)$ $[m_\nu=0.15~\mathrm{eV}]$} & 17\% & var. & $1.17'$ & 1.0 \\
                  & 100\% & p.c. (mult.) & $0.586'$ & mult. \\
  \hline
  \multirow{2}{*}{$f_6(R)$ $[m_\nu=0~\mathrm{eV}]$} & 12\% & var. & $1.17'$ & 1.0 \\
                  & 39\% & p.c. ($>5\sigma$) & $18.8'$ & 1.5 \\
  \hline
  \multirow{2}{*}{$f_6(R)$ $[m_\nu=0.06~\mathrm{eV}]$} & 4.7\% & var. & $0.586'$ & 1.5 \\
                  & 41\% & p.c. ($>5\sigma$) & $0.586'$ & 2.0 \\
  \hline
  \multirow{2}{*}{$f_6(R)$ $[m_\nu=0.1~\mathrm{eV}]$} & 12\% & var. & $2.34'$ & 2.0 \\
                  & 66\% & p.c. ($>5\sigma$) & $0.586'$ & 2.0 \\
  \hline
\end{tabularx}}
\label{tab:max_disc_eff}
\end{table}

We have carried out the same analysis for the other MG models, namely $f_{R0}=-10^{-4}$ with $m_\nu=(0, 0.3)~\mathrm{eV}$ and $f_{R0}=-10^{-6}$ with $m_\nu=(0, 0.06, 0.1)~\mathrm{eV}$. A summary of the results is presented in Table~\ref{tab:max_disc_eff}, and $f_5(R)$ is included for comparison. Shown are the maximum discrimination efficiencies from $\Lambda$CDM attained for each model along with the statistic, aperture $\vartheta$, and source redshift $z_s$ that produces it. To highlight the difference between second- and higher-order statistics, results for the variance and for the best discriminating statistic are shown for each model. For all models except $f_4(R)$ $[m_\nu=0~\mathrm{eV}]$, the unique maximum efficiency is achieved with peaks counted either above a $3\sigma$ or $5\sigma$ threshold. On the other hand, the $f_4(R)$ case with $m_\nu=0~\mathrm{eV}$ reaches 100\% with all of the statistics for multiple combinations of $\vartheta$ and $z_s$. This is expected given that this model diverges from $\Lambda$CDM most significantly already at the power spectrum level.

\section{Conclusion}\label{sec:conclusion}
The cosmological data do not yet point to a unique model within which all observations can be explained, with several cosmologies still fitting the data. The standard $\Lambda$CDM model indeed accommodates the broad range of current observational probes that measure structure growth and the universal expansion across cosmic time. The fundamental nature of the late-time acceleration, however, remains unclear---in particular whether it is caused by a fluid (dark energy) component, a cosmological constant $\Lambda$, or instead by a modification to general relativity at large scales. Many modified gravity models, such as the $f(R)$ family, are still viable given the data.

We have explored in this paper several particular $f(R)$ models that mimic $\Lambda$CDM in terms of their background evolution and their matter power spectra at $z=0$. The models can be written in terms of two free parameters, $n$ and $f_{R0}$, which determine the density and scale at which the modified gravitational interaction takes effect. We fixed $n=1$ throughout and considered $|f_{R0}|$ values within the range $10^{-4}$ to $10^{-6}$.

By design, the models we chose are difficult to distinguish from $\Lambda$CDM based on observations at linear scales. Furthermore, using $N$-body simulations, we showed that the amplitude of the matter power spectrum is larger relative to $\Lambda$CDM with increasing $|f_{R0}|$, but this can be in turn reduced by the inclusion of massive neutrinos (up to $m_\nu=0.3~\mathrm{eV}$) in the model. This is because massive neutrinos suppress the growth of structure on scales smaller than their free-streaming length, thereby compensating the increased clustering due to larger $|f_{R0}|$. For certain combinations then of $f_{R0}$ and $m_\nu$, the model well reproduces $\Lambda$CDM not only on linear scales, but also into the non-linear regime.

Our primary goal has been to determine whether there are weak-lensing observables that are more efficient than standard second-order statistics in discriminating between GR and MG. We first verified that the convergence power spectrum $P_\kappa(\ell)$ and aperture mass variance $\langle M_\mathrm{ap}^2 \rangle$, both two-point measurements, of our simulated cosmologies behaved as expected and consistently with each other. As with the matter power spectra, deviations from $\Lambda$CDM of the MG models decreased as $|f_{R0}|$ decreased. Then, focusing on $f_5(R)$, we showed that the discrepancy diminished further by including $m_\nu$ up to 0.15 eV over the angular scales considered.

The non-Gaussian weak-lensing observables we have studied are the aperture mass skewness (third order), kurtosis (fourth order), and peak counts as a function of map filtering scale and source galaxy redshift. We considered redshifts up to $z=2.0$ and aperture scales in the range from $0.293'$ up to $18.8'$. Provided that noise is properly taken into account, a full treatment of which is beyond the scope of this work (though see the Appendix), our results may be potentially interesting for future wide-field galaxy surveys like Euclid.

For $f_5(R)$ with $m_\nu=0.15~\mathrm{eV}$, the $f_5(R)$ model most closely mimicking $\Lambda$CDM, there exist filtering scales for which the skewness and kurtosis measurements deviate more from $\Lambda$CDM compared to the variance, indicating a greater chance of using these observables to discriminate between the models. This was also the case for peak counts above thresholds of $3\sigma$ and $5\sigma$. However, a further study of the discrimination efficiency based on the false discovery rate (FDR) formalism showed that only peak counts offered any reliable power to distinguish the MG model from $\Lambda$CDM. The discrimination efficiency calculation does this by quantifying the overlap between the actual distributions of observables computed over the 256 model realisations. We conclude therefore that we are less likely to mistake a true MG model for $\Lambda$CDM by measuring peak counts compared to second- and higher-order moments of the aperture mass.

A particularly interesting feature of peak counts we have found is that they can be less sensitive to differences in neutrino mass than to the model of gravity. The example in Fig.~\ref{fig:histograms} demonstrates this in the significant overlap among the three $f_5(R)$ histograms which are each in turn nearly fully disjoint from the $\Lambda$CDM histogram. This behaviour was seen as well for the other sets of models with the same value of $f_{R0}$ but differing $m_\nu$. In addition to offering more discrimination power compared to aperture mass moments, at least for the models we have considered, peaks may generally be a more robust observable for distinguishing between standard and modified gravity. 

To better understand the sensitivity of peak counts to neutrinos, it would be useful to systematically study variations in the halo mass function as a function of $m_\nu$ and redshift for the same $f(R)$ model. As high signal-to-noise weak-lensing peaks are thought to trace the most massive DM halos in the universe, peak counts can probe the high-mass tail of the mass function. Some results in \cite{HCB.etal.2018}, who used the same {\small DUSTGRAIN}-{\em pathfinder} simulations as we have here, are already instructive on this front (cf. their Figs.~8 and 9). For example, at $z=0$, the $f_5(R)$ model without neutrinos shows an increase in DM halo abundance of between 10\% and 20\% relative to GR over the mass range $13.3 \leq \log_{10}(M_{200\mathrm{m}}\,h\,/\,M_\odot) \leq 15.0$, where $M_{200m}$ ($M_\odot$) is the halo (solar) mass. The same model with $m_\nu=0.15~\mathrm{eV}$ shows a maximum increase of only up to about 12\% over the same mass range, a reduction we expect from the MG--neutrino degeneracy, but interestingly the halo abundance prediction becomes consistent with GR for the most massive halos around $10^{15} \,M_\odot/h$. This is not the case for $f_4(R)$ with $m_\nu=0.3~\mathrm{eV}$, where the GR abundance is recovered a low but not high masses, and results for all models evolve with redshift as well. A detailed study of the mass function and its correlations with the peak count signal over the full parameter space we have considered would therefore be enlightening, however it is beyond the scope of this work.

In terms of source redshift dependence, we have found generally that higher $z_s$ observations tend to provide greater discrimination efficiencies between models for every statistic, although we did find exceptions for particular filtering scales. In no case, however, did $z_s=0.5$ outperform $z_s=2.0$, which we can understand as the high redshift convergence maps carrying more significant cosmological information that is accessible by these statistics. Physically, since light travels through (and is distorted by) more structures along the line of sight, we expect better sensitivity for higher redshift sources. Moreover, as we considered the different redshift planes independently, we have not exploited the evolution of observables with $z$, which likely constitutes a further signal that could be used to distinguish the models. 

We found similar but not precisely equivalent results for the $f_4(R)$ and $f_6(R)$ models (both with $m_\nu=0$ and $m_\nu>0$) compared to $f_5(R)$ (cf. Table~\ref{tab:max_disc_eff}). They are similar in the sense that peak counts remain the best observable tested for breaking degeneracies with $\Lambda$CDM for any model with $m_\nu>0$. On the other hand, no filtering scale nor source galaxy redshift provided a larger discrimination efficiency than 66\% when $m_\nu>0$, significantly lower than was seen for $f_5(R)$ models. This reflects the strong degeneracy that can persist for suitably chosen combinations of MG parameters and neutrino mass sums. Without neutrinos, the $f_4(R)$ model is easily distinguishable from $\Lambda$CDM, while $f_6(R)$ is not.

As we have not sought in this paper to find an optimal statistic, one may reasonably wonder whether there exist other $M_\mathrm{ap}$ observables (e.g. derived from different filter functions, filtering scales, statistics, etc.) that would be better at breaking degeneracies not only between MG and GR, but also within MG models themselves. A likely way to improve results would be to use the full distribution of peaks as a function of signal-to-noise, rather than the simple threshold cut we have employed. We leave a dedicated study of this question to future work.

\begin{acknowledgements}
AP acknowledges support by an Enhanced Eurotalents Fellowship, a Marie Sk\l{}odowska-Curie Actions Programme co-funded by the European Commission and Commissariat {\`a} l'{\'e}nergie atomique et aux {\'e}nergies alternatives (CEA). AP wishes to thank M. Kilbinger, S. Pires, D. Elbaz, and S. Casas for many useful discussions while preparing this paper. VP and MB thank L. Lombriser and K. Koyama for useful discussions on $f(R)$ models.
CG and MB acknowledge support from the Italian Ministry for Education, University and Research (MIUR) through the SIR individual grant SIMCODE, project number RBSI14P4IH, from the grant MIUR PRIN 2015 ``Cosmology and Fundamental Physics: illuminating the Dark Universe with Euclid" and from the agreement ASI n.I/023/12/0 ``Attivit\`a relative alla fase B2/C per la missione Euclid".
The authors wish to acknowledge the European Community through the grant DEDALE (contract no. 665044) within the H2020 Framework Programme of the European Commission, the Euclid Collaboration, the European Space Agency and the support of the Centre National d'Etudes Spatiales.
The {\small DUSTGRAIN}-{\em pathfinder} simulations discussed in this work have been performed and analysed on the Marconi supercomputing machine at Cineca thanks to the PRACE project SIMCODE1 (grant nr. 2016153604) and on the computing facilities of the Computational Center for Particle and Astrophysics (C2PAP) and of the Leibniz Supercomputer Centre (LRZ) under the project ID pr94ji.
\end{acknowledgements}

\bibliographystyle{aa}
\bibliography{refs}

\begin{thebibliography}{105}
\expandafter\ifx\csname natexlab\endcsname\relax\def\natexlab#1{#1}\fi

\bibitem[{{Amendola}(2000)}]{Amendola.2000}
{Amendola}, L. 2000, \prd, 62, 043511

\bibitem[{{Amendola} {et~al.}(2018){Amendola}, {Appleby}, {Avgoustidis},
  {Bacon}, {Baker}, {Baldi}, {Bartolo}, {Blanchard}, {Bonvin}, {Borgani},
  {Branchini}, {Burrage}, {Camera}, {Carbone}, {Casarini}, {Cropper}, {de
  Rham}, {Dietrich}, {Di Porto}, {Durrer}, {Ealet}, {Ferreira}, {Finelli},
  {Garc{\'{\i}}a-Bellido}, {Giannantonio}, {Guzzo}, {Heavens}, {Heisenberg},
  {Heymans}, {Hoekstra}, {Hollenstein}, {Holmes}, {Hwang}, {Jahnke},
  {Kitching}, {Koivisto}, {Kunz}, {La Vacca}, {Linder}, {March}, {Marra},
  {Martins}, {Majerotto}, {Markovic}, {Marsh}, {Marulli}, {Massey}, {Mellier},
  {Montanari}, {Mota}, {Nunes}, {Percival}, {Pettorino}, {Porciani},
  {Quercellini}, {Read}, {Rinaldi}, {Sapone}, {Sawicki}, {Scaramella},
  {Skordis}, {Simpson}, {Taylor}, {Thomas}, {Trotta}, {Verde}, {Vernizzi},
  {Vollmer}, {Wang}, {Weller}, \& {Zlosnik}}]{Euclid.2016}
{Amendola}, L., {Appleby}, S., {Avgoustidis}, A., {et~al.} 2018, Living Reviews
  in Relativity, 21, 2

\bibitem[{{Anderson} {et~al.}(2014){Anderson}, {Aubourg}, {Bailey}, {Beutler},
  {Bhardwaj}, {Blanton}, {Bolton}, {Brinkmann}, {Brownstein}, {Burden},
  {Chuang}, {Cuesta}, {Dawson}, {Eisenstein}, {Escoffier}, {Gunn}, {Guo}, {Ho},
  {Honscheid}, {Howlett}, {Kirkby}, {Lupton}, {Manera}, {Maraston}, {McBride},
  {Mena}, {Montesano}, {Nichol}, {Nuza}, {Olmstead}, {Padmanabhan},
  {Palanque-Delabrouille}, {Parejko}, {Percival}, {Petitjean}, {Prada},
  {Price-Whelan}, {Reid}, {Roe}, {Ross}, {Ross}, {Sabiu}, {Saito}, {Samushia},
  {S{\'a}nchez}, {Schlegel}, {Schneider}, {Scoccola}, {Seo}, {Skibba},
  {Strauss}, {Swanson}, {Thomas}, {Tinker}, {Tojeiro}, {Maga{\~n}a}, {Verde},
  {Wake}, {Weaver}, {Weinberg}, {White}, {Xu}, {Y{\`e}che}, {Zehavi}, \&
  {Zhao}}]{BOSS.2014}
{Anderson}, L., {Aubourg}, {\'E}., {Bailey}, S., {et~al.} 2014, \mnras, 441, 24

\bibitem[{{Baker} {et~al.}(2017){Baker}, {Bellini}, {Ferreira}, {Lagos},
  {Noller}, \& {Sawicki}}]{BBF.etal.2017}
{Baker}, T., {Bellini}, E., {Ferreira}, P.~G., {et~al.} 2017, Physical Review
  Letters, 119, 251301

\bibitem[{{Baldi} {et~al.}(2014){Baldi}, {Villaescusa-Navarro}, {Viel},
  {Puchwein}, {Springel}, \& {Moscardini}}]{BVNV.etal.2014}
{Baldi}, M., {Villaescusa-Navarro}, F., {Viel}, M., {et~al.} 2014, \mnras, 440,
  75

\bibitem[{{Benjamini} \& {Hochberg}(1995)}]{BH.1995}
{Benjamini}, Y. \& {Hochberg}, Y. 1995, Journal of the Royal Statistical
  Society. Series B (Methodological), 57, 289

\bibitem[{{Bernardeau} {et~al.}(1997){Bernardeau}, {van Waerbeke}, \&
  {Mellier}}]{BvWM.1997}
{Bernardeau}, F., {van Waerbeke}, L., \& {Mellier}, Y. 1997, \aap, 322, 1

\bibitem[{{Betoule} {et~al.}(2014){Betoule}, {Kessler}, {Guy}, {Mosher},
  {Hardin}, {Biswas}, {Astier}, {El-Hage}, {Konig}, {Kuhlmann}, {Marriner},
  {Pain}, {Regnault}, {Balland}, {Bassett}, {Brown}, {Campbell}, {Carlberg},
  {Cellier-Holzem}, {Cinabro}, {Conley}, {D'Andrea}, {DePoy}, {Doi}, {Ellis},
  {Fabbro}, {Filippenko}, {Foley}, {Frieman}, {Fouchez}, {Galbany}, {Goobar},
  {Gupta}, {Hill}, {Hlozek}, {Hogan}, {Hook}, {Howell}, {Jha}, {Le Guillou},
  {Leloudas}, {Lidman}, {Marshall}, {M{\"o}ller}, {Mour{\~a}o}, {Neveu},
  {Nichol}, {Olmstead}, {Palanque-Delabrouille}, {Perlmutter}, {Prieto},
  {Pritchet}, {Richmond}, {Riess}, {Ruhlmann-Kleider}, {Sako}, {Schahmaneche},
  {Schneider}, {Smith}, {Sollerman}, {Sullivan}, {Walton}, \&
  {Wheeler}}]{BKG.etal.2014}
{Betoule}, M., {Kessler}, R., {Guy}, J., {et~al.} 2014, \aap, 568, A22

\bibitem[{{Bettoni} {et~al.}(2017){Bettoni}, {Ezquiaga}, {Hinterbichler}, \&
  {Zumalac{\'a}rregui}}]{BEH.etal.2017}
{Bettoni}, D., {Ezquiaga}, J.~M., {Hinterbichler}, K., \& {Zumalac{\'a}rregui},
  M. 2017, \prd, 95, 084029

\bibitem[{{Boubekeur} {et~al.}(2014){Boubekeur}, {Giusarma}, {Mena}, \&
  {Ram{\'{\i}}rez}}]{BGM.etal.2014}
{Boubekeur}, L., {Giusarma}, E., {Mena}, O., \& {Ram{\'{\i}}rez}, H. 2014,
  \prd, 90, 103512

\bibitem[{{Castro} {et~al.}(2018){Castro}, {Quartin}, {Giocoli}, {Borgani}, \&
  {Dolag}}]{CQG.etal.2018}
{Castro}, T., {Quartin}, M., {Giocoli}, C., {Borgani}, S., \& {Dolag}, K. 2018,
  \mnras, 478, 1305

\bibitem[{{Clowe} {et~al.}(2006){Clowe}, {Schneider}, {Arag{\'o}n-Salamanca},
  {Bremer}, {De Lucia}, {Halliday}, {Jablonka}, {Milvang-Jensen}, {Pell{\'o}},
  {Poggianti}, {Rudnick}, {Saglia}, {Simard}, {White}, \&
  {Zaritsky}}]{CSA.etal.2006}
{Clowe}, D., {Schneider}, P., {Arag{\'o}n-Salamanca}, A., {et~al.} 2006, \aap,
  451, 395

\bibitem[{{Couchot} {et~al.}(2017){Couchot}, {Henrot-Versill{\'e}},
  {Perdereau}, {Plaszczynski}, {Rouill{\'e} d'Orfeuil}, {Spinelli}, \&
  {Tristram}}]{CHP.etal.2017}
{Couchot}, F., {Henrot-Versill{\'e}}, S., {Perdereau}, O., {et~al.} 2017, \aap,
  606, A104

\bibitem[{{Creminelli} \& {Vernizzi}(2017)}]{CV.2017}
{Creminelli}, P. \& {Vernizzi}, F. 2017, Physical Review Letters, 119, 251302

\bibitem[{{Crisostomi} \& {Koyama}(2018)}]{CK.2017}
{Crisostomi}, M. \& {Koyama}, K. 2018, \prd, 97, 084004

\bibitem[{{Crittenden} {et~al.}(2002){Crittenden}, {Natarajan}, {Pen}, \&
  {Theuns}}]{CNP.etal.2002}
{Crittenden}, R.~G., {Natarajan}, P., {Pen}, U.-L., \& {Theuns}, T. 2002, \apj,
  568, 20

\bibitem[{{DES Collaboration} {et~al.}(2017){DES Collaboration}, {Abbott},
  {Abdalla}, {Alarcon}, {Aleksi{\'c}}, {Allam}, {Allen}, {Amara}, {Annis},
  {Asorey}, {Avila}, {Bacon}, {Balbinot}, {Banerji}, {Banik}, {Barkhouse},
  {Baumer}, {Baxter}, {Bechtol}, {Becker}, {Benoit-L{\'e}vy}, {Benson},
  {Bernstein}, {Bertin}, {Blazek}, {Bridle}, {Brooks}, {Brout}, {Buckley-Geer},
  {Burke}, {Busha}, {Capozzi}, {Carnero Rosell}, {Carrasco Kind}, {Carretero},
  {Castander}, {Cawthon}, {Chang}, {Chen}, {Childress}, {Choi}, {Conselice},
  {Crittenden}, {Crocce}, {Cunha}, {D'Andrea}, {da Costa}, {Das}, {Davis},
  {Davis}, {De Vicente}, {DePoy}, {DeRose}, {Desai}, {Diehl}, {Dietrich},
  {Dodelson}, {Doel}, {Drlica-Wagner}, {Eifler}, {Elliott}, {Elsner},
  {Elvin-Poole}, {Estrada}, {Evrard}, {Fang}, {Fernandez}, {Fert{\'e}},
  {Finley}, {Flaugher}, {Fosalba}, {Friedrich}, {Frieman},
  {Garc{\'{\i}}a-Bellido}, {Garcia-Fernandez}, {Gatti}, {Gaztanaga}, {Gerdes},
  {Giannantonio}, {Gill}, {Glazebrook}, {Goldstein}, {Gruen}, {Gruendl},
  {Gschwend}, {Gutierrez}, {Hamilton}, {Hartley}, {Hinton}, {Honscheid},
  {Hoyle}, {Huterer}, {Jain}, {James}, {Jarvis}, {Jeltema}, {Johnson},
  {Johnson}, {Kacprzak}, {Kent}, {Kim}, {King}, {Kirk}, {Kokron}, {Kovacs},
  {Krause}, {Krawiec}, {Kremin}, {Kuehn}, {Kuhlmann}, {Kuropatkin}, {Lacasa},
  {Lahav}, {Li}, {Liddle}, {Lidman}, {Lima}, {Lin}, {MacCrann}, {Maia},
  {Makler}, {Manera}, {March}, {Marshall}, {Martini}, {McMahon}, {Melchior},
  {Menanteau}, {Miquel}, {Miranda}, {Mudd}, {Muir}, {M{\"o}ller}, {Neilsen},
  {Nichol}, {Nord}, {Nugent}, {Ogando}, {Palmese}, {Peacock}, {Peiris},
  {Peoples}, {Percival}, {Petravick}, {Plazas}, {Porredon}, {Prat}, {Pujol},
  {Rau}, {Refregier}, {Ricker}, {Roe}, {Rollins}, {Romer}, {Roodman},
  {Rosenfeld}, {Ross}, {Rozo}, {Rykoff}, {Sako}, {Salvador}, {Samuroff},
  {S{\'a}nchez}, {Sanchez}, {Santiago}, {Scarpine}, {Schindler}, {Scolnic},
  {Secco}, {Serrano}, {Sevilla-Noarbe}, {Sheldon}, {Smith}, {Smith}, {Smith},
  {Soares-Santos}, {Sobreira}, {Suchyta}, {Tarle}, {Thomas}, {Troxel},
  {Tucker}, {Tucker}, {Uddin}, {Varga}, {Vielzeuf}, {Vikram}, {Vivas},
  {Walker}, {Wang}, {Wechsler}, {Weller}, {Wester}, {Wolf}, {Yanny}, {Yuan},
  {Zenteno}, {Zhang}, {Zhang}, \& {Zuntz}}]{DES.WL.2017}
{DES Collaboration}, {Abbott}, T.~M.~C., {Abdalla}, F.~B., {et~al.} 2017, ArXiv
  e-prints [\eprint[arXiv]{1708.01530}]

\bibitem[{{Dietrich} \& {Hartlap}(2010)}]{DH.2010}
{Dietrich}, J.~P. \& {Hartlap}, J. 2010, \mnras, 402, 1049

\bibitem[{{Dima} \& {Vernizzi}(2018)}]{DV.2018}
{Dima}, A. \& {Vernizzi}, F. 2018, \prd, 97, 101302

\bibitem[{{Ezquiaga} \& {Zumalac{\'a}rregui}(2017)}]{EZ.2017}
{Ezquiaga}, J.~M. \& {Zumalac{\'a}rregui}, M. 2017, Physical Review Letters,
  119, 251304

\bibitem[{{Fan} {et~al.}(2010){Fan}, {Shan}, \& {Liu}}]{FSL.2010}
{Fan}, Z., {Shan}, H., \& {Liu}, J. 2010, \apj, 719, 1408

\bibitem[{{Fluri} {et~al.}(2018){Fluri}, {Kacprzak}, {Sgier},
  {R{\'e}fr{\'e}gier}, \& {Amara}}]{FKS.etal.2018}
{Fluri}, J., {Kacprzak}, T., {Sgier}, R., {R{\'e}fr{\'e}gier}, A., \& {Amara},
  A. 2018, ArXiv e-prints [\eprint[arXiv]{1803.08461}]

\bibitem[{{Giocoli} {et~al.}(2018{\natexlab{a}}){Giocoli}, {Baldi}, \&
  {Moscardini}}]{GBM.2018}
{Giocoli}, C., {Baldi}, M., \& {Moscardini}, L. 2018{\natexlab{a}}, ArXiv
  e-prints [\eprint[arXiv]{1806.04681}]

\bibitem[{{Giocoli} {et~al.}(2017){Giocoli}, {Di Meo}, {Meneghetti}, {Jullo},
  {de la Torre}, {Moscardini}, {Baldi}, {Mazzotta}, \&
  {Metcalf}}]{GDMM.etal.2017}
{Giocoli}, C., {Di Meo}, S., {Meneghetti}, M., {et~al.} 2017, \mnras, 470, 3574

\bibitem[{{Giocoli} {et~al.}(2016){Giocoli}, {Jullo}, {Metcalf}, {de la Torre},
  {Yepes}, {Prada}, {Comparat}, {G{\"o}ttlober}, {Kyplin}, {Kneib}, {Petkova},
  {Shan}, \& {Tessore}}]{GJM.etal.2016}
{Giocoli}, C., {Jullo}, E., {Metcalf}, R.~B., {et~al.} 2016, \mnras, 461, 209

\bibitem[{{Giocoli} {et~al.}(2015){Giocoli}, {Metcalf}, {Baldi}, {Meneghetti},
  {Moscardini}, \& {Petkova}}]{GMB.etal.2015}
{Giocoli}, C., {Metcalf}, R.~B., {Baldi}, M., {et~al.} 2015, \mnras, 452, 2757

\bibitem[{{Giocoli} {et~al.}(2018{\natexlab{b}}){Giocoli}, {Moscardini},
  {Baldi}, {Meneghetti}, \& {Metcalf}}]{GMB.etal.2018}
{Giocoli}, C., {Moscardini}, L., {Baldi}, M., {Meneghetti}, M., \& {Metcalf},
  R.~B. 2018{\natexlab{b}}, \mnras, 478, 5436

\bibitem[{{Hagstotz} {et~al.}(2018){Hagstotz}, {Costanzi}, {Baldi}, \&
  {Weller}}]{HCB.etal.2018}
{Hagstotz}, S., {Costanzi}, M., {Baldi}, M., \& {Weller}, J. 2018, ArXiv
  e-prints [\eprint[arXiv]{1806.07400}]

\bibitem[{{Hamana} {et~al.}(2003){Hamana}, {Miyazaki}, {Shimasaku}, {Furusawa},
  {Doi}, {Hamabe}, {Imi}, {Kimura}, {Komiyama}, {Nakata}, {Okada}, {Okamura},
  {Ouchi}, {Sekiguchi}, {Yagi}, \& {Yasuda}}]{HMS.etal.2003}
{Hamana}, T., {Miyazaki}, S., {Shimasaku}, K., {et~al.} 2003, \apj, 597, 98

\bibitem[{{Hamana} {et~al.}(2012){Hamana}, {Oguri}, {Shirasaki}, \&
  {Sato}}]{HOS.etal.2012}
{Hamana}, T., {Oguri}, M., {Shirasaki}, M., \& {Sato}, M. 2012, \mnras, 425,
  2287

\bibitem[{{He}(2013)}]{He.2013}
{He}, J.-h. 2013, \prd, 88, 103523

\bibitem[{{Hetterscheidt} {et~al.}(2007){Hetterscheidt}, {Simon}, {Schirmer},
  {Hildebrandt}, {Schrabback}, {Erben}, \& {Schneider}}]{HSS.etal.2007}
{Hetterscheidt}, M., {Simon}, P., {Schirmer}, M., {et~al.} 2007, \aap, 468, 859

\bibitem[{{Heymans} {et~al.}(2013){Heymans}, {Grocutt}, {Heavens}, {Kilbinger},
  {Kitching}, {Simpson}, {Benjamin}, {Erben}, {Hildebrandt}, {Hoekstra},
  {Mellier}, {Miller}, {Van Waerbeke}, {Brown}, {Coupon}, {Fu},
  {Harnois-D{\'e}raps}, {Hudson}, {Kuijken}, {Rowe}, {Schrabback}, {Semboloni},
  {Vafaei}, \& {Velander}}]{CFHTLenS.WL.2013}
{Heymans}, C., {Grocutt}, E., {Heavens}, A., {et~al.} 2013, \mnras, 432, 2433

\bibitem[{{Higuchi} \& {Shirasaki}(2016)}]{HS.2016}
{Higuchi}, Y. \& {Shirasaki}, M. 2016, \mnras, 459, 2762

\bibitem[{{Hildebrandt} {et~al.}(2017){Hildebrandt}, {Viola}, {Heymans},
  {Joudaki}, {Kuijken}, {Blake}, {Erben}, {Joachimi}, {Klaes}, {Miller},
  {Morrison}, {Nakajima}, {Verdoes Kleijn}, {Amon}, {Choi}, {Covone}, {de
  Jong}, {Dvornik}, {Fenech Conti}, {Grado}, {Harnois-D{\'e}raps}, {Herbonnet},
  {Hoekstra}, {K{\"o}hlinger}, {McFarland}, {Mead}, {Merten}, {Napolitano},
  {Peacock}, {Radovich}, {Schneider}, {Simon}, {Valentijn}, {van den Busch},
  {van Uitert}, \& {Van Waerbeke}}]{KIDS.WL.2017}
{Hildebrandt}, H., {Viola}, M., {Heymans}, C., {et~al.} 2017, \mnras, 465, 1454

\bibitem[{{Hu} {et~al.}(2016){Hu}, {Raveri}, {Rizzato}, \&
  {Silvestri}}]{HRR.etal.2016}
{Hu}, B., {Raveri}, M., {Rizzato}, M., \& {Silvestri}, A. 2016, \mnras, 459,
  3880

\bibitem[{{Hu} \& {Sawicki}(2007)}]{HS.2007}
{Hu}, W. \& {Sawicki}, I. 2007, \prd, 76, 064004

\bibitem[{{Jain} \& {Seljak}(1997)}]{JS.1997}
{Jain}, B. \& {Seljak}, U. 1997, \apj, 484, 560

\bibitem[{{Jain} {et~al.}(2013){Jain}, {Vikram}, \& {Sakstein}}]{JVS.2013}
{Jain}, B., {Vikram}, V., \& {Sakstein}, J. 2013, \apj, 779, 39

\bibitem[{{Jarvis} {et~al.}(2004){Jarvis}, {Bernstein}, \& {Jain}}]{JBJ.2004}
{Jarvis}, M., {Bernstein}, G., \& {Jain}, B. 2004, \mnras, 352, 338

\bibitem[{{Jarvis} {et~al.}(2003){Jarvis}, {Bernstein}, {Fischer}, {Smith},
  {Jain}, {Tyson}, \& {Wittman}}]{JBF.etal.2003}
{Jarvis}, M., {Bernstein}, G.~M., {Fischer}, P., {et~al.} 2003, \aj, 125, 1014

\bibitem[{{Kacprzak} {et~al.}(2016){Kacprzak}, {Kirk}, {Friedrich}, {Amara},
  {Refregier}, {Marian}, {Dietrich}, {Suchyta}, {Aleksi{\'c}}, {Bacon},
  {Becker}, {Bonnett}, {Bridle}, {Chang}, {Eifler}, {Hartley}, {Huff},
  {Krause}, {MacCrann}, {Melchior}, {Nicola}, {Samuroff}, {Sheldon}, {Troxel},
  {Weller}, {Zuntz}, {Abbott}, {Abdalla}, {Armstrong}, {Benoit-L{\'e}vy},
  {Bernstein}, {Bernstein}, {Bertin}, {Brooks}, {Burke}, {Carnero Rosell},
  {Carrasco Kind}, {Carretero}, {Castander}, {Crocce}, {D'Andrea}, {da Costa},
  {Desai}, {Diehl}, {Evrard}, {Neto}, {Flaugher}, {Fosalba}, {Frieman},
  {Gerdes}, {Goldstein}, {Gruen}, {Gruendl}, {Gutierrez}, {Honscheid}, {Jain},
  {James}, {Jarvis}, {Kuehn}, {Kuropatkin}, {Lahav}, {Lima}, {March},
  {Marshall}, {Martini}, {Miller}, {Miquel}, {Mohr}, {Nichol}, {Nord},
  {Plazas}, {Romer}, {Roodman}, {Rykoff}, {Sanchez}, {Scarpine}, {Schubnell},
  {Sevilla-Noarbe}, {Smith}, {Soares-Santos}, {Sobreira}, {Swanson}, {Tarle},
  {Thomas}, {Vikram}, {Walker}, {Zhang}, \& {DES Collaboration}}]{DES.SV.2016}
{Kacprzak}, T., {Kirk}, D., {Friedrich}, O., {et~al.} 2016, \mnras, 463, 3653

\bibitem[{{Kaiser} \& {Squires}(1993)}]{KS.1993}
{Kaiser}, N. \& {Squires}, G. 1993, \apj, 404, 441

\bibitem[{{Kaiser} {et~al.}(1994){Kaiser}, {Squires}, \&
  {Fahlman}}]{KSF.etal.1994}
{Kaiser}, N., {Squires}, G., \& {Fahlman}, G.~{Woods}, D. 1994, Clusters of
  galaxies, Proc. of the XIVth Moriond Astrophysics Meeting (M{\'e}ribel,
  France, 12-19 March 1994) p269

\bibitem[{{Khoury} \& {Weltman}(2004)}]{KW.2004}
{Khoury}, J. \& {Weltman}, A. 2004, \prd, 69, 044026

\bibitem[{{Kilbinger}(2015)}]{Kilbinger.2014}
{Kilbinger}, M. 2015, Reports on Progress in Physics, 78, 086901

\bibitem[{{Kilbinger} {et~al.}(2013){Kilbinger}, {Fu}, {Heymans}, {Simpson},
  {Benjamin}, {Erben}, {Harnois-D{\'e}raps}, {Hoekstra}, {Hildebrandt},
  {Kitching}, {Mellier}, {Miller}, {Van Waerbeke}, {Benabed}, {Bonnett},
  {Coupon}, {Hudson}, {Kuijken}, {Rowe}, {Schrabback}, {Semboloni}, {Vafaei},
  \& {Velander}}]{KFH.etal.2013}
{Kilbinger}, M., {Fu}, L., {Heymans}, C., {et~al.} 2013, \mnras, 430, 2200

\bibitem[{{Kilbinger} \& {Schneider}(2005)}]{KS.2005}
{Kilbinger}, M. \& {Schneider}, P. 2005, \aap, 442, 69

\bibitem[{{Koyama}(2016)}]{Koyama.2016}
{Koyama}, K. 2016, Reports on Progress in Physics, 79, 046902

\bibitem[{{Kratochvil} {et~al.}(2010){Kratochvil}, {Haiman}, \&
  {May}}]{KHM.2010}
{Kratochvil}, J.~M., {Haiman}, Z., \& {May}, M. 2010, \prd, 81, 043519

\bibitem[{{Kruse} \& {Schneider}(1999)}]{KS.1999}
{Kruse}, G. \& {Schneider}, P. 1999, \mnras, 302, 821

\bibitem[{{Kruse} \& {Schneider}(2000)}]{KS.2000}
{Kruse}, G. \& {Schneider}, P. 2000, \mnras, 318, 321

\bibitem[{{Laureijs} {et~al.}(2011){Laureijs}, {Amiaux}, {Arduini},
  {Augu{\`e}res}, {Brinchmann}, {Cole}, {Cropper}, {Dabin}, {Duvet}, {Ealet},
  \& et~al.}]{Euclid.2011}
{Laureijs}, R., {Amiaux}, J., {Arduini}, S., {et~al.} 2011, ArXiv e-prints
  [\eprint[arXiv]{1110.3193}]

\bibitem[{{Leonard} {et~al.}(2012){Leonard}, {Pires}, \& {Starck}}]{LPS.2012}
{Leonard}, A., {Pires}, S., \& {Starck}, J.-L. 2012, \mnras, 423, 3405

\bibitem[{{Lewis} {et~al.}(2000){Lewis}, {Challinor}, \& {Lasenby}}]{CAMB}
{Lewis}, A., {Challinor}, A., \& {Lasenby}, A. 2000, \apj, 538, 473

\bibitem[{{Lin} \& {Kilbinger}(2015)}]{LK.2015}
{Lin}, C.-A. \& {Kilbinger}, M. 2015, \aap, 576, A24

\bibitem[{{Liu} \& {Haiman}(2016)}]{LH.2016}
{Liu}, J. \& {Haiman}, Z. 2016, \prd, 94, 043533

\bibitem[{{Liu} {et~al.}(2016){Liu}, {Li}, {Zhao}, {Chiu}, {Fang}, {Pan},
  {Wang}, {Du}, {Yuan}, {Fu}, \& {Fan}}]{LLZ.etal.2016}
{Liu}, X., {Li}, B., {Zhao}, G.-B., {et~al.} 2016, Physical Review Letters,
  117, 051101

\bibitem[{{Lombriser}(2014)}]{Lombriser.2014}
{Lombriser}, L. 2014, Annalen der Physik, 526, 259

\bibitem[{{Lombriser} \& {Lima}(2017)}]{LL.2017}
{Lombriser}, L. \& {Lima}, N.~A. 2017, Physics Letters B, 765, 382

\bibitem[{{Lombriser} {et~al.}(2012){Lombriser}, {Schmidt}, {Baldauf},
  {Mandelbaum}, {Seljak}, \& {Smith}}]{LSB.etal.2012}
{Lombriser}, L., {Schmidt}, F., {Baldauf}, T., {et~al.} 2012, \prd, 85, 102001

\bibitem[{{Lombriser} \& {Taylor}(2016)}]{LT.2016}
{Lombriser}, L. \& {Taylor}, A. 2016, \jcap, 3, 031

\bibitem[{{LSST Science Collaboration} {et~al.}(2009){LSST Science
  Collaboration}, {Abell}, {Allison}, {Anderson}, {Andrew}, {Angel}, {Armus},
  {Arnett}, {Asztalos}, {Axelrod}, \& et~al.}]{LSST.2009}
{LSST Science Collaboration}, {Abell}, P.~A., {Allison}, J., {et~al.} 2009,
  ArXiv e-prints [\eprint[arXiv]{0912.0201}]

\bibitem[{{Marian} {et~al.}(2012){Marian}, {Smith}, {Hilbert}, \&
  {Schneider}}]{MSH.etal.2012}
{Marian}, L., {Smith}, R.~E., {Hilbert}, S., \& {Schneider}, P. 2012, \mnras,
  423, 1711

\bibitem[{{Martinelli} {et~al.}(2009){Martinelli}, {Melchiorri}, \&
  {Amendola}}]{MMA.2009}
{Martinelli}, M., {Melchiorri}, A., \& {Amendola}, L. 2009, \prd, 79, 123516

\bibitem[{{Martinet} {et~al.}(2015){Martinet}, {Bartlett}, {Kiessling}, \&
  {Sartoris}}]{MBK.etal.2015}
{Martinet}, N., {Bartlett}, J.~G., {Kiessling}, A., \& {Sartoris}, B. 2015,
  \aap, 581, A101

\bibitem[{{Martinet} {et~al.}(2018){Martinet}, {Schneider}, {Hildebrandt},
  {Shan}, {Asgari}, {Dietrich}, {Harnois-D{\'e}raps}, {Erben}, {Grado},
  {Heymans}, {Hoekstra}, {Klaes}, {Kuijken}, {Merten}, \&
  {Nakajima}}]{MSH.etal.2018}
{Martinet}, N., {Schneider}, P., {Hildebrandt}, H., {et~al.} 2018, \mnras, 474,
  712

\bibitem[{{Maturi} {et~al.}(2011){Maturi}, {Fedeli}, \&
  {Moscardini}}]{MFM.2011}
{Maturi}, M., {Fedeli}, C., \& {Moscardini}, L. 2011, \mnras, 416, 2527

\bibitem[{{Motohashi} {et~al.}(2013){Motohashi}, {Starobinsky}, \&
  {Yokoyama}}]{MSY.2013}
{Motohashi}, H., {Starobinsky}, A.~A., \& {Yokoyama}, J. 2013, Physical Review
  Letters, 110, 121302

\bibitem[{{Navarro} {et~al.}(1996){Navarro}, {Frenk}, \& {White}}]{NFW.1996}
{Navarro}, J.~F., {Frenk}, C.~S., \& {White}, S.~D.~M. 1996, \apj, 462, 563

\bibitem[{{Peel} {et~al.}(2017){Peel}, {Lin}, {Lanusse}, {Leonard}, {Starck},
  \& {Kilbinger}}]{PLL.etal.2017}
{Peel}, A., {Lin}, C.-A., {Lanusse}, F., {et~al.} 2017, \aap, 599, A79

\bibitem[{{Petri} {et~al.}(2016){Petri}, {Haiman}, \& {May}}]{PHM.2016}
{Petri}, A., {Haiman}, Z., \& {May}, M. 2016, \prd, 93, 063524

\bibitem[{{Petri} {et~al.}(2017){Petri}, {Haiman}, \& {May}}]{PHM.2017}
{Petri}, A., {Haiman}, Z., \& {May}, M. 2017, \prd, 95, 123503

\bibitem[{{Pettorino}(2013)}]{Pettorino.2013}
{Pettorino}, V. 2013, \prd, 88, 063519

\bibitem[{{Pettorino} \& {Baccigalupi}(2008)}]{PB.2008}
{Pettorino}, V. \& {Baccigalupi}, C. 2008, \prd, 77, 103003

\bibitem[{{Pires} {et~al.}(2012){Pires}, {Leonard}, \& {Starck}}]{PLS.2012}
{Pires}, S., {Leonard}, A., \& {Starck}, J.-L. 2012, \mnras, 423, 983

\bibitem[{{Pires} {et~al.}(2009){Pires}, {Starck}, {Amara},
  {R{\'e}fr{\'e}gier}, \& {Teyssier}}]{PSA.etal.2009}
{Pires}, S., {Starck}, J.-L., {Amara}, A., {R{\'e}fr{\'e}gier}, A., \&
  {Teyssier}, R. 2009, \aap, 505, 969

\bibitem[{{Planck Collaboration} {et~al.}(2016{\natexlab{a}}){Planck
  Collaboration}, {Ade}, {Aghanim}, {Arnaud}, {Ashdown}, {Aumont},
  {Baccigalupi}, {Banday}, {Barreiro}, {Bartlett}, \&
  et~al.}]{Planck.Cosmo.2016}
{Planck Collaboration}, {Ade}, P.~A.~R., {Aghanim}, N., {et~al.}
  2016{\natexlab{a}}, \aap, 594, A13

\bibitem[{{Planck Collaboration} {et~al.}(2016{\natexlab{b}}){Planck
  Collaboration}, {Ade}, {Aghanim}, {Arnaud}, {Ashdown}, {Aumont},
  {Baccigalupi}, {Banday}, {Barreiro}, {Bartolo}, \& et~al.}]{Planck.DE.2016}
{Planck Collaboration}, {Ade}, P.~A.~R., {Aghanim}, N., {et~al.}
  2016{\natexlab{b}}, \aap, 594, A14

\bibitem[{{Puchwein} {et~al.}(2013){Puchwein}, {Baldi}, \&
  {Springel}}]{PBS.2013}
{Puchwein}, E., {Baldi}, M., \& {Springel}, V. 2013, \mnras, 436, 348

\bibitem[{{Roncarelli} {et~al.}(2007){Roncarelli}, {Moscardini}, {Borgani}, \&
  {Dolag}}]{RMB.etal.2007}
{Roncarelli}, M., {Moscardini}, L., {Borgani}, S., \& {Dolag}, K. 2007, \mnras,
  378, 1259

\bibitem[{{Sakstein}(2015)}]{Sakstein.2015}
{Sakstein}, J. 2015, Physical Review Letters, 115, 201101

\bibitem[{{Sakstein} \& {Jain}(2017)}]{SJ.2017}
{Sakstein}, J. \& {Jain}, B. 2017, Physical Review Letters, 119, 251303

\bibitem[{{Sch{\"a}fer} {et~al.}(2012){Sch{\"a}fer}, {Heisenberg},
  {Kalovidouris}, \& {Bacon}}]{SHK.etal.2012}
{Sch{\"a}fer}, B.~M., {Heisenberg}, L., {Kalovidouris}, A.~F., \& {Bacon},
  D.~J. 2012, \mnras, 420, 455

\bibitem[{{Schirmer} {et~al.}(2007){Schirmer}, {Erben}, {Hetterscheidt}, \&
  {Schneider}}]{SEH.etal.2007}
{Schirmer}, M., {Erben}, T., {Hetterscheidt}, M., \& {Schneider}, P. 2007,
  \aap, 462, 875

\bibitem[{{Schneider}(1996)}]{Schneider.1996}
{Schneider}, P. 1996, \mnras, 283, 837

\bibitem[{{Schneider} {et~al.}(1998){Schneider}, {van Waerbeke}, {Jain}, \&
  {Kruse}}]{SvWJ.etal.1998}
{Schneider}, P., {van Waerbeke}, L., {Jain}, B., \& {Kruse}, G. 1998, \mnras,
  296, 873

\bibitem[{{Schrabback} {et~al.}(2010){Schrabback}, {Hartlap}, {Joachimi},
  {Kilbinger}, {Simon}, {Benabed}, {Brada{\v c}}, {Eifler}, {Erben},
  {Fassnacht}, {High}, {Hilbert}, {Hildebrandt}, {Hoekstra}, {Kuijken},
  {Marshall}, {Mellier}, {Morganson}, {Schneider}, {Semboloni}, {van Waerbeke},
  \& {Velander}}]{SHJ.etal.2010}
{Schrabback}, T., {Hartlap}, J., {Joachimi}, B., {et~al.} 2010, \aap, 516, A63

\bibitem[{{Shan} {et~al.}(2018){Shan}, {Liu}, {Hildebrandt}, {Pan}, {Martinet},
  {Fan}, {Schneider}, {Asgari}, {Harnois-D{\'e}raps}, {Hoekstra}, {Wright},
  {Dietrich}, {Erben}, {Getman}, {Grado}, {Heymans}, {Klaes}, {Kuijken},
  {Merten}, {Puddu}, {Radovich}, \& {Wang}}]{SLH.etal.2018}
{Shan}, H., {Liu}, X., {Hildebrandt}, H., {et~al.} 2018, \mnras, 474, 1116

\bibitem[{{Shan} {et~al.}(2014){Shan}, {Kneib}, {Comparat}, {Jullo},
  {Charbonnier}, {Erben}, {Makler}, {Moraes}, {Van Waerbeke}, {Courbin},
  {Meylan}, {Tao}, \& {Taylor}}]{SKC.etal.2014}
{Shan}, H.~Y., {Kneib}, J.-P., {Comparat}, J., {et~al.} 2014, \mnras, 442, 2534

\bibitem[{{Shirasaki} {et~al.}(2017){Shirasaki}, {Nishimichi}, {Li}, \&
  {Higuchi}}]{SNL.etal.2017}
{Shirasaki}, M., {Nishimichi}, T., {Li}, B., \& {Higuchi}, Y. 2017, \mnras,
  466, 2402

\bibitem[{{Smith}(2009)}]{Smith.2009}
{Smith}, T.~L. 2009, ArXiv e-prints [\eprint[arXiv]{0907.4829}]

\bibitem[{{Springel}(2005)}]{Springel.2005}
{Springel}, V. 2005, \mnras, 364, 1105

\bibitem[{{Starck} {et~al.}(2007){Starck}, {Fadili}, \& {Murtagh}}]{SFM.2007}
{Starck}, J.-L., {Fadili}, J., \& {Murtagh}, F. 2007, IEEE Transactions on
  Image Processing, 16, 297

\bibitem[{{van Waerbeke}(1998)}]{vanWaerbeke.1998}
{van Waerbeke}, L. 1998, \aap, 334, 1

\bibitem[{{van Waerbeke}(2000)}]{vanWaerbeke.2000}
{van Waerbeke}, L. 2000, \mnras, 313, 524

\bibitem[{{Van Waerbeke} {et~al.}(2001){Van Waerbeke}, {Mellier}, {Radovich},
  {Bertin}, {Dantel-Fort}, {McCracken}, {Le F{\`e}vre}, {Foucaud},
  {Cuillandre}, {Erben}, {Jain}, {Schneider}, {Bernardeau}, \&
  {Fort}}]{vWMR.etal.2001}
{Van Waerbeke}, L., {Mellier}, Y., {Radovich}, M., {et~al.} 2001, \aap, 374,
  757

\bibitem[{{Vicinanza} {et~al.}(2018){Vicinanza}, {Cardone}, {Maoli},
  {Scaramella}, \& {Er}}]{VCM.etal.2018}
{Vicinanza}, M., {Cardone}, V.~F., {Maoli}, R., {Scaramella}, R., \& {Er}, X.
  2018, \prd, 97, 023519

\bibitem[{{Viel} {et~al.}(2010){Viel}, {Haehnelt}, \& {Springel}}]{VHS.2010}
{Viel}, M., {Haehnelt}, M.~G., \& {Springel}, V. 2010, \jcap, 6, 015

\bibitem[{{Vikram} {et~al.}(2014){Vikram}, {Sakstein}, {Davis}, \&
  {Neil}}]{VSD.etal.2014}
{Vikram}, V., {Sakstein}, J., {Davis}, C., \& {Neil}, A. 2014, ArXiv e-prints
  [\eprint[arXiv]{1407.6044}]

\bibitem[{{Villaescusa-Navarro} {et~al.}(2017){Villaescusa-Navarro},
  {Banerjee}, {Dalal}, {Castorina}, {Scoccimarro}, {Angulo}, \&
  {Spergel}}]{VNBD.etal.2017}
{Villaescusa-Navarro}, F., {Banerjee}, A., {Dalal}, N., {et~al.} 2017, ArXiv
  e-prints [\eprint[arXiv]{1708.01154}]

\bibitem[{{Wright} {et~al.}(2017){Wright}, {Winther}, \& {Koyama}}]{WWK.2017}
{Wright}, B.~S., {Winther}, H.~A., \& {Koyama}, K. 2017, \jcap, 10, 054

\bibitem[{{Yang} {et~al.}(2011){Yang}, {Kratochvil}, {Wang}, {Lim}, {Haiman},
  \& {May}}]{YKW.etal.2011}
{Yang}, X., {Kratochvil}, J.~M., {Wang}, S., {et~al.} 2011, \prd, 84, 043529

\bibitem[{{Zennaro} {et~al.}(2017){Zennaro}, {Bel}, {Villaescusa-Navarro},
  {Carbone}, {Sefusatti}, \& {Guzzo}}]{ZBVN.etal.2017}
{Zennaro}, M., {Bel}, J., {Villaescusa-Navarro}, F., {et~al.} 2017, \mnras,
  466, 3244

\bibitem[{{Zhang} {et~al.}(2003){Zhang}, {Pen}, {Zhang}, \&
  {Dubinski}}]{ZPZ.etal.2003}
{Zhang}, T.-J., {Pen}, U.-L., {Zhang}, P., \& {Dubinski}, J. 2003, \apj, 598,
  818

\end{thebibliography}

\begin{appendix}\label{appendix}
\section{Impact of galaxy shape noise}
We are not able to measure the true shear field in practice; instead we measure galaxy ellipticities $\epsilon$, which in the absence of systematic errors, represent an unbiased measurement of the reduced shear field $\boldsymbol{g}(\boldsymbol{\theta})=\boldsymbol{\gamma}(\boldsymbol{\theta}) / [1 - \kappa(\boldsymbol{\theta})]$ when averaged over many galaxies. A primary source of noise in weak lensing analyses is therefore due to the non-circular intrinsic shapes of galaxies. In this Appendix, we study the effect of galaxy shape noise on our results by reproducing Figs.~\ref{fig:histograms} and \ref{fig:polar_fR5} with such representative noise included.

Given that we work with convergence maps directly from our simulations, we generate noisy versions of these maps by simply adding a noise term: $\kappa_N(\boldsymbol{\theta})=\kappa(\boldsymbol{\theta})+N(\boldsymbol{\theta})$. The noise $N(\boldsymbol{\theta})$ is modeled as a Gaussian random field with zero mean and variance given by \citep{vanWaerbeke.2000}
\be
\sigma^2_\mathrm{pix}=\frac{\sigma_\epsilon^2}{2}\frac{1}{n_\mathrm{gal}\,A_\mathrm{pix}},
\ee
where $\sigma^2_\epsilon=\sigma^2_{\epsilon_1}+\sigma^2_{\epsilon_2}$ is the total galaxy ellipticity variance, $n_\mathrm{gal}$ is the number density of galaxies, and $A_\mathrm{pix}$ is the pixel area. Upcoming weak lensing surveys should achieve $n_\mathrm{gal}\approx 30~\mathrm{arcmin}^{-2}$ and $\sigma_\epsilon\approx 0.4$, corresponding to a standard deviation of about 0.28 per component of ellipticity. 

\begin{figure*}
	\includegraphics[width=\columnwidth]{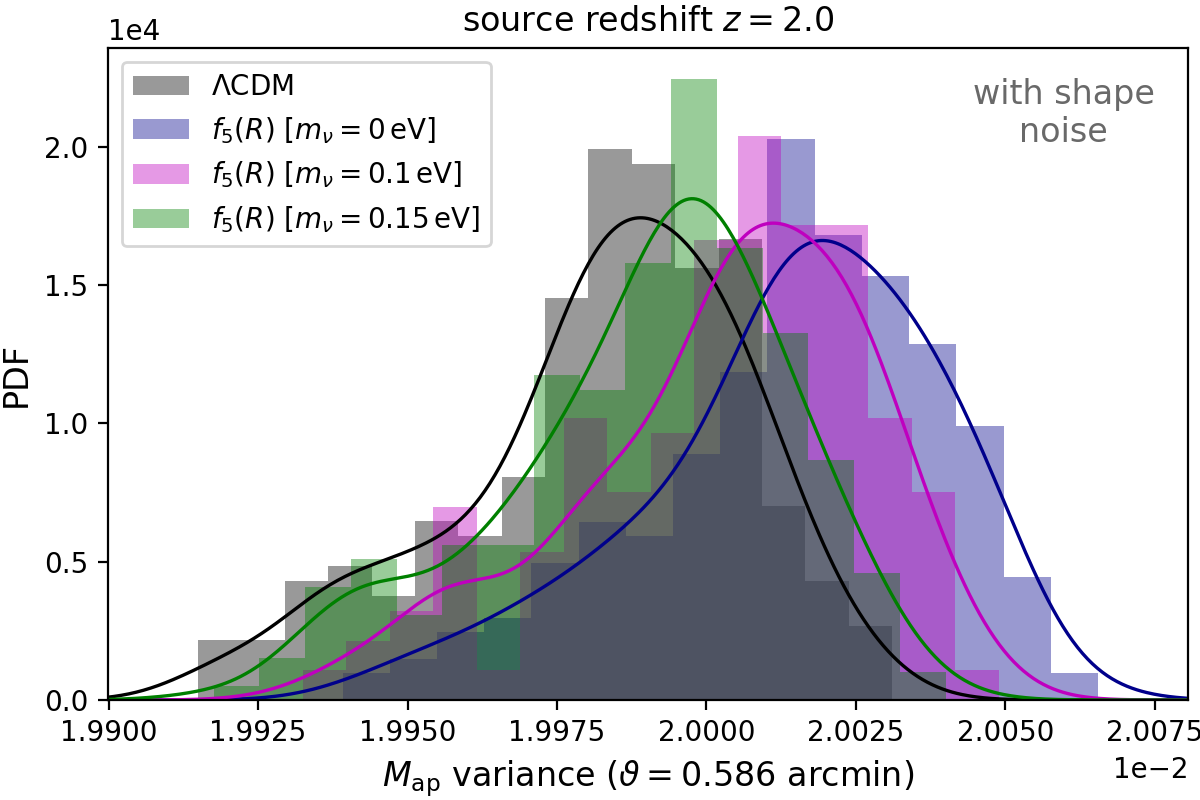}
    \includegraphics[width=\columnwidth]{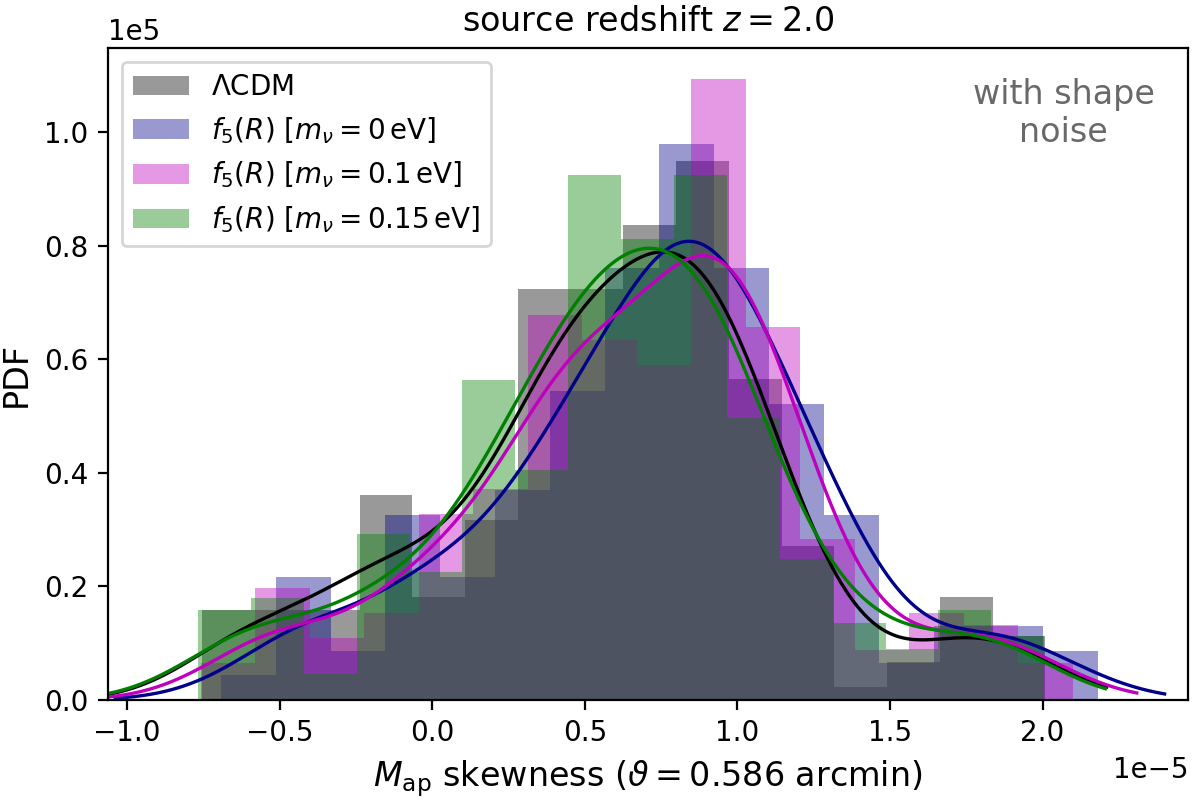}\\
    \includegraphics[width=\columnwidth]{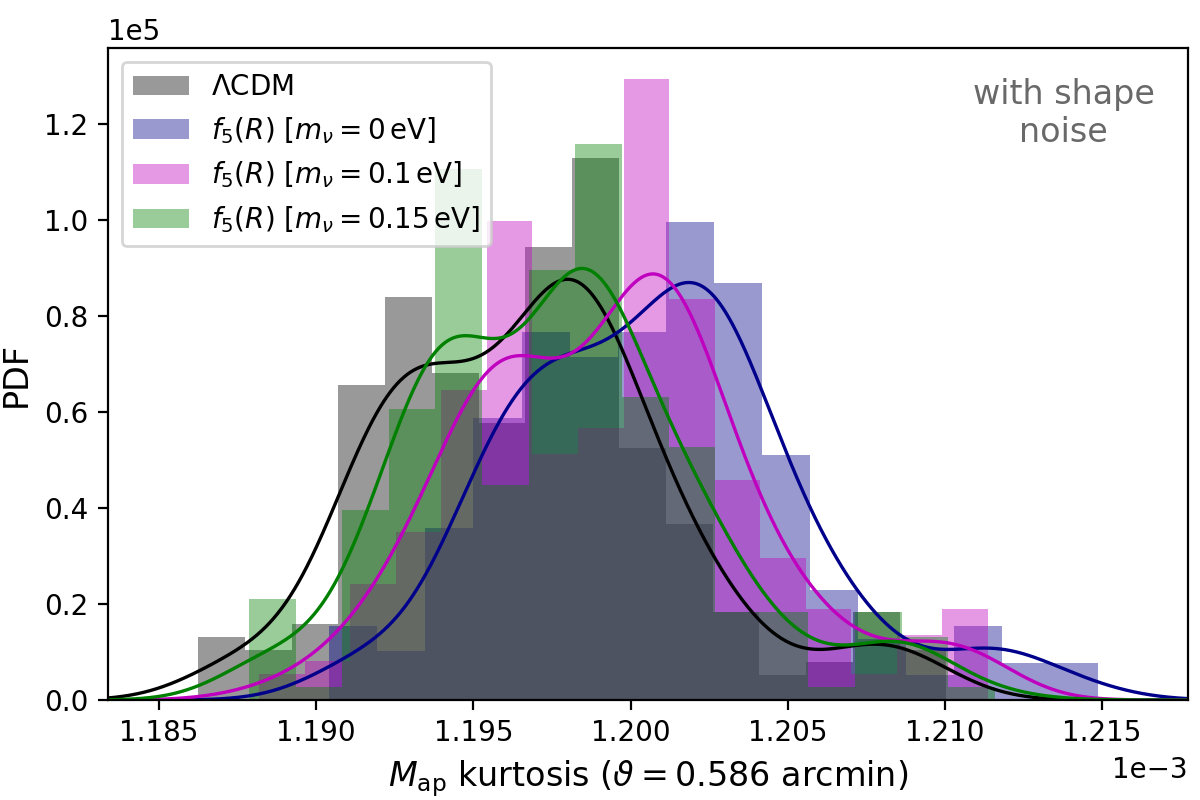}
    \includegraphics[width=\columnwidth]{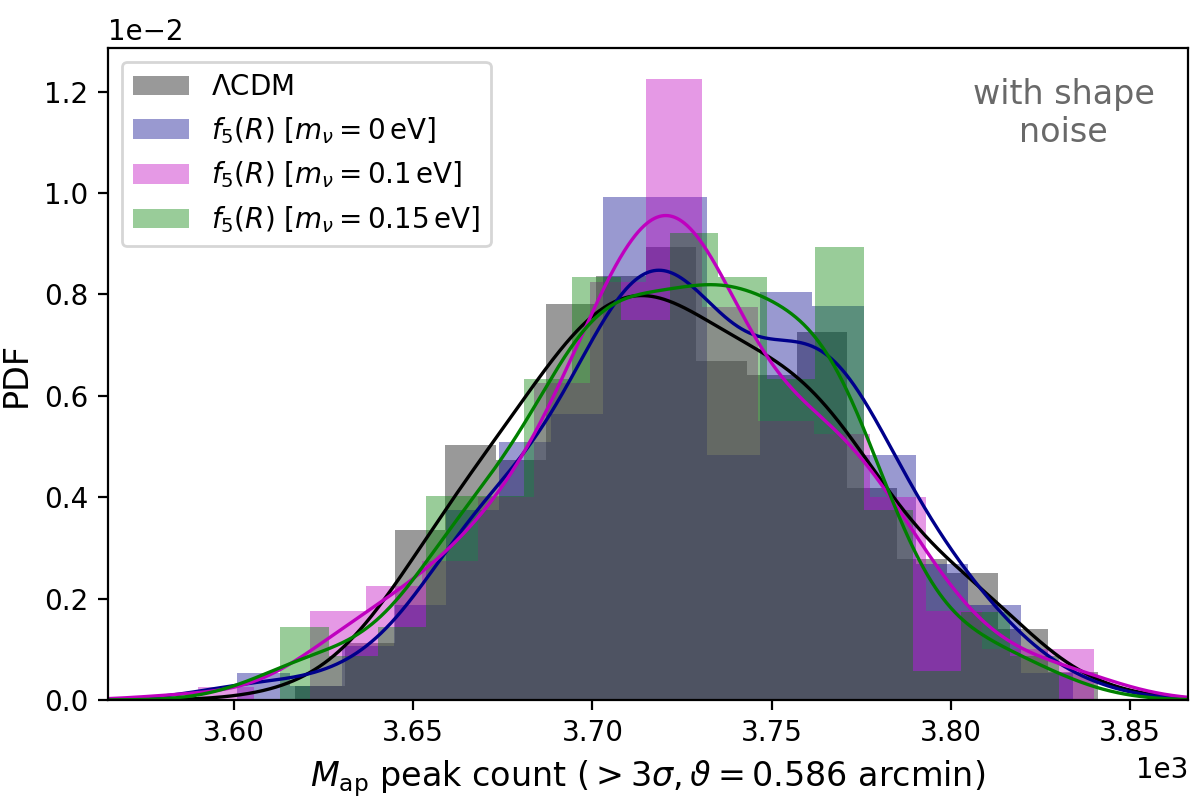}
    \caption{Histograms of aperture mass statistics corresponding to the distributions in Fig.~\ref{fig:histograms} with galaxy shape noise included. The filtering scale is again $\vartheta=0.586'$, and sources are at redshift $z_s=2.0$. The profiles of the noisy distributions have all changed compared to their noiseless counterparts. Overall, the $f_5(R)$ distributions are much more difficult to distinguish from $\Lambda$CDM, as well as from each other, than in the noiseless case at this scale. In particular, this level of noise washes out the discrimination power of peak counts above $3\sigma$.}
    \label{fig:histograms_noisy}
\end{figure*}

\begin{figure*}
	\includegraphics[width=\columnwidth]{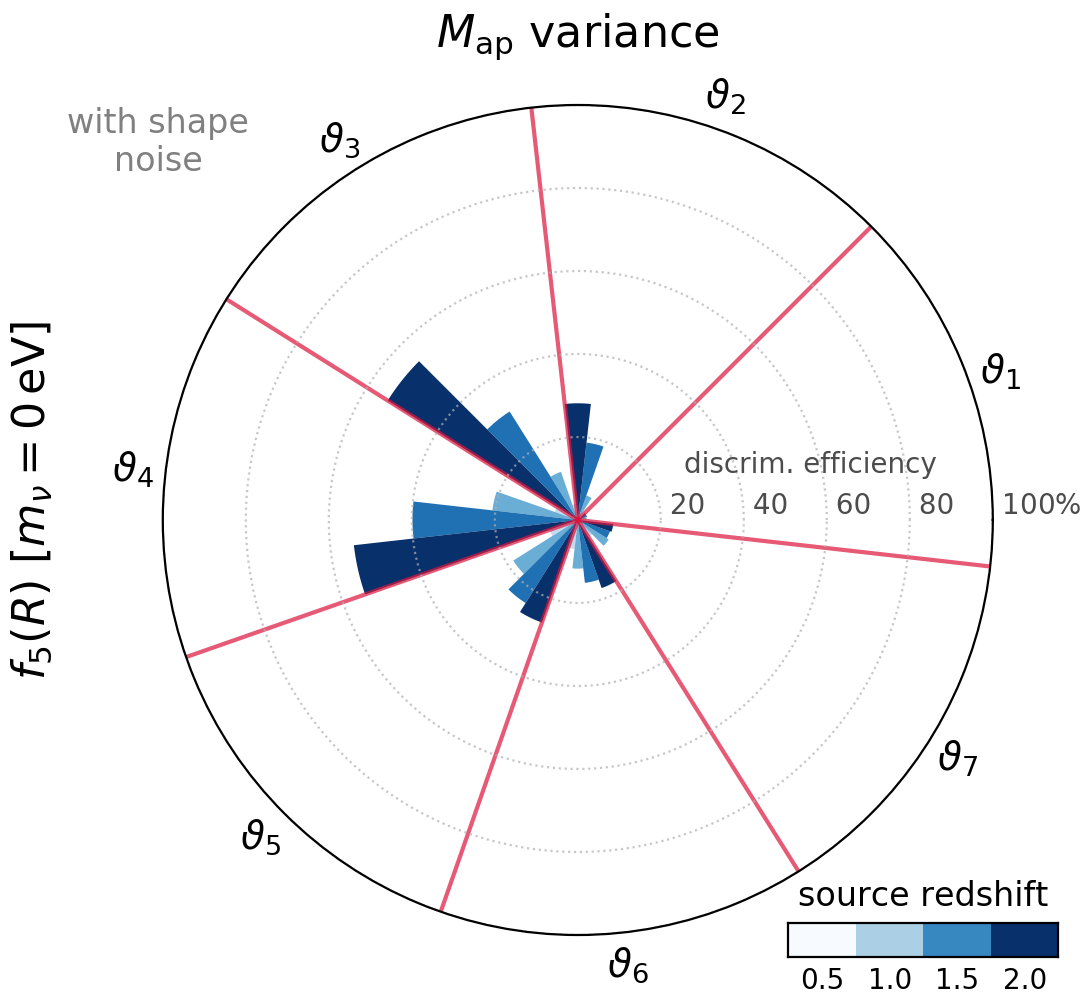}
    \includegraphics[width=\columnwidth]{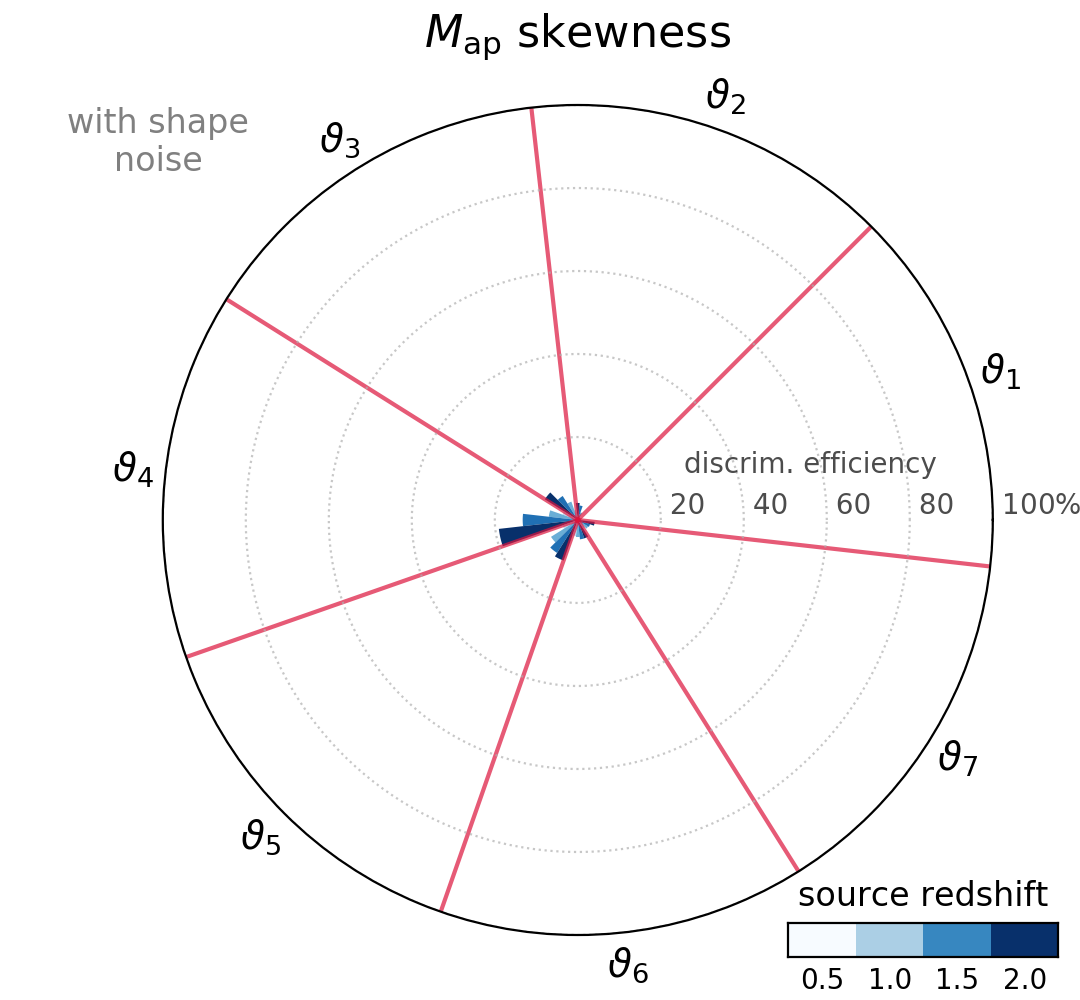}\\
    \includegraphics[width=\columnwidth]{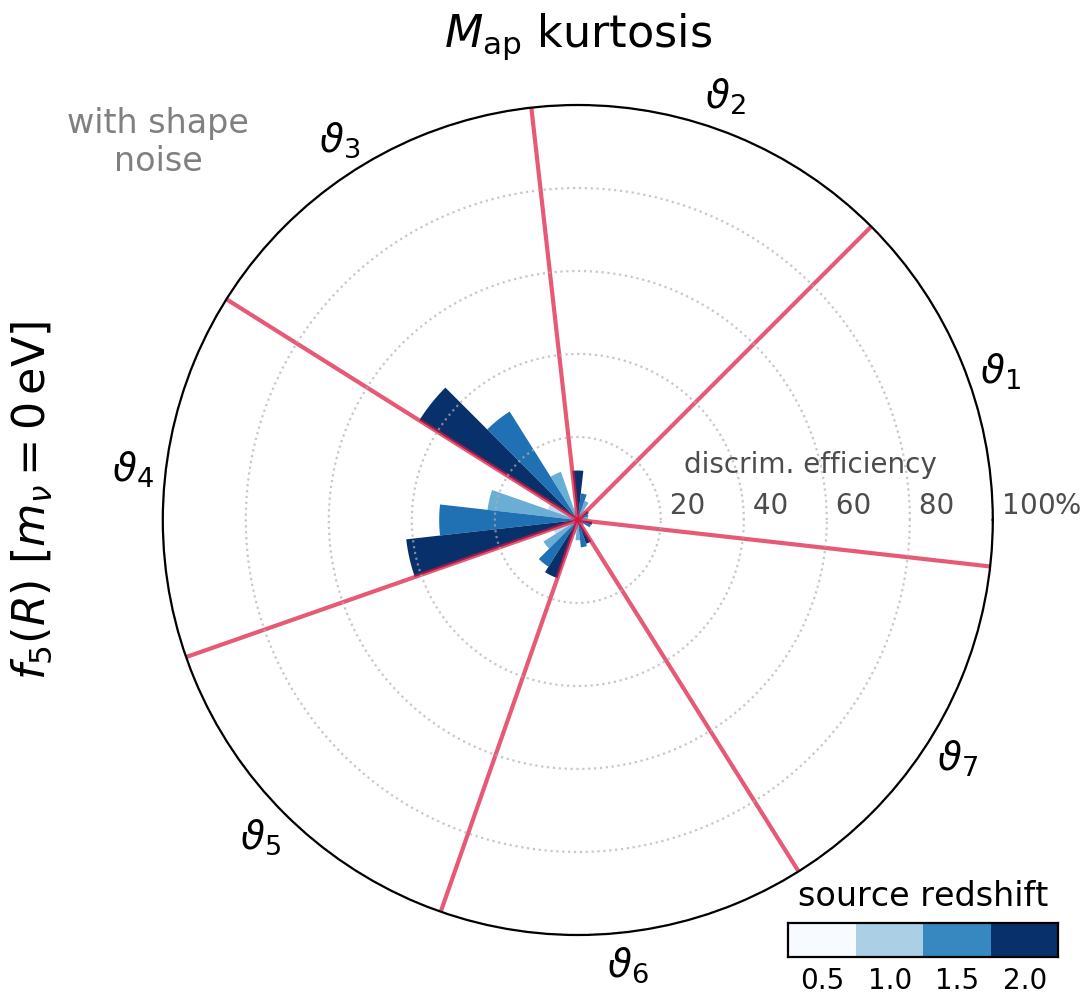}
    \includegraphics[width=\columnwidth]{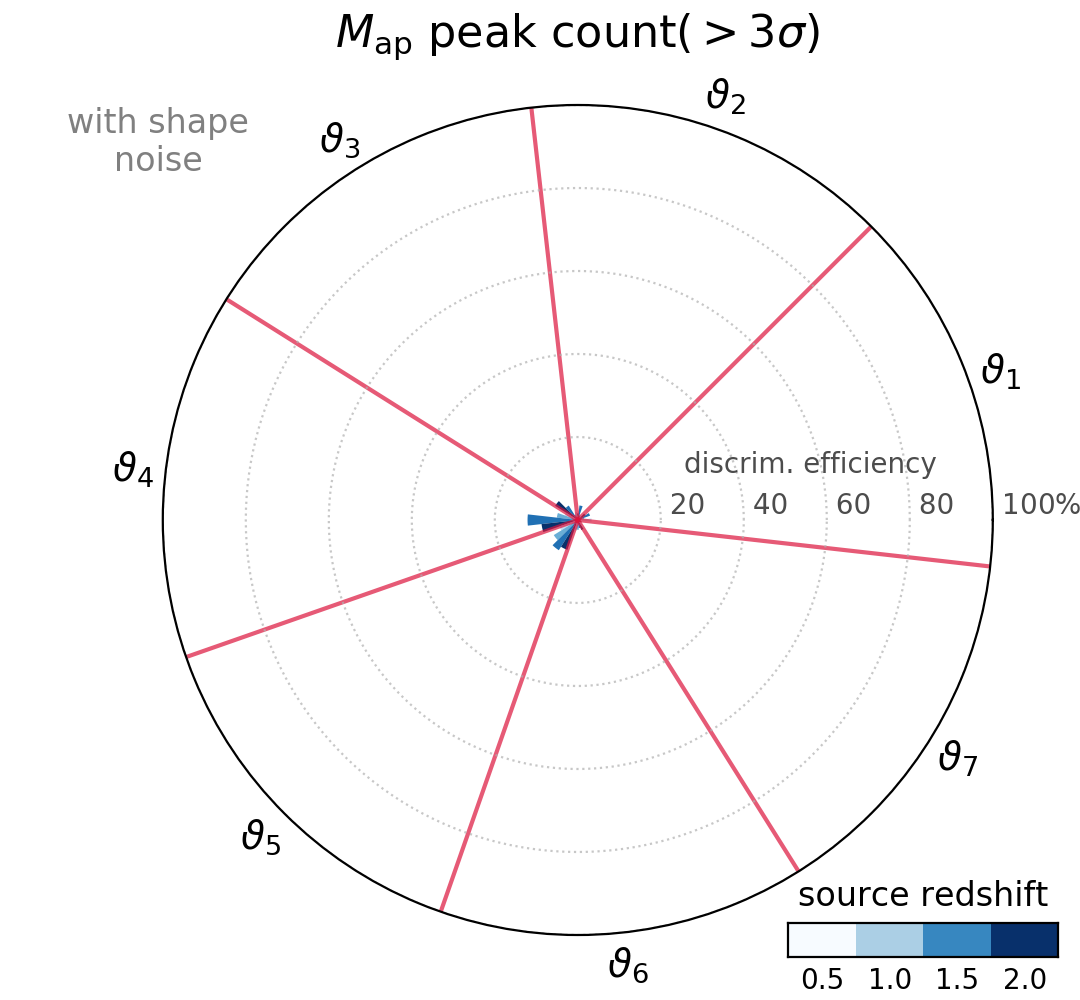}
    \caption{Reproduction of Fig.~\ref{fig:polar_fR5} including galaxy shape noise for the $f_5(R)$ model without neutrinos. The effect of such noise is to reduce the discrimination efficiency from $\Lambda$CDM across all aperture scales, source redshifts, and statistics shown (with the exception of certain kurtosis scales as discussed in the text). In particular, the noise dominates the measured skewness and peak count distributions at each scale so that both statistics essentially lose all previous discrimination power between $f_5(R)$ and $\Lambda$CDM, although this would likely be mitigated by a denoising procedure.}
    \label{fig:polar_fR5_noisy}
\end{figure*}

Our methodology produces lensing maps for sources at fixed redshift, namely $z_s=0.5$, 1.0, 1.5, and 2.0. In real observations, of course, galaxies are distributed in redshift, and the effective number density of a survey reflects this distribution. We can imagine that our lines of sight come from a survey in which the galaxies have been divided into four bins such that each contains the same number of galaxies and that our source planes lie at the centres. Assuming an overall galaxy number density of $30~\mathrm{arcmin}^{-2}$, this gives $n_\mathrm{gal}=7.5~\mathrm{arcmin}^{-2}$ per bin and $\sigma_\mathrm{pix}=0.7$.

Histograms corresponding to those in Fig.~\ref{fig:histograms} of the noisy observables under the above conditions are shown in Fig.~\ref{fig:histograms_noisy}. In terms of the variance (upper left), the distributions have all shifted towards larger mean values compared to their noiseless versions. They also now exhibit a prominent leftward skewness, in contrast to the approximately Gaussian distributions for the noiseless case. The  skewness and kurtosis distributions (upper right and lower left, respectively) deviate significantly from Gaussians, the kurtosis especially now exhibiting strong asymmetry. Peak counts above $3\sigma$ (lower right) for the four models now have fully overlapping distributions, converging to a central value of around one third of that found in the noiseless case. Overall, the effect of shape noise is to wash out the distinctions between the models seen previously at this scale.

Discrimination efficiencies $\mathcal{E}$ corresponding to Fig.~\ref{fig:polar_fR5} for the noisy maps of $f_5(R)$ are shown in Fig.~\ref{fig:polar_fR5_noisy}. As the overlap of the histograms suggest for $\vartheta_2=0.586'$, the skewness and peak counts lose all discrimination power not only at this filtering scale, but at all others as well. The noise dominates the signal for these statistics such that neither is capable of distinguishing $f_5(R)$ from $\Lambda$CDM any longer. In contrast, the variance and kurtosis retain some discrimination power (although nowhere exceeding 55\%), in particular for filter scales $\geq\vartheta_3$. The noisy variance plot is qualitatively similar to its counterpart in Fig.~\ref{fig:polar_fR5} except that each bar has been scaled towards the centre. The kurtosis exhibits a somewhat different behaviour in that while $\vartheta_1$ and $\vartheta_2$ give lower $\mathcal{E}$ values, scales $\geq\vartheta_3$ actually gain in discrimination power compared to the noiseless case.

What is not directly apparent from these figures is the way that the discrimination power changes as a function of noise level. We have verified that lower noise values, for example $\sigma_\mathrm{pix}=0.2$ and $0.4$, produce discrimination efficiencies at intermediate values between the noiseless case and $\sigma_\mathrm{pix}=0.7$. The tendency of $\mathcal{E}$ to decrease monotonically with increasing noise holds for scales $\vartheta_1$ and $\vartheta_2$ across all statistics we have considered. On the other hand, for scales $\geq\vartheta_3$ and for the non-Gaussian statistics, $\mathcal{E}$ can achieve a maximum with increasing noise before eventually decreasing again to zero for large enough $\sigma_\mathrm{pix}$. What is seen in Fig.~\ref{fig:polar_fR5_noisy} is that the noise level is not high enough to fully suppress $\mathcal{E}$ using kurtosis as it is for skewness and peak counts. However, both these latter statistics have scales at which the maximum discrimination efficiency is largest with some non-zero noise level $\sigma_\mathrm{pix}<0.7$. To summarise, some amount of Gaussian noise can actually improve the distinction between models on some scales, but it depends sensitively on the initial shapes and separations of the noiseless histograms.

We have considered here one important source of noise present in any weak-lensing analysis, but many others exist as well in real data. We leave a more complete and realistic treatment of noise for future work. We also recall that we have not attempted to denoise the maps before computing statistics, a step which would likely compensate to some extent the loss of discrimination power seen in Figs.~\ref{fig:histograms_noisy} and \ref{fig:polar_fR5_noisy}. Finally, we do not show noisy results for $f_5(R)$ with $m_\nu=0.15~\mathrm{eV}$ that would correspond to Fig.~\ref{fig:polar_fR5_0.15eV}, since the discrimination efficiency does not exceed 10\% for any statistic at any scale or source redshift.
\end{appendix}

\end{document}